\documentclass[twocolumn]{aastex701}
\usepackage{xspace}

\usepackage{float}
\usepackage{enumitem}
\usepackage{amsmath}
\usepackage{longtable}

\graphicspath{{./}{.}}

\newcommand{\expname}{\texttt{EXP}\xspace}
\newcommand{\pyexp}{\texttt{pyEXP}\xspace}
\newcommand{\pynbody}{\texttt{pynbody}\xspace}
\newcommand{\zzero}{$z = 0$\xspace}
\newcommand{\lmass}{$\log(M/M_{\odot})$\xspace}
\newcommand{\lmstar}{$\log(M_{\star}/M_{\odot})$\xspace}

\newcommand{\msol}{$M_{\odot}$\xspace}

\newcommand{\logsfr}[1][]{$\log\,\mathrm{SFR}_{#1}$\xspace}
\newcommand{\logssfr}[1][]{$\log\,\mathrm{sSFR}_{#1}$\xspace}

\newcommand{\cyc}{$\zeta_{\rm SFR}$\xspace}
\newcommand{\psdwelch}{${\rm PSD}_{\rm Welch}$\xspace}

\newcommand{\galaxyname}[1]{\texttt{#1}\xspace}

\usepackage[dvipsnames]{xcolor}
\definecolor{gedablue}{RGB}{48,97,243}
\definecolor{gedared}{RGB}{255,0,0}
\definecolor{gedaorange}{RGB}{255,69,0}
\newcommand{\diskcolor}[1]{\textcolor{gedaorange}{#1}}
\newcommand{\nodiskcolor}[1]{#1}

\begin{document}

\title{Galactic Dynamics Drives Cyclic Star Formation in Dipole Dwarf Galaxies}

\author[0000-0003-1509-9966]{Robel Geda}
\affiliation{Department of Astrophysical Sciences, Princeton University, 4 Ivy Lane, Princeton, NJ 08540, USA}
\email{robel@princeton.edu}

\author[0000-0001-7831-4892]{Akaxia Cruz}
\affiliation{Center for Computational Astrophysics, Flatiron Institute, New York, NY 10010, USA}
\affiliation{Department of Physics, Princeton University, Princeton, NJ 08540, USA}
\affiliation{Department of Astrophysical Sciences, Princeton University, 4 Ivy Lane, Princeton, NJ 08540, USA}
\email{}

\author[0000-0002-0372-3736]{Alyson Brooks}
\affiliation{Center for Computational Astrophysics, Flatiron Institute, New York, NY 10010, USA}
\affiliation{Department of Physics \& Astronomy, Rutgers, the State University of New Jersey, Piscataway, NJ, 08854, USA}
\email{}

\author[0000-0002-5612-3427]{Jenny E. Greene}
\affiliation{Department of Astrophysical Sciences, Princeton University, 4 Ivy Lane, Princeton, NJ 08540, USA}
\email{}

\author[0000-0001-5522-5029]{Carrie Filion}
\affiliation{Center for Computational Astrophysics, Flatiron Institute, New York, NY 10010, USA}
\email{}

\author[0000-0003-0872-7098]{Adrian~M.~Price-Whelan}
\affiliation{Center for Computational Astrophysics, Flatiron Institute, New York, NY 10010, USA}
\email{}

\author[0000-0001-5510-2803]{Thomas R. Quinn}
\affiliation{Astronomy Department, University of Washington, Box 351580, Seattle, WA, 98195-1580, USA}
\email{}

\author[0000-0001-8745-0263]{James Wadsley}
\affiliation{Department of Physics and Astronomy, McMaster University, 1280 Main Street West, Hamilton, Ontario, L8S 4M1, Canada}
\email{}

\author[0000-0002-9642-7193]{Ben Keller}
\affiliation{Department of Physics and Materials Science, University of Memphis, 3720 Alumni Avenue, Memphis, TN 38152, USA}
\email{}

\begin{abstract}
Dwarf galaxies are dark matter-dominated systems that are sensitive to feedback and display bursty star formation histories. This makes them ideal probes of the processes that regulate star formation. This work investigates the timescales of star formation rate (SFR) variability in isolated dwarf galaxies using high-resolution cosmological zoom-in simulations of $50$ dwarf galaxies drawn from the Marvelous Massive Dwarfs and Extended Marvelous Dwarfs simulation suites ($7.3 <$ \lmstar $< 9.1$). We measure star formation histories in $10$ Myr bins and characterize their variability with power spectral densities (PSDs). At the population level, the PSDs of both disk and non-disk galaxies follow a broken power-law, with a break timescale near $100$ Myr. Individually, our galaxies show both smooth and bursty star formation. A subset of disk galaxies show quasi-periodic SFR oscillations about their mean, which we refer to as cyclic star formation. The dominant timescale of these oscillations scales linearly with half-light radius and is approximated by the orbital period at that radius. Cyclic galaxies have Magellanic Spiral (Sm) morphologies and host lopsided dipole gas disks. In these systems, the bulk of the gas in the lopsided disk moves between its maximum and minimum distance from the galactic center, compressing the star-forming gas once per radial orbit and driving the SFR cycle. Cyclic star formation adds a narrow Lorentzian component to the PSD, peaked at radial orbital frequency of the disk ($\sim 200-500$ Myr), which is comparable to burst timescales inferred from observations.
\end{abstract}

\section{Introduction}
\label{sec:Introduction}

Dwarf galaxies (\lmstar $< 9.5$) are dark matter dominated systems with shallow potential wells, which leaves them sensitive to feedback, mergers, and environment \citep[e.g.,][]{1997PASP..109..745B, 2019ApJ...886...74M, 2022NatAs...6..659B, 2022ApJ...940....8M, 2025MNRAS.541.2180R, Cruz2026}. These processes operate on different timescales and leave their imprints on the star formation histories (SFH) that dwarf galaxies display. Disentangling the contribution of each process requires constraining the variability in star formation rates (SFR), and the timescales over which they take place.  

Observations find that dwarfs form stars in bursts \citep{2009ARA&A..47..371T, 2010ApJ...724...49M, 2010ApJ...721..297M, 2011ApJ...739....5W, 2014MNRAS.441.2717K, 2014ApJ...789...96A, 2022MNRAS.511.4464A, 2023ApJ...950..125M}, and constrain that variability through tracers averaging over different timescales \citep[e.g.,][]{Weisz2012}. H$\alpha$ and far ultraviolet (FUV) emission trace star formation over $\sim 10$ Myr and $100$ Myr, respectively. \cite{Weisz2012} and \cite{Emami2019} both argue for significant bursts in low-mass galaxies (\lmstar $\lesssim 9$), with burst timescales of $\sim 100$--$300$ Myr that lengthen with stellar mass. With the advent of JWST there is now also evidence for burstiness on $\sim 100$ Myr timescales in cosmic-noon dwarfs \citep{Mintz2026} and in typical galaxies at cosmic dawn \citep[e.g.][]{2024Natur.629...53L, 2024MNRAS.533.1111E, 2025ApJ...979..193C, 2025MNRAS.544.4551S}. At a population level, and at a given epoch, galaxies observed at different phases of a burst cycle scatter about the star forming main sequence (MS, $M_\star$--SFR relation), with dwarf galaxies occupying a mass regime where the MS scatter grows \citep[e.g.,][]{KadoFong2024}.

On the theoretical side, the extended regulator model provides a physical framework for this variability \citep{Caplar2019, Tacchella2020, Iyer2024}. In this model the SFR is set by a gas reservoir that is fed by inflow and drained by star formation and outflows \citep{Tacchella2020}. Fluctuations in SFR due to fluctuations in inflow are damped on short timescales because the gas reservoir averages out fluctuations shorter than its depletion time. This leaves the short-timescale burstiness to processes internal to the galaxy, such as feedback cycles, the formation and destruction of individual molecular clouds, and dynamical disturbances. These short-timescale processes all act on roughly a galactic dynamical time and leave the SFR uncorrelated beyond it. The two families of processes result in a SFH power spectral density (PSD) with three power-law regimes separated by breaks around the galactic dynamical time and at the timescale on which the reservoir responds to inflow. The PSD rises at timescale below the galactic dynamical timescale, is flat between the dynamical timescale and the inflow timescale \citep{Iyer2020}, and steeply rises again above it \citep{Tacchella2020}. The flat regime arises because the two timescales are widely separated, leaving a range over which neither process imprints any structure (white noise). \cite{Iyer2020} measured the PSDs of SFHs across a wide range of simulations, with different galaxy formation or feedback models, and found them to be well described by broken power-laws, with slopes of $\sim 1$ to $3$ below the dynamical break and a flattening towards longer timescales.

Simulations of dwarf galaxies with different prescriptions for baryonic physics predict bursty star formation in low-mass systems, but differ on the timescales and mechanisms that play a role \citep{Stinson2007}. In the \texttt{FIRE-2} simulations, dwarf galaxies ($7 \lesssim $ \lmstar $\lesssim 9.6$) form stars in bursts on $\sim 100$ Myr timescales, their cold gas is dispersion dominated, and long lived stellar disks are rare \citep{ElBadry16, Hopkins2023, Benavides2025}. SFR variability occurs because, after each burst, star forming gas is expelled to larger radii causing a period of relatively low star formation. The gas returns to the central region due to gravitation, triggering the next burst. This causes ``breathing modes" \citep{Stinson2007} that couple the dwarf galaxy's size to its sSFR \citep{ElBadry16, patel18}. In contrast, the Marvelous Massive Dwarfs (MMD) simulations produce compact galaxies \citep{Cruz2026}, which can grow in size due to secular buildup of stellar mass in angular-momentum supported disks at relatively stable sSFRs \citep{Keith2025, Geda2026}. 

In this work, we examine high spatial and mass resolution cosmological zoom-in simulations of $50$ isolated dwarf galaxies ($7.3 <$ \lmstar $< 9.1$) from the MMD and Extended Marvelous Dwarfs suites \citep{Cruz2026}. These galaxies lie in the regime where bursty star formation is expected and where disks begin to form, making them an ideal and timely sample for this study. We apply the PSD framework introduced by \cite{Tacchella2020} to probe SFR timescales and identify their physical causes. In particular, we constrain the overall PSD of our sample as a function of morphology and probe galaxies that display periodic SFR variability. We focus our analysis on timescales below the inflow timescale (fluctuation timescales below $<1$ Gyr), which set the overall SFR variability \citep{Caplar2019, Tacchella2020, Iyer2024}. 

This paper is organized as follows. Section \ref{sec:Methods} describes the simulations, the SFH and PSD measurements, the basis function expansions, the sample, and the high cadence reruns. Section \ref{sec:Power Spectral Density Distributions} presents the PSDs of our simulated galaxy sample, separates bursty from cyclic SFHs, and examines how the dominant timescale scales with galaxy properties. Section \ref{sec:Cyclic SFH and Disk Dynamics} links periodic star formation to the morphology and orbital dynamics of gas disks. Section \ref{sec:Discussion} places cyclic PSDs in the context of SFH models and discusses their drivers and observational signatures. 

\section{Methods}
\label{sec:Methods}

\subsection{Simulations}
\label{sec:Simulation}

The Marvelous Massive Dwarfs \citep[][]{Cruz2026, Keith2025, 2025MNRAS.541.2180R, 2025ApJ...993..230P, Geda2026} combined with the Extended Marvelous Dwarf \citep{Cruz2026} simulations, referred to as ``Massive Dwarfs" throughout this paper, are a suite of high resolution cosmological zoom-in simulations of individual dwarf galaxies from the \texttt{Romulus25} simulation volume \citep[$(25\ {\rm Mpc})^3$,][]{Tremmel2017}. The zoom-in simulations are run up to \zzero and target isolated star-forming dwarf galaxies spanning $7.0 \leq$ \lmstar $\leq 9.1$ at $z = 0.1$. We consider a galaxy isolated if no companion with \lmstar $\geq 10$ resides within $1.5$ Mpc at $z=0.1$ \citep{2012ApJ...757...85G}. 
 
The zoom-in simulations were carried out with the N-body$+$SPH code \texttt{ChaNGa} \citep{Jetley2008, Jetley2010, Menon2015}, incorporating and expanding on hydrodynamic methods from \texttt{GASOLINE2} \citep{Wadsley2004, Wadsley2017}. Within the zoom-in region, the gravitational force softening length is $87$ pc, with initial particle masses of $3.3 \times 10^{3}$~\msol, $994$~\msol, and $1.8 \times 10^{4}$~\msol for gas, stars, and dark matter, respectively. Full particle data outputs are saved at $\sim$$300$ Myr cadence. We rerun the last $\sim 1$ Gyr of a subset of galaxies of dynamical interest at higher cadence ($\sim$$10$ Myr) which we describe in Section \ref{sec:high_cadence_reruns} and introduce in Section \ref{sec:Dynamics of Cyclic Galaxies}. The simulations adopt a Planck 2013 cosmology \citep[$\Omega_{m}=0.3086$, $\Omega_{\Lambda}=0.6914$, $h = 0.67$, $\sigma_{8}=0.77$,][]{2014A&A...571A..16P}.

Star formation occurs stochastically in gas that has cooled below $1000$ K and reached number densities above $0.1\ m_h\ {\rm cm}^{-3}$ \citep{2012MNRAS.425.3058C}. The formation probability $p_*$ follows a modified Schmidt law \citep{1992ApJ...391..502K, 2006MNRAS.373.1074S} scheme in which the molecular hydrogen fraction  $X_{{\rm H}_2}$ regulates when stars can form \citep{2012MNRAS.425.3058C}:

\begin{equation}
p_* = \frac{m_{\rm g}}{m_{\rm s}} \left(1 - {\rm exp}\left[- 
\frac{c^{*}_{0} X_{{\rm H}_2} \Delta t}{t_{\rm ff}} \right]\right),
\end{equation}

\noindent where $c^{*}_{0} = 0.1$ is the star formation efficiency, $\Delta t$ ($\sim1$ Myr) is the timestep, and $t_{\rm ff}$ is the local free-fall time. In practice, this molecular hydrogen dependence drives star formation into gas well above the density floor ($n \gtrsim 100\ m_h\ {\rm cm}^{-3}$, where $m_h$ is the mass of hydrogen). Cooling follows \cite{2012MNRAS.425.3058C}, with metal-line cooling and metal diffusion from \cite{2010MNRAS.407.1581S}, and non-equilibrium ${\rm H}_{2}$ chemistry tracked throughout. All stellar masses and SFRs are derived assuming a \citet{2001MNRAS.322..231K} IMF.  SN feedback follows the multiphase superbubble model of \cite{2014MNRAS.442.3013K}, depositing $10^{51}$ erg per event into a temporarily two-phase medium. Thermal conduction regulates mass flow from the cold to the hot phase via evaporation, with each phase carrying independent mass, energy, and non-equilibrium ionization states, and cooling rates computed separately under pressure equilibrium \citep{2014MNRAS.442.3013K}.
 
The Massive Dwarfs include BH seeding and growth, with a seed mass of $10^{5}$\msol, lower than achieved in \texttt{Romulus25} and set by the resolution of dark matter to properly resolve dynamical orbits. The BHs are seeded according to high-redshift gas conditions (high density and low gas temperature regions, with $n >1.5 \times10^4$~cm$^{-3}$ and $T < 2 \times 10^4$~K), motivated by direct collapse seeding models \citep{Wise2019, Regan2020, Regan2023}, rather than assuming a constant halo occupation fraction of unity \citep{Tremmel2015, Tremmel2017, 2019MNRAS.482.2913B}. To suppress fragmentation into stars, BH formation is restricted to gas particles with low metallicity ($Z < 10^{-4}$) and low molecular hydrogen fraction ($X_{\mathrm{H}_2} < 10^{-4}$). Accretion onto BHs uses a boosted Bondi scheme \citep{Booth2009}, and the associated superbubble feedback \citep{Tremmel2015, Tremmel2017} follows \cite{2014MNRAS.442.3013K}.

Halo finding, centering, and merger tree construction follow the methods described in \cite{Geda2026}. Briefly, we use Amiga's Halo Finder \citep[AHF,][]{2001MNRAS.325..845K, 2004MNRAS.351..399G} to identify halos and bound particles, with galaxy shrinking-sphere centers \citep[SSC,][]{2003MNRAS.338...14P} and orientations determined using \pynbody\footnote{\url{https://github.com/pynbody/pynbody}}. Halo masses are defined following \cite{1998ApJ...495...80B}, as the mass enclosed within a sphere whose mean density equals the redshift dependent virial overdensity derived from the spherical collapse model. Halo merger tree construction and sub-halo  tracking across snapshots follows the ``voter and graveyard'' method \citep{2022MNRAS.510..959I, 10.1093/mnras/staf016}. 

To compare with observational stellar masses derived from photometric measurements, we correct the sum of the star particle masses by applying a factor of $0.6$ to all galaxy stellar masses as described in \cite{2013ApJ...766...56M}, who show that photometrically inferred stellar masses recover approximately $60\%$ of the total stellar mass in star particles for galaxies in our mass scale. At each snapshot, galaxies are oriented face-on using the angular-momentum (AM) vector of bound gas within $5$ kpc of the center. We measure the galaxy's half light radius ($R_{\rm e}$) in the Johnson $R$ band in the face-on configuration at each snapshot using \pynbody. The full list of simulations before sample selection can be found in Table \ref{table:overview} in Appendix \ref{appendix:Table of Galaxies}.

\subsection{Star Formation Rate Measurements}
\label{sec:Star Formation Rates Measurements}

We examine how each galaxy assembled its \zzero stellar mass through its SFH.  Every star particle in the simulations contains a record of its formation time and initial mass, resulting in a maximum temporal resolution finer ($\sim 1$ Myr) than our fiducial snapshot resolution ($\sim 300$ Myr). From these records, we build the SFH of each galaxy's \zzero stellar population by constructing histograms of the formation times weighted by star particle initial mass. Motivated by observational measurements of instantaneous star formation rates (SFRs) derived from H$\alpha$ and FUV emissions \citep[e.g.,][]{Weisz2012,Emami2019}, we adopt temporal bins of $\Delta t= 10$ Myr (${\rm SFR}_{10}$) and $\Delta t=100$ Myr (${\rm SFR}_{100}$) respectively. We apply the same time range ($1.5-13.8$ Gyr) and bins ($\Delta t$) for all galaxies when constructing their SFHs.

Though merger trees allow us to separate in-situ star formation from ex-situ accretion on an individual star particle basis, we consider every star particle in the galaxy at \zzero when constructing its SFH. We adopt this definition because observationally inferred SFHs recover the cumulative stellar population with little means of isolating in-situ star formation rates across cosmic time. Furthermore, we note that in practice the difference between aggregate and in-situ SFR for our simulated galaxies is negligible after $\sim 6$ Gyr of cosmic time because mergers that introduce significant new stellar mass decline substantially at later epochs. In particular, after $6$ Gyr of cosmic time, satellites account for only $0.5\%$ of the final stellar mass on average, with a standard deviation of $1\%$ and a maximum of $8\%$ (a single low mass outlier) across the sample.

\subsection{Power Spectral Density Measurements}
\label{sec:Power Spectral Density Measurements}

To quantify the characteristic timescales of SFR variability, we adopt the PSD formalism proposed by \cite{Caplar2019}. Specifically, we compute the PSD of each galaxy's log star formation time series \citep[which is log-normal,][]{Iyer2020} using the standard discrete Fourier transform, computed via Fast Fourier transform (FFT). The FFT yields the full complex spectrum from which we extract PSDs at each positive frequency ($f_k$,  the $k$-th frequency component) or corresponding fluctuation timescale ($\tau_k \equiv 1/f_k$). Throughout this work we use $f_k$ exclusively rather than angular frequency ($\omega_k = 2\pi f_k$) to avoid confusion. 

In addition to the PSD, we make use of both the amplitude ($\mathcal{A}_k$) and power ($\mathcal{P}_k$) spectra in quantifying the sizes and significance of oscillations and fluctuations.\footnote{For a standard FFT with spacing $\Delta f$: $\text{PSD}_k = \frac{\mathcal{P}_k}{\Delta f} = \frac{\mathcal{A}_k^2}{2 \Delta f}$} Writing the time series as $\delta(t)$, we decompose it into a sum of oscillations:

\begin{equation}
\label{eq:delta(t)}
    \delta(t) = \sum_k \delta_k(t), \qquad
    \delta_k(t) = \mathcal{A}_k\cos\left(2\pi f_k t + \phi_k\right),
\end{equation}
\noindent where $\delta_k(t)$ is the contribution of frequency $f_k$, with amplitude $\mathcal{A}_k$ and phase $\phi_k$.

Motivated by \cite{Iyer2020}, we adopt Welch's method \citep{1967ITAuE..15...70W}, which reduces spectral density variance by averaging over overlapping windowed segments. The adopted parameters are listed in Table \ref{tab:welch_params}; in particular, setting the segment length equal to the full time series recovers the frequency resolution of the standard FFT while retaining the benefits of windowing and de-trending \citep{1978IEEEP..66...51H}.

\subsection{Basis Function Expansions}
\label{sec:Basis Function Expansions}

In the same way that we use PSDs to characterize the timescales that contribute to a galaxy's SFH, we use basis function expansions (BFEs) to characterize the spatial structure (i.e., shape) of each galaxy's stellar and gas distributions. BFEs represent the density and potential fields of a galaxy as a linear combination of orthogonal functions. Specifically, we use the empirical, bi-orthogonal basis function approach of \cite{Weinberg1999} and \cite{Petersen2022}, as implemented in the \expname package and its Python interface \pyexp\footnote{\url{https://github.com/EXP-code/EXP}} \citep{Petersen2024}. The expansion coefficients provide a compact summary of the mass distribution. In this work we use the expansions only to quantify how far the stellar and gas disks depart from axisymmetry (Section \ref{sec:Morphologies of Cyclic Galaxies}). Because the coefficients also describe the potential, the same expansions can be used to compute the radial and azimuthal orbital frequencies of these non-axisymmetric disks directly, which we plan to explore in future work.

\subsubsection{Cylindrical Basis}
\label{sec:Cylinder Basis}

In this work, we make use of the \texttt{cylinder} basis of \expname. \expname represents disk densities through defining a rotation from a spherical basis into a user-defined disk density \citep{Weinberg1999, Petersen2022}. The resulting basis functions are meridional planes in $R$ and $z$ (\lq radial orders’, denoted with $n$), multiplied by $e^{im\phi}$ (\lq azimuthal order', denoted $m$), weighed by the (complex) coefficients, $C_{mn}$. We include both vertically symmetric and vertically anti-symmetric terms in our expansion. The lowest-order function represents the target density, such that the azimuthal harmonic $m{=}1$ terms describe a lopsided disk (i.e. dipole), $m{=}2$ terms describe bars or two-armed spirals, et cetera. We adopt as the target density an exponential disk with an isothermal vertical profile \citep{1942ApJ....95..329S}:

\begin{equation}
\label{eq:target_density}
    \rho_{0}(R, z) = \frac{M_{\rm d}}{4 \pi a^{2} h}\,
    \exp\left(-\frac{R}{a}\right) {\rm sech}^{2}\left(\frac{z}{h}\right),
\end{equation}

\noindent where $M_{\rm d}$ is the total disk mass, $a$ the scale length, and $h$ the scale height. Equation \ref{eq:target_density} is not fit to the data. It defines the zeroth-order, equilibrium state of the expansion, and the higher-order terms describe departures from it, such as lopsided or barred disks. 

We center each galaxy on its shrinking-sphere center and rotate it face-on using the AM vector of the bound gas within $5$ kpc of the center (Section \ref{sec:Simulation}). We build a separate basis for each galaxy, setting $a$ and $h$ to the stellar half-mass radius and half-mass height. We then compute the coefficients of the stellar and cold gas ($\log(T/{\rm K}) < 3.8$) components from particles within $20$ kpc of the center, which is a volume that contains the bulk of the disk in every galaxy in our sample. The full \texttt{cylinder} basis parameters are listed in Table \ref{tab:exp_params} in Appendix \ref{appendix:BFE Parameters}.

\begin{figure*}[t]
\centering
\includegraphics[width=\textwidth]{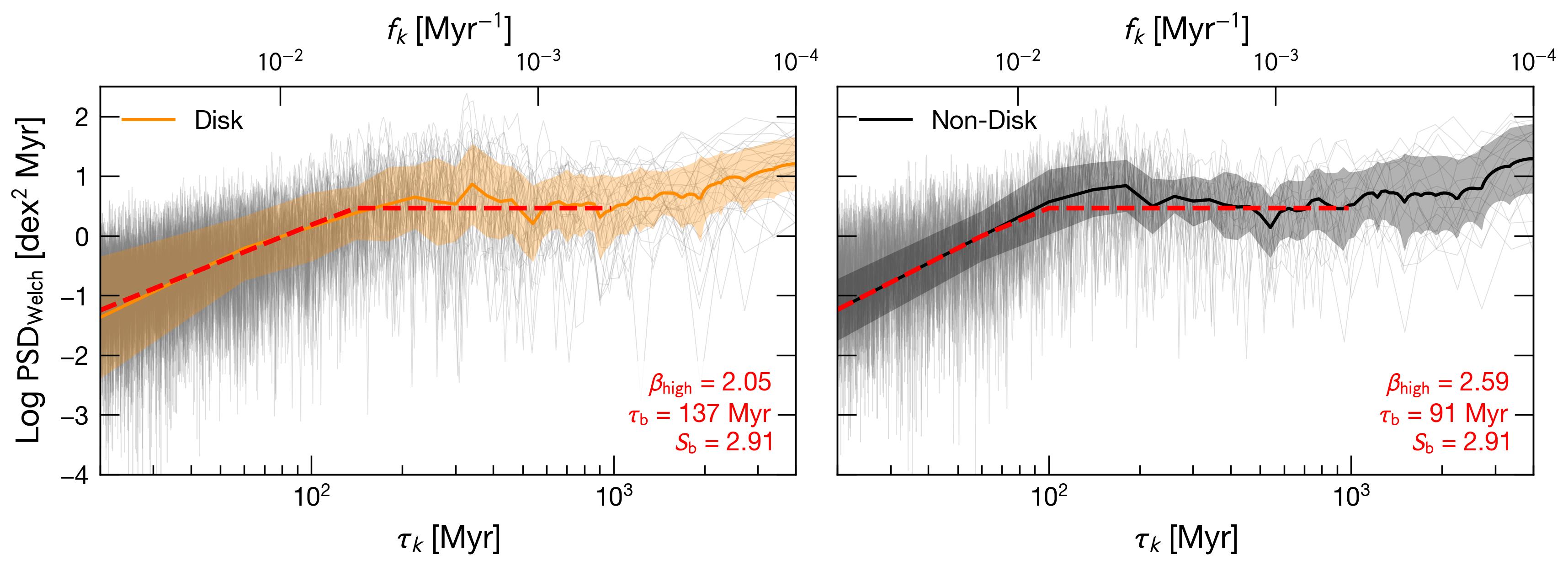}
\caption{\label{fig:sfr_power_spec}
Welch power spectral density of \logsfr[10] as a function of fluctuation timescale $\tau_k$ (bottom axis) and frequency $f_k$ (top axis) for disk (left) and non-disk (right) galaxies. Gray lines show the \psdwelch of individual galaxies. Solid lines show the mean of each subsample (for $\tau_k \leq 1$ Gyr) and shaded bands show the $1\sigma$ scatter. Red dashed lines show broken power-law fits to the mean \psdwelch and $1\sigma$ scatter, with the high-frequency slope $\beta_{\rm high}$, break timescale $\tau_{\rm b}$, and break amplitude $S_{\rm b}$ listed in each panel. SFHs are constructed in $\Delta t = 10$ Myr bins from the formation time of all star particles residing in each galaxy at $z=0$. The two subsamples show broadly similar distributions, with a characteristic broken power-law that resembles a damped random walk ($\beta_{\rm low} = 0,\ \beta_{\rm high} = 2$). 
}
\end{figure*}

\subsubsection{Coefficient Powers and Amplitudes}
\label{sec:Coefficient Power}

We summarize each expansion by the power in each azimuthal order $m$, summed over radial order $n$:

\begin{equation}
\label{eq:power}
    P_m = \sum_{n=0}^{n_{\rm max}-1} \left| c_{mn} \right|^{2},
\end{equation}

\noindent where $c_{mn}$ are the expansion coefficients. We measure deviations from axisymmetry using the amplitude in each harmonic normalized by the axisymmetric amplitude:

\begin{equation}
\label{eq:normalized_amplitude}
    \frac{A_{m}}{A_{m=0}} \equiv  \left( \frac{P_m}{P_0} \right)^{1/2}.
\end{equation}

\noindent We report $A_{m=1}/A_{m=0}$ as a measure of lopsidedness (the dipole amplitude). The power in each harmonic is an approximate summary of the expansion, but it is sufficient for the comparisons made in this work.

\begin{figure*} 
\includegraphics[width=\textwidth]{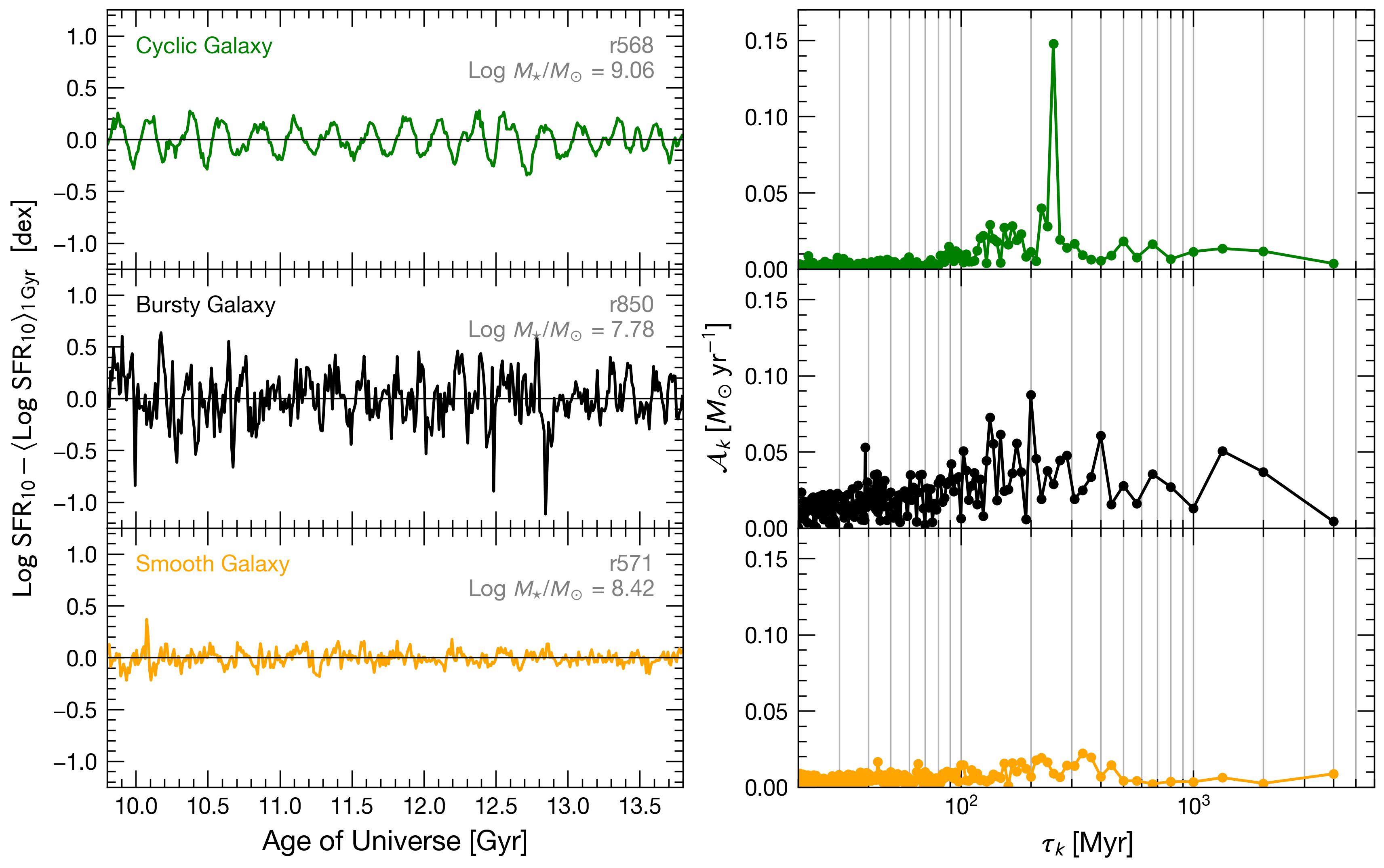}
\centering
\caption{\label{fig:bursty_vs_cyclic}
Comparison of a cyclic (green, disk galaxy), bursty (black, non-disk galaxy), and smooth (orange, disk galaxy) SFH over the last $4$ Gyr prior to \zzero ($\Delta t = 10\,\mathrm{Myr}$ bins). \textit{Left:} $\log\,\mathrm{SFR}$ after subtracting a 1~Gyr boxcar moving mean. \textit{Right:} FFT amplitude spectrum of the moving mean subtracted SFH as a function of fluctuation timescale. The cyclic galaxy exhibits a sharply peaked spectrum dominated by a single mode, indicating quasi-periodic SFR oscillations at a well-defined timescale. In the second row, though the bursty galaxy spans a wider mean subtracted SFR distribution, it lacks a dominating timescale that imprints structure into the SFH (stochastic and aperiodic variability). The smooth SFH in the bottom panels also shows stochastic and aperiodic variability, but lacks strong variability (low amplitudes across its frequency spectrum).}
\end{figure*}

\subsection{Simulated Galaxy Sample}
\label{sec:Galaxy Sample}

\subsubsection{Final Sample}
\label{sec:Final Sample}
Our simulation suite consists of $53$ galaxies, with $41$ from the Massive Dwarfs and $12$ from the Extended Marvelous Dwarfs suite. All galaxies across both suites share the same simulation setup detailed in Section \ref{sec:Simulation}. We exclude from our final sample two galaxies, \galaxyname{r615} and \galaxyname{r918}, which recently experienced a major merger. We exclude an additional galaxy, \galaxyname{r1223}, from the analysis as its star formation history was not available. Excluding the three galaxies, the final sample consists of $50$ galaxies with a total mass range of $10.2 \leq$ \lmass $\leq 11.0$ and stellar mass range of $7.3 <$ \lmstar $< 9.1$.

\subsubsection{Disk Galaxy Classification}
\label{sec:Disk Galaxy Classification}
We adopt the kinematic disk classification of \cite{Cruz2026} and \cite{Geda2026}. For each galaxy, we compute the circularity parameter $\hat{j}_z = j_z/|\vec{j}|$, where $\vec{j}$ is the specific angular momentum of a stellar particle and $j_z$ is its component along the $z$ axis. A galaxy is classified as disk-dominated (hereafter a disk galaxy) if the median $\hat{j}_z$ of its stellar population exceeds $0.5$, and as non-disk otherwise. Constructing subsamples of our final sample on this basis, we identify $32$ disk galaxies ($64\%$) and $18$ non-disk galaxies ($36\%$). Both the disk and non-disk sub-samples span the full stellar mass range listed above, both with a mean stellar mass of \lmstar $=8.2$. 

\subsubsection{High-Cadence Snapshot Reruns}
\label{sec:high_cadence_reruns}
To study the orbital dynamics of our simulated galaxies in detail, we re-simulate the final $\sim1$ Gyr of $12$ galaxies of dynamical interest from the original suite, saving snapshots at a cadence of $\sim 10$ Myr. Each rerun restarts from the corresponding snapshot of the original simulation and evolves to \zzero with the same setup as the original suite (code, resolution, physics, and halo finder). We specifically begin the reruns only $1$ Gyr before \zzero to limit the growth of numerical divergence from the ``butterfly effect'' \citep{2019MNRAS.482.2244K, 2019ApJ...871...21G}.

\section{Power Spectral Density Distributions}
\label{sec:Power Spectral Density Distributions}

Galaxies experience star formation variability (fluctuations) on a range of timescales \citep{Caplar2019, Tacchella2020}. In particular, SFR variability occurs on timescales that correspond to the physical processes that govern star formation \citep[see Figure 1 of][for a summary of timescales]{Iyer2020}. In this section, we examine the PSDs of the SFHs of galaxies in our final sample to identify and understand important star formation timescales. First we compare the mean power spectra of the disk and non-disk sub-samples, to see whether morphology affects the distribution of fluctuation power across timescales in Section \ref{sec:The PSD of Disks versus Non-Disks}. In Section \ref{sec:Bursty versus Cyclic Star Formation} we examine the strength of the periodicity in SFR using the PSDs of individual galaxies. Lastly, we analyze the periods at which the spectra peak and examine how they scale with stellar mass and galaxy size in Section \ref{sec:Dominant Fluctuation Timescales}. Throughout this section, although we present the full spectrum, we will focus on physical processes that regulate SFR from $20$ Myr up to $1$ Gyr scales, which correspond to fluctuation timescales up to and including orbital dynamical timescales \citep{2019MNRAS.484..915M, 2019ApJ...879...11K, Tacchella2020, Iyer2020}. 

\begin{figure}[t]
\includegraphics[width=8.5cm]{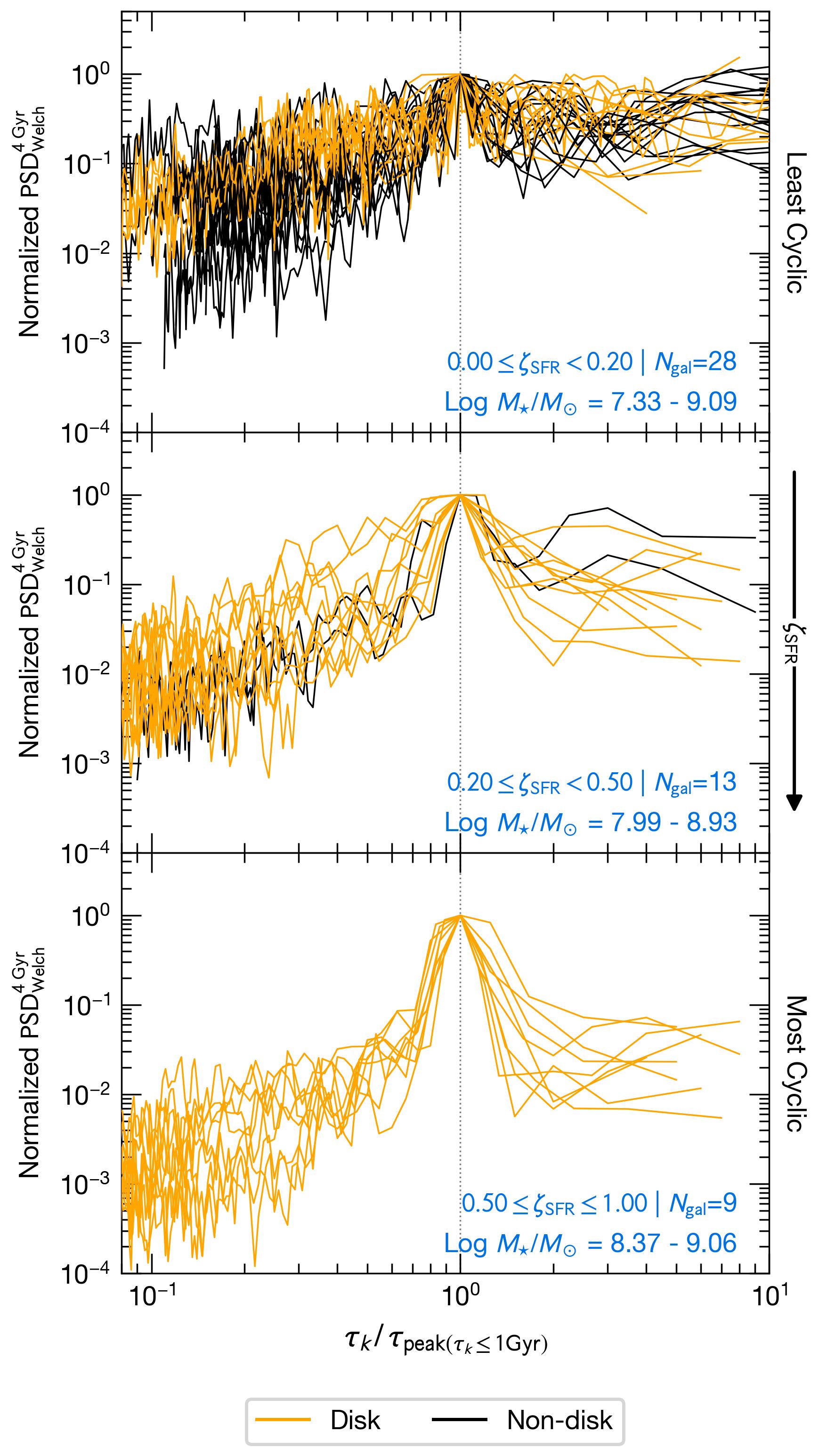}
\centering
\caption{\label{fig:norm_amp_spec_cyclic}
Normalized \psdwelch of the raw SFHs of all galaxies in our final sample during the last $4$ Gyr leading up to \zzero, split into three bins of \cyc (top to bottom, $0.00$--$0.20$, $0.20$--$0.50$, and $0.5$--$1.0$). Orange lines show disk galaxies and black lines non-disk galaxies. $\tau_{\rm peak}$ is the timescale of peak PSD power at $\tau_k \leq 1$ Gyr, while \cyc is measured from the amplitude spectrum of the mean-subtracted SFH. With increasing \cyc, the frequencies of cyclic SFH against the mean appear as larger bumps in the raw \psdwelch.}
\end{figure}

\subsection{The PSD of Disks versus Non-Disks}
\label{sec:The PSD of Disks versus Non-Disks}

Motivated by \cite{Iyer2020}, we start by examining the Welch PSDs
(\psdwelch), which averages the Fourier power computed from overlapping segments of a given SFH. This method is ideal for sub-sample comparisons since it suppresses random error on the power at each timescale, which a direct transform may suffer from. The Welch PSDs of individual galaxies in our final sample are computed using the full SFH of each galaxy binned at $10$ Myr intervals (i.e., $\log {\rm SFR_{10}}$). Section \ref{sec:Power Spectral Density Measurements} gives the details of how we compute \psdwelch.

\begin{figure*} 
\includegraphics[width=\textwidth]{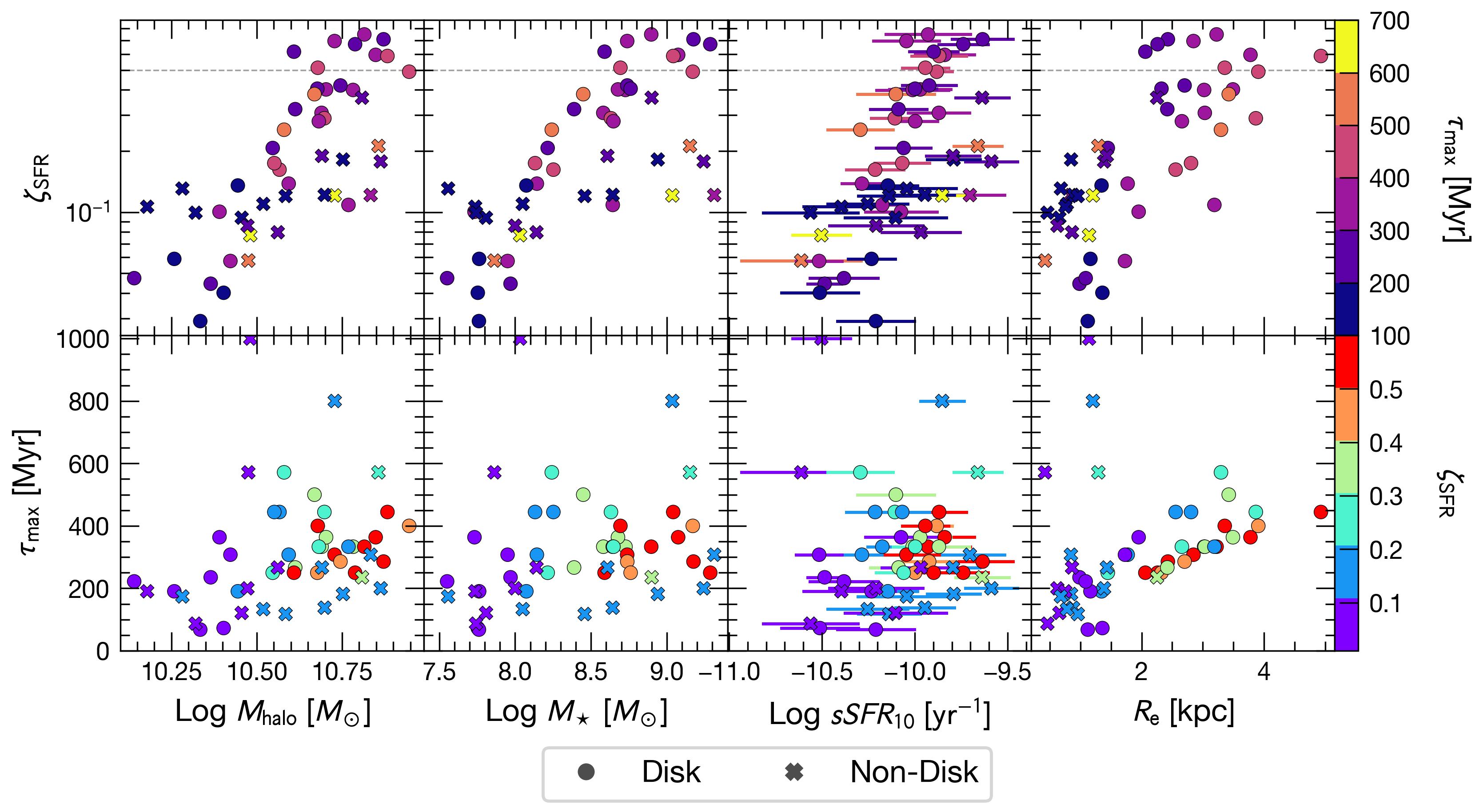}
\centering
\caption{\label{fig:peak_tau_vs_three_panels}
\cyc and $\tau_{\rm max}$ of the mean-subtracted SFHs of individual galaxies, measured over a window of $4$ Gyr ending at \zzero. Each column shows both quantities as a function of total mass, stellar mass, \logssfr[10], and half-light radius. The size and masses are measured at \zzero. The \logssfr[10] panels show the mean and $1$ standard deviation spread for the same mean-subtracted SFH over a window of $4$ Gyr. Both \cyc and $\tau_{\rm max}$ show a trend as a function of mass and \logssfr[10], particularly for the disk sample, though with some scatter. $\tau_{\rm max}$ shows a strong linear relation with size, especially for the disk-subsample.
}
\end{figure*}

Figure \ref{fig:sfr_power_spec} compares the mean \psdwelch and corresponding 1-sigma spread of the disk and non-disk sub-samples as a function of fluctuation timescale. Both sub-samples show similar PSDs, with broken power-law distributions for timescales below $\tau_k < 1$ Gyr, followed by an increase in power at longer timescales. To quantify the distribution below $\tau_k \leq 1$ Gyr, we use \texttt{astropy}\footnote{\url{https://docs.astropy.org/en/v7.1.1/modeling}} to fit the mean and 1 standard deviation spread of each sub-sample's distribution within that timescale using a broken power-law model ($S_{\rm BPL}$):

\begin{equation}
\label{eq:bpl}
    S_{\rm BPL}(f_k) = \left \{
                 \begin{array}{ll}
                   S_{\rm b}\ (f_k / f_{b}) ^ {-\beta_{\rm low}} & : f_{k} < f_{b} \\
                   S_{\rm b}\ (f_k / f_{b}) ^ {-\beta_{\rm high}} & :  f_{k} > f_{b} \\
                 \end{array}
               \right.,
\end{equation}

\noindent where $S_{\rm BPL}(f_k)$ is the PSD value at frequency $f_k$,  $f_{b}$ is the break frequency (break timescale is $\tau_{\rm b} \equiv 1/f_{\rm b}$), and $S_{\rm b} = S_{\rm BPL}(f_{b})$ is the normalization; ${\beta_{\rm low}}$ is the low frequency slope (longer timescales), while ${\beta_{\rm high}}$ is the high frequency slope (short timescales). 

Both sub-samples result in a flat slope on the long timescales (${\beta_{\rm low}} < 0.01$), with a reduced chi squared ($\chi^2_\nu$) of $1.53$ for disks and $1.96$ for non-disks. On short timescales, the disk results in a slope of ${\beta_{\rm high}} = 2.05$ and non-disk sub-samples results in ${\beta_{\rm high}} = 2.59$. The disk galaxy sub-sample results in a slightly higher break timescale ($\tau_{\rm b} = 137$ Myr) than non-disk sub-sample ($\tau_{\rm b} = 91$ Myr). We note that these broken power-law fits indicate that the overall shape of the PSD within $\tau_k \leq 1$ Gyr resembles a Lorentzian distribution centered at $f_k =0$ (i.e. $S(f_k) \propto 1 / (1+f_k^2)$). This distribution, which is also the power spectrum of a damped random walk, is characterized in log-space by Brownian noise ($\beta = 2$) on short timescales and white noise ($\beta = 0$) on long timescales \citep{Caplar2019, Iyer2020}. 

The overall average behavior of the MMD measured spectra, regardless of morphology, is in good agreement with both analytic expectations and previous simulation studies. \cite{Tacchella2020} showed (see their Section 3.2 and Appendix C) that star formation sustained by the stochastic formation and disruption of molecular clouds produce a damped random walk ($\beta_{\rm low} = 0,\ \beta_{\rm high} = 2$). Our break timescales of $\sim 100$ Myr fall within the range of characteristic timescales predicted by their framework. Our spectra also compare well with the PSDs measured by \cite{Iyer2020} across a large set of cosmological simulations with different galaxy evolution models, including the Marvel suite \citep{2021ApJ...906...96A} which use the same code and similar physics to the MMD simulation suite. In particular, the MMDs exhibit similar broken power-law slopes to those measured for the zoom-in simulations with varied galaxy formation models examined in \cite{Iyer2020} ($\beta_{\rm low} = 0,\ \beta_{\rm high} \approx 1.5\ \text{-}\ 2.5$). However there is also interesting information in the individual power spectra of MMD galaxies that show some of them have power at preferred frequencies, which we explore in the next section.

\subsection{Bursty versus Cyclic Star Formation}
\label{sec:Bursty versus Cyclic Star Formation}

In Section \ref{sec:The PSD of Disks versus Non-Disks} we analyzed the overall shape of our PSDs for our disk and non-disk sub-samples over cosmic time. In this section, we examine the variability of SFR against the mean (baseline) SFR for individual galaxies. We consider fluctuations ($\Delta$\logsfr) over the final $4$ Gyr of the simulation. To isolate and measure variability, we subtract out a $1$ Gyr moving average from the $\log {\rm SFR_{10}}$ time series (``mean-subtracted SFR", $\log \mathrm{SFR}_{10}(t)  - \langle \log \mathrm{SFR}_{10} \rangle_{1{\rm Gyr}}(t)$). This background subtraction acts as a filter that removes secular evolution of $\log \mathrm{SFR}_{10}$ (e.g., stellar mass of galaxy) and isolates the fluctuations about the slowly varying mean. 

While all galaxies in our final sample show similar amplitudes of variability ($\mathcal{A}_k$, defined in Section \ref{sec:Power Spectral Density Measurements}) around the mean, a subset of the sources show quasi-periodic behavior (correlated in time, resembling sinusoidal oscillations), with all the periodic variability occurring over a very narrow range of frequency. We will call these systems ``cyclic". We also see galaxies with unstructured strong fluctuations against the mean (uncorrelated in time, resembling white noise), which we will refer to as ``bursty". In their PSDs or amplitude spectra, cyclic galaxies are dominated by a single frequency or timescale while purely bursty galaxies display relatively flat slopes, lacking a favored frequency. 

Figure \ref{fig:bursty_vs_cyclic} presents the SFH of an example  cyclic (\galaxyname{r568}),  bursty (\galaxyname{r850}), and smooth (\galaxyname{r571}) galaxy, from top to bottom. The left panel shows the mean-subtracted SFH within the last $4$ Gyr of the simulation. For intuition, we include the corresponding discrete Fourier amplitude spectrum ($\mathcal{A}_k$), plotted linearly, in the right panel. The cyclic galaxy (\lmstar $= 9.06$ at \zzero) displays regular, nearly sinusoidal oscillations whose amplitude spectrum is dominated by a single narrow peak at a well defined timescale. In contrast the bursty galaxy (\lmstar $= 7.78$ at \zzero), despite spanning a wider $\Delta\log{\rm SFR}_{10}(t)$ range about the mean than the cyclic galaxy, shows irregular stochastic fluctuations, which results in an amplitude spectrum distributed broadly across timescales and no strongly preferred frequency. We note that not all cyclic galaxies remain cyclic for as long as \galaxyname{r568}. 

Several diagnostics can quantify how cyclic a SFH is relative to its mean. We define a metric based on the power in the sinusoid, $\zeta_{\rm SFR}$, which we call cyclicity. Using the discrete power spectrum of the mean-subtracted SFR, we adopt the ratio of the power in the peak bin ($\mathcal{P}_{\rm k_{max}}$) and its two adjacent bins, which recovers the power leaked out of the peak by the finite time baseline, to the power summed over all frequencies ($\sum_{k} \mathcal{P}_{k}$, excluding the DC term $k{=}0$) as our definition of the relative \cyc of our final sample \citep{1929RSPSA.125...54F}\footnote{Note that, though neither of the following are used in this work, Equation \ref{eq:zeta} is similar to Fisher's periodicity test \citep{1929RSPSA.125...54F}, and by the Wiener-Khinchin theorem, $\zeta_{\rm SFR}$ tracks the height of the first peak after the first rebound of the autocorrelation function for a sinusoidal signal.}:

\begin{equation} 
\label{eq:zeta}
    \zeta_{\rm SFR} = \frac{\displaystyle\sum_{k = k_{\rm max}-1}^{k_{\rm max}+1} \mathcal{P}_{k}}
                           {\displaystyle\sum_{k} \mathcal{P}_{k}} 
\end{equation} 

\noindent where higher values correspond to more cyclic SFHs. This definition of \cyc is convenient in that it is  dimensionless and invariant under a rescaling of the SFR $\mathcal{A}_k$.  It can be interpreted as the fraction of the total power, or equivalently the fraction of the SFR variance, carried by a dominant frequency. Lastly, for each galaxy in our final sample, we identify the dominant fluctuation timescale ($\tau_{\rm max}$) as the timescale at which the mean-subtracted spectrum peaks ($\tau_{\rm max} = \tau_k(\mathcal{P}_{\rm max}) = \tau_k(\mathcal{A}_{\rm max})$). For reference, the galaxies in Figure~\ref{fig:bursty_vs_cyclic} have $\zeta_{\rm SFR} = 0.68$ (\galaxyname{r568}), $0.09$ (\galaxyname{r850}), and $0.11$ (\galaxyname{r571}).

To probe and benchmark the properties associated with cyclic star formation, we consider galaxies in which the dominant timescale contributes over half of the SFR variance (\cyc $\gtrsim 0.5$). We refer to these galaxies as our ``most cyclic galaxies" ($9$ galaxies, $18\%$ of our final sample). Figure \ref{fig:norm_amp_spec_cyclic} shows the normalized \psdwelch of the raw $4$ Gyr SFH (i.e., not the mean-subtracted) of every galaxy in our final sample, split into three bins of \cyc. The x-axis is normalized by the timescale of peak \psdwelch power within $\tau_k \leq 1$ Gyr. The peak Welch timescale is near the discrete Fourier peak timescale of the mean-subtracted SFH ($\tau_{\rm peak}({\rm PSD}_{\rm Welch}) \approx \tau_{\rm max}$). \cyc itself is measured separately from the amplitude spectrum of the mean-subtracted SFH (which probes fluctuation against the mean). As \cyc increases, the cyclic fluctuations about the mean SFR emerge as a progressively more prominent bump in the raw \psdwelch. Most importantly, we note that the highest values of \cyc are occupied by disk galaxies. 

Galaxies move in and out of cyclic star formation episodes. The \psdwelch in Figure \ref{fig:norm_amp_spec_cyclic} identifies galaxies that experienced a  cyclic phase at some point within the $4$ Gyr window considered in this section (i.e., $\zeta_{\rm SFR} = \zeta_{\rm SFR}^{4{\rm Gyr}}$), and the bump is most prominent for those with the strongest and longest episodes. Over the full $13.8$ Gyr window of Figure \ref{fig:sfr_power_spec}, the same bumps are diluted in power. This is because cyclic star formation only lasts for part of a galaxy's history, typically at late times, so averaging over the full SFH mixes in long stretches with no cycling. Therefore \cyc is also a measure of how significant the cyclic episodes are relative to the SFH window considered. Furthermore, stacking multiple galaxies suppresses the bump further, since $\tau_{\rm max}$ varies from  galaxy to galaxy ($\tau_{\rm max} \sim 200$--$500$ Myr for \cyc $\gtrsim 0.5$ galaxies) and the  misaligned peaks average out. In Figure \ref{fig:sfr_power_spec}, this misalignment appears instead as slightly increased scatter in the stacked \psdwelch on $\tau_k \sim 100$ Myr timescales. To understand what sets $\tau_{\rm max}$ for individual galaxies, we analyze the distribution of \cyc and $\tau_{\rm max}$ as a function of stellar mass and galaxy size in  Section \ref{sec:Dominant Fluctuation Timescales}.

\subsection{Scaling of Dominant Fluctuation Timescales}
\label{sec:Dominant Fluctuation Timescales}

Figure \ref{fig:peak_tau_vs_three_panels} plots \cyc and $\tau_{\rm max}$ against each galaxy's \zzero total mass, stellar mass, ${\rm sSFR}_{10}$, and size ($R_{\rm e}$), with the disk and non-disk sub-samples shown with circle and ``x" markers, respectively. We note that for galaxies with low \cyc values or that lack a dominating frequency (flat spectra), $\tau_{\rm max}$ can be poorly constrained due to shot noise (especially in the non-disk sub-sample). Despite this limitation, we include all $\tau_{\rm max}$ in this section for completeness. Table \ref{table:pearson_corr} shows the Pearson correlation coefficients \citep{1896RSPTA.187..253P} for the log-log distributions in Figure \ref{fig:peak_tau_vs_three_panels}.

\begin{table}[ht]
\centering
\caption{Pearson correlation coefficients ($r$) and $p$-values between galaxy properties ($X$) and SFH periodicity measures ($Y$), shown in Figure \ref{fig:peak_tau_vs_three_panels}, computed in log-log space for disk and non-disk galaxies.}
\label{table:pearson_corr}
\begin{tabular}{llcccc}
\hline \hline
 & & \multicolumn{2}{c}{Disk ($N=32$)} & \multicolumn{2}{c}{Non-disk ($N=18$)} \\
\cline{3-4} \cline{5-6}
\multicolumn{1}{c}{$X$} & \multicolumn{1}{c}{$Y$} & $r$ & $p$ & $r$ & $p$ \\
\multicolumn{1}{c}{(1)} & \multicolumn{1}{c}{(2)} & (3) & (4) & (5) & (6) \\
\hline
$\log M_{\rm halo}$ & $\log \zeta_{\rm SFR}$ & $0.86$ & $<0.001$ & $0.60$ & $0.009$ \\
$\log M_\star$ & $\log \zeta_{\rm SFR}$ & $0.89$ & $<0.001$ & $0.63$ & $0.005$ \\
$\log sSFR_{10}$ & $\log \zeta_{\rm SFR}$ & $0.85$ & $<0.001$ & $0.78$ & $<0.001$ \\
$\log R_{\rm e}$ & $\log \zeta_{\rm SFR}$ & $0.78$ & $<0.001$ & $0.78$ & $<0.001$ \\
\hline
$\log M_{\rm halo}$ & $\log \tau_{\rm max}$ & $0.52$ & $0.003$ & $0.28$ & $0.256$ \\
$\log M_\star$ & $\log \tau_{\rm max}$ & $0.45$ & $0.010$ & $0.28$ & $0.257$ \\
$\log sSFR_{10}$ & $\log \tau_{\rm max}$ & $0.33$ & $0.061$ & $0.05$ & $0.848$ \\
$\log R_{\rm e}$ & $\log \tau_{\rm max}$ & $0.74$ & $<0.001$ & $0.30$ & $0.232$ \\
\hline
\end{tabular}
\end{table}

\begin{figure*} 
\includegraphics[width=\textwidth]{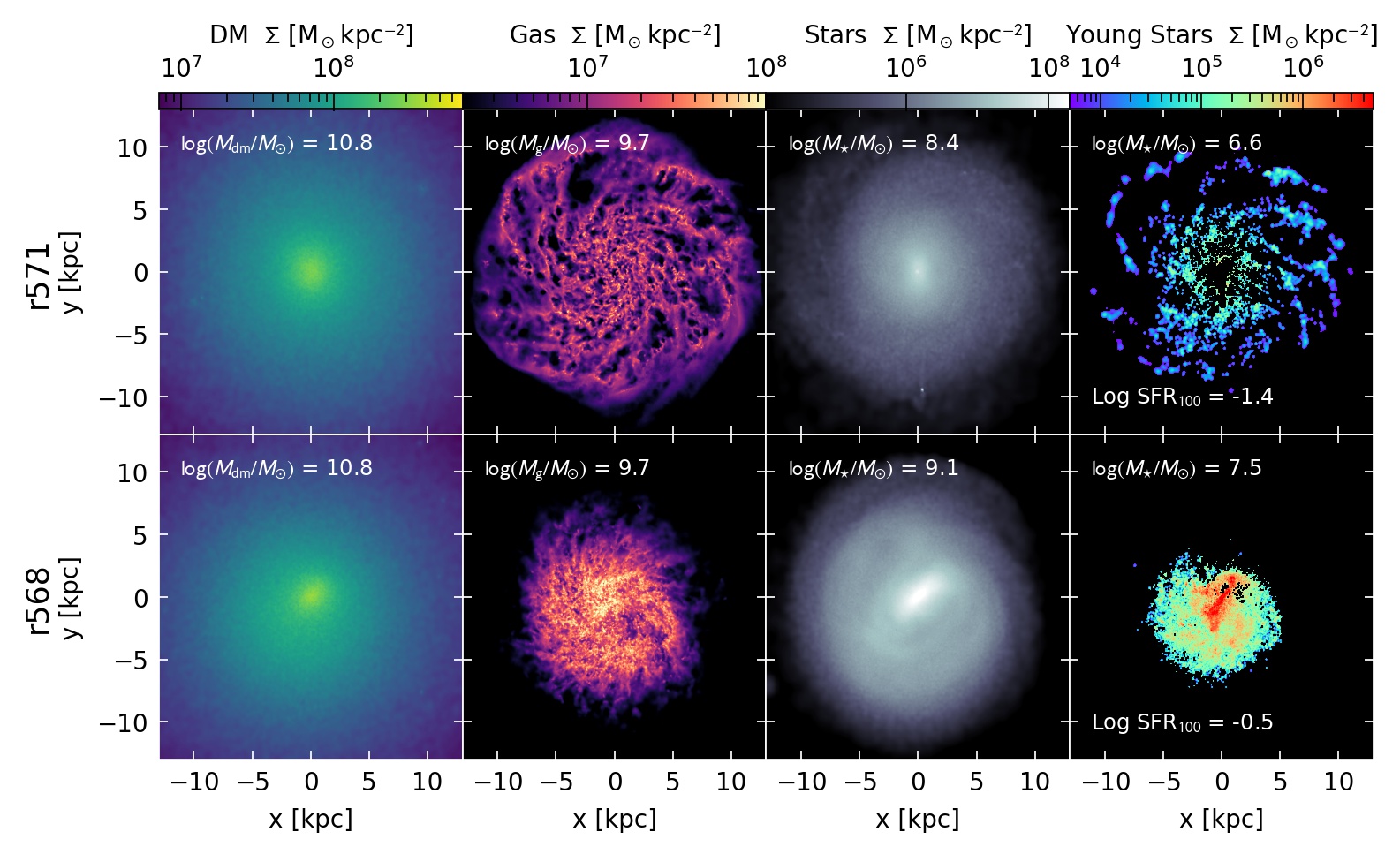}
\centering
\caption{\label{fig:two_galaxies_maps}
Face-on projected density maps of two galaxies; a stable disk (\galaxyname{r571}, top row) with a smooth SFH and a lopsided disk (\galaxyname{r568}, bottom row) with a cyclic SFH. The columns show projected density maps for dark matter, gas, stars, and young stars (age $<100$ Myr) within a $13$ kpc box at \zzero. Each panel contains the total mass of each component within the halo (even if not visualized here). The young stars panel includes the total $\log {\rm SFR}_{100}$ of the galaxy at \zzero in the bottom left, which corresponds to the same age of the population visualized in that panel.   
}
\end{figure*}

\textbf{Galaxy Mass:} The top row of Figure \ref{fig:peak_tau_vs_three_panels} shows that \cyc increases with both halo and stellar mass following an approximate power law, albeit with scatter (Table \ref{table:pearson_corr}). At fixed mass, disk galaxies have systematically higher \cyc than the non-disk sub-sample, and the most cyclic systems (\cyc $\gtrsim 0.5$) are exclusively disks with \lmstar $> 8.3$. The second row shows a weaker but significant power-law correlation between $\tau_{\rm max}$ and mass for disks ($r \simeq 0.5$). For the non-disk sub-sample, no significant trend is detected. The correlation between \cyc and disk stellar mass may be driven by star formation rate, as high-\cyc galaxies form stars at systematically higher rates than low-\cyc galaxies.

\textbf{Galaxy sSFR:} We consider whether a cyclic SFR implies a high sSFR or large variations in sSFR, which are sometimes referred to as burstiness in the literature.  We consider the variability of \logssfr[10] over the same $4$ Gyr window. The third column of Figure \ref{fig:peak_tau_vs_three_panels} plots \cyc and $\tau_{\rm max}$ against the mean \logssfr[10], with horizontal bars marking the $1\sigma$ spread within the window. Although \cyc correlates with the mean sSFR ($r \simeq 0.8$, Table \ref{table:pearson_corr}), we attribute this largely to the star-forming main sequence, since the mean sSFR tracks stellar mass in our sample. The $1\sigma$ spread in \logssfr[10], by contrast, shows no clear dependence on \cyc. We note that the lowest mass galaxies span the widest raw range of \logssfr[10] (not shown in Figure \ref{fig:peak_tau_vs_three_panels}). The most cyclic galaxies are therefore not necessarily the ``burstiest" galaxies in our sample.

\begin{figure}[t]
\includegraphics[width=7cm]{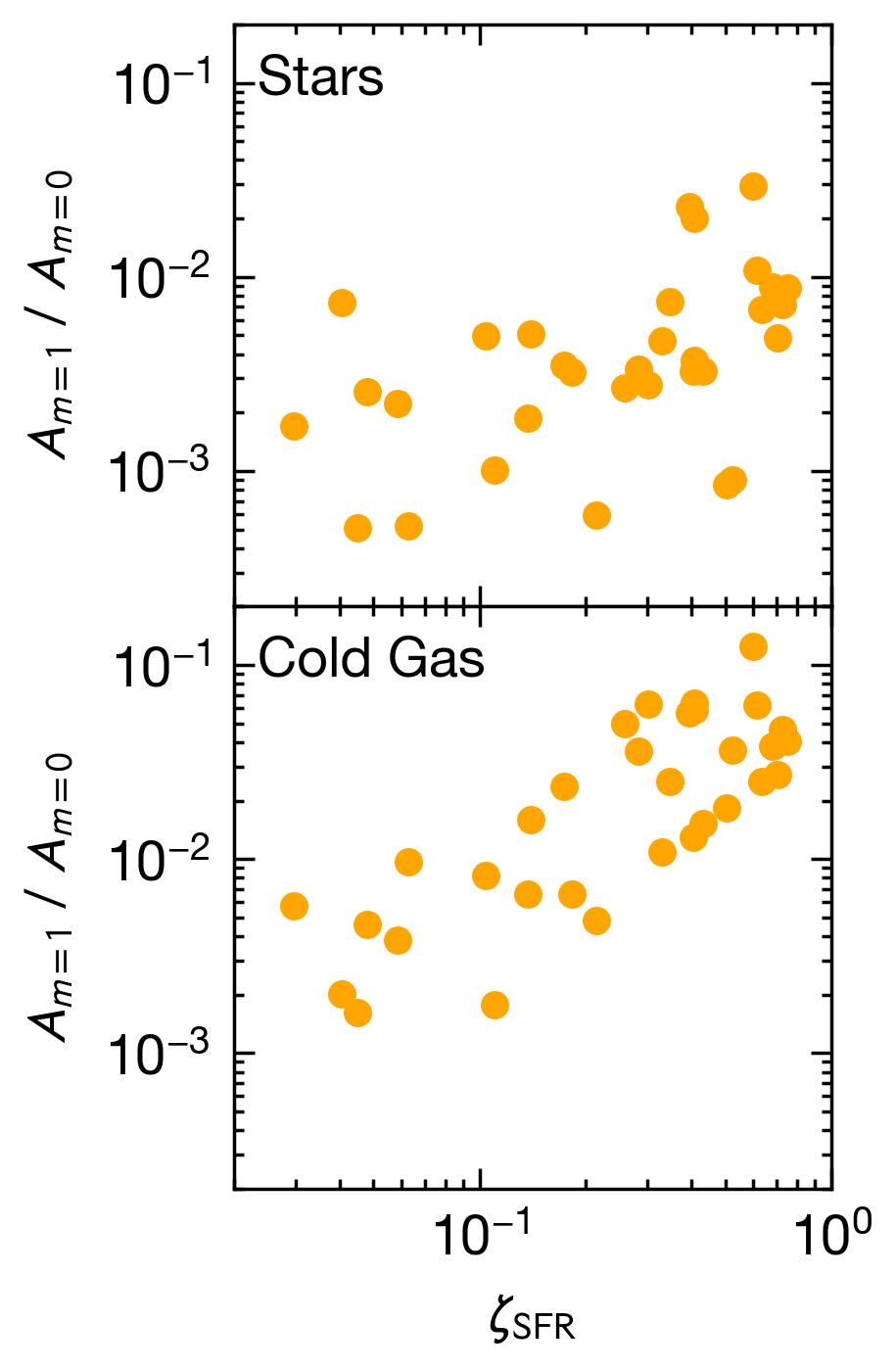}
\centering
\caption{\label{fig:bfe_disks}
Normalized dipole amplitude $A_{m=1}/A_{m=0}$ (Equation \ref{eq:normalized_amplitude}) of the stellar (top) and cold gas (bottom) components of each galaxy in the disk sub-sample as a function of \cyc, measured at \zzero. The lopsidedness of both components increases with \cyc, and at fixed \cyc the cold gas is systematically more lopsided than the stars.
}
\end{figure}

\textbf{Galaxy Size:} The last column of Figure \ref{fig:peak_tau_vs_three_panels} shows \cyc and $\tau_{\rm max}$ as a function of $R_{\rm e}$. \cyc increases with size in both sub-samples (Pearson $r = 0.78$). We note that, over the stellar mass range of our sample ($7.3 \lesssim$ \lmstar $\lesssim 9.1$), $R_{\rm e}$ correlates only weakly with $M_\star$ due to large scatter in mass-size at the dwarf galaxy regime \citep{Cruz2026}. For $\tau_{\rm max}$ as function of $R_{\rm e}$, a linear trend is apparent for disk galaxies (Pearson $r = 0.74$). This trend indicates that more spatially extended disk galaxies have $\tau_{\rm max}$ on longer timescales (i.e., $\tau_{\rm max} \propto R_{\rm e}$). The overall $\tau_{\rm max}$-$R_{\rm e}$ distribution suggests a relationship between SFH and galaxy dynamics, which we will explore in Section \ref{sec:Cyclic SFH and Disk Dynamics}. 

\begin{figure*}[t]
\includegraphics[width=12cm]{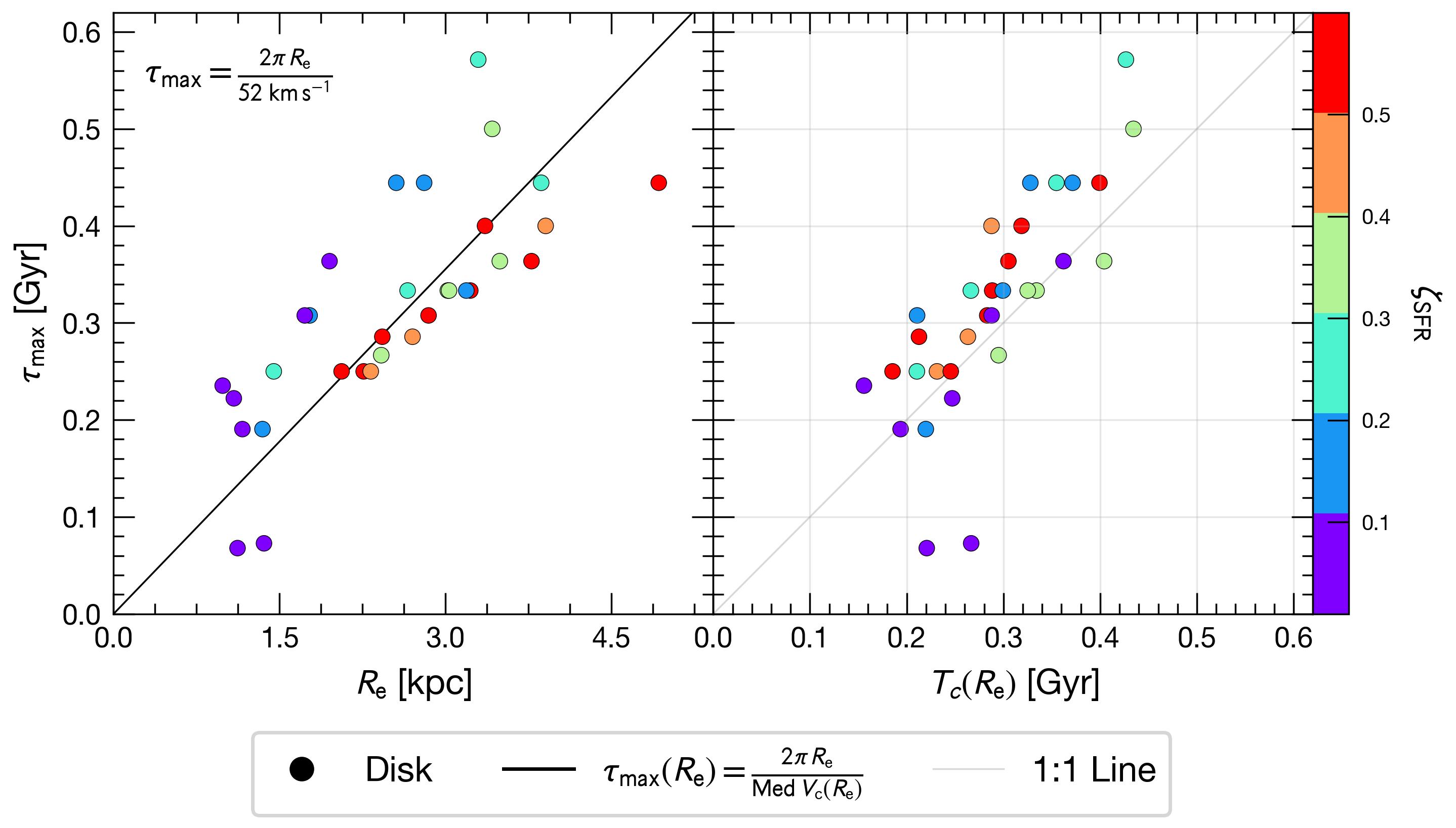}
\centering
\caption{\label{fig:tau_vs_re_tdyn}
The dominant SFR fluctuation timescale $\tau_{\rm max}$ for the disk sub-sample, colored by \cyc. The left panel shows $\tau_{\rm max}$ against effective radius $R_{\rm e}$. The solid black line is the prediction from the disk sub-sample median circular speed measured at $R_{\rm e}$. The right panel shows $\tau_{\rm max}$ against the circular angular orbital period $T_{\rm c}(R_{\rm e})$ for individual galaxies, with the $1{:}1$ line in gray. These results indicate that $\tau_{\rm max}$ is comparable to the angular orbital period at $R_{\rm e}$.
}
\end{figure*}

\section{Cyclic SFH and Disk Dynamics}
\label{sec:Cyclic SFH and Disk Dynamics}

In this Section, we link cyclic star formation to galaxy properties to better understand their causes. A size dependence of the dominant SFR fluctuation timescales ($\tau_{\rm max} \propto R_{\rm e}$, see Figure \ref{fig:peak_tau_vs_three_panels}) is what is expected if cyclic SFR fluctuation timescales are tied to the dynamical timescales of  galaxies. In particular, the relation is most apparent for disk galaxies. In this Section, to constrain the relationship between cyclic star formation and disk dynamics, we first explore the morphologies of cyclic galaxies (Section \ref{sec:Morphologies of Cyclic Galaxies}). We then compare $\tau_{\rm max}$ to the two dynamical timescales of a disk galaxy; the angular orbital period (Section \ref{sec:Angular Orbital Frequencies}) and the radial period (Section \ref{sec:Radial Orbital Frequencies}).

\subsection{Morphologies of Cyclic Galaxies}
\label{sec:Morphologies of Cyclic Galaxies}

A key detail associated with galaxies that express the strongest cyclic SFHs is that they are disk galaxies (see Section \ref{sec:Bursty versus Cyclic Star Formation} and Figure \ref{fig:norm_amp_spec_cyclic}). The galaxies in our disk sub-sample feature a wide range of disk morphologies, ranging from classical disks similar to \galaxyname{M33} to lopsided disks resembling the Large Magellanic Cloud (LMC) or \galaxyname{NGC 55}. On inspection, we find the stellar and gaseous disks of galaxies with the strongest cyclic SFHs are lopsided. These galaxies can be classified as Magellanic Spiral (Sm) or Magellanic irregulars, though not all of them host a bar. 

To visualize the differences in morphology, Figure \ref{fig:two_galaxies_maps} shows the face-on projected density maps of dark matter, gas, stars, and young stars (age $<100$ Myr) at \zzero for two galaxies; a stable disk (\galaxyname{r571}, top row) with a smooth SFH (neither bursty nor cyclic) and a lopsided disk (\galaxyname{r568}, bottom row) with a cyclic SFH. Both galaxies have similar dark matter ($\log(M_{\rm dm}/M_{\odot})= 10.8$) and total gas ($\log(M_g/M_{\odot})=9.7$) masses. In \galaxyname{r571} the gas and stars are distributed in a symmetric disk, with the young stars tracing the smooth surface density of the gas. In the case of \galaxyname{r568}, the gas shows an asymmetrical distribution, with the new stars following a similar lopsided distribution.

To quantify the degree to which the stellar and gas disks are lopsided, we use the cylindrical BFEs (Section \ref{sec:Basis Function Expansions}), applied to the stellar and cold gas ($\log T/K< 3.8$) components of the galaxy (see Figure \ref{fig:exp_cold_gas_example} in Appendix \ref{appendix:Parameters} for an example expansion). We use the cold gas because this gas phase traces the star forming component and is largely confined to the galaxy's disk \citep{Geda2026}. Figure \ref{fig:bfe_disks} shows the normalized dipole amplitude, $A_{m=1}/A_{m=0}$, of the stellar and cold gas components for each galaxy in the disk sub-sample as a function of \cyc. In both components, $A_{m=1}/A_{m=0}$ increases with \cyc, with scatter. The most cyclic galaxies (\cyc $\gtrsim 0.5$) have systematically larger $m{=}1$ amplitudes in their disks than the less cyclic disks, which indicates increasing lopsidedness. 

Comparing the cold gas and stellar lopsidedness, the cold gas shows systematically increasing $A_{m=1}/A_{m=0}$ as a function \cyc. This is expected because the cold gas is collisional and responds not only to gravity but also to pressure, feedback, and the angular momentum injected into the disk \citep{Geda2026}, whereas the stars are collisionless. The stellar disk also includes old stars that formed before the disk settled, so it has a weaker imprint of the disk's morphology. The cold gas is therefore the component that takes up the shape of the disk and responds on short timescales to perturbations and angular momentum injected during disk formation \citep{Geda2026}. This motivates its use as the tracer of disk dynamics and lopsidedness.

\subsection{Dynamics of Cyclic Galaxies}
\label{sec:Dynamics of Cyclic Galaxies}

In this Section we revisit the $\tau_{\rm max}$--$R_{\rm e}$ relation from Section \ref{sec:Dominant Fluctuation Timescales} and the lopsidedness of cyclic galaxies from Section \ref{sec:Morphologies of Cyclic Galaxies} through the lens of disk dynamics. We propose that the periodicity in cyclic SFHs comes from the sloshing of a lopsided disk, which periodically compresses and rarefies the star forming gas relative to its mean density. We explore these connections through relevant dynamical timescales and how they relate to $\tau_{\rm max}$. We begin with how $\tau_{\rm max}$ scales with idealized angular period and then turn to the radial period of the disk, which sets how often the star-forming gas is compressed.

\subsubsection{Angular Orbital Frequencies}
\label{sec:Angular Orbital Frequencies}

To translate $R_{\rm e}$ into an angular dynamical timescale, we approximate the circular orbital period of a parcel of mass in the disk at $R_{\rm e}$, assuming an axisymmetric mass distribution enclosed within $R_{\rm e}$, as:

\begin{equation}
\label{eq:torb}
     T_{\rm c}(R_{\rm e}) = \frac{2\pi R_{\rm e}}{V_{\rm c}(R_{\rm e})} = \frac{2\pi}{\Omega_{c}(R_{\rm e})},
\end{equation} 

\noindent where $T_{\rm c}$ is the circular orbital time, $V_{\rm c}(R_{\rm e})$ the circular velocity determined by the axisymmetric mass enclosed within $R_{\rm e}$, and $\Omega_{\rm c}(R_{\rm e}) = V_{\rm c}(R_{\rm e}) / R_{\rm e}$ is the corresponding circular angular frequency at $R_{\rm e}$, in radians per unit time. 

Figure \ref{fig:tau_vs_re_tdyn} shows the $\tau_{\rm max}$-size (left) and $\tau_{\rm max}-T_{\rm c}(R_{\rm e})$ (right) distributions for our disk sub-sample, colored by \cyc (with red being most cyclic). In the left panel, we show the relation resulting from Equation \ref{eq:torb} using the median $V_{\rm c}(R_{\rm e})$ of our disk sample (${\rm Med}\ V_{\rm c}(R_{\rm e}) = 52 ~{\rm km\,s^{-1}}$) and assuming a boundary condition of $\tau_{\rm max}(0) = T_{\rm c}(0) = 0$. This result and the $T_{\rm c}(R_{\rm e})$ values of individual galaxies (shown in the right panel) indicate that $\tau_{\rm max}$ is comparable to the angular orbital period at $R_{\rm e}$.

\subsubsection{Radial Oscillation of the Lopsided Gas Disk}
\label{sec:Radial Orbital Frequencies}

In addition to its angular frequency, a lopsided gas disk also oscillates radially, with the off-center overdensity moving inward and outward relative to the galactic center as it completes one orbit. We characterize this motion by its period $T_{\rm r}$, where $T_{\rm r}$ is the time between successive innermost passages of the bulk cold gas. As stated above, this radial frequency is of particular interest for cyclic star formation because it periodically compresses and rarefies the star-forming gas relative to its mean density. We re-simulate the last $1$ Gyr at high snapshot cadence ($\sim 10$ Myr) for $11$ of our most cyclic galaxies (ranked by \cyc) and $1$ of our most stable (\galaxyname{r571}) disk galaxies (see Section \ref{sec:high_cadence_reruns}). Throughout this work we have aligned each galaxy face-on using the AM vector of the corresponding snapshot (Section \ref{sec:Simulation}). In this section we instead use the \zzero AM vector as a fixed reference frame so that position angles can be compared across snapshots.

\begin{figure*}[p] 
\includegraphics[width=\textwidth]{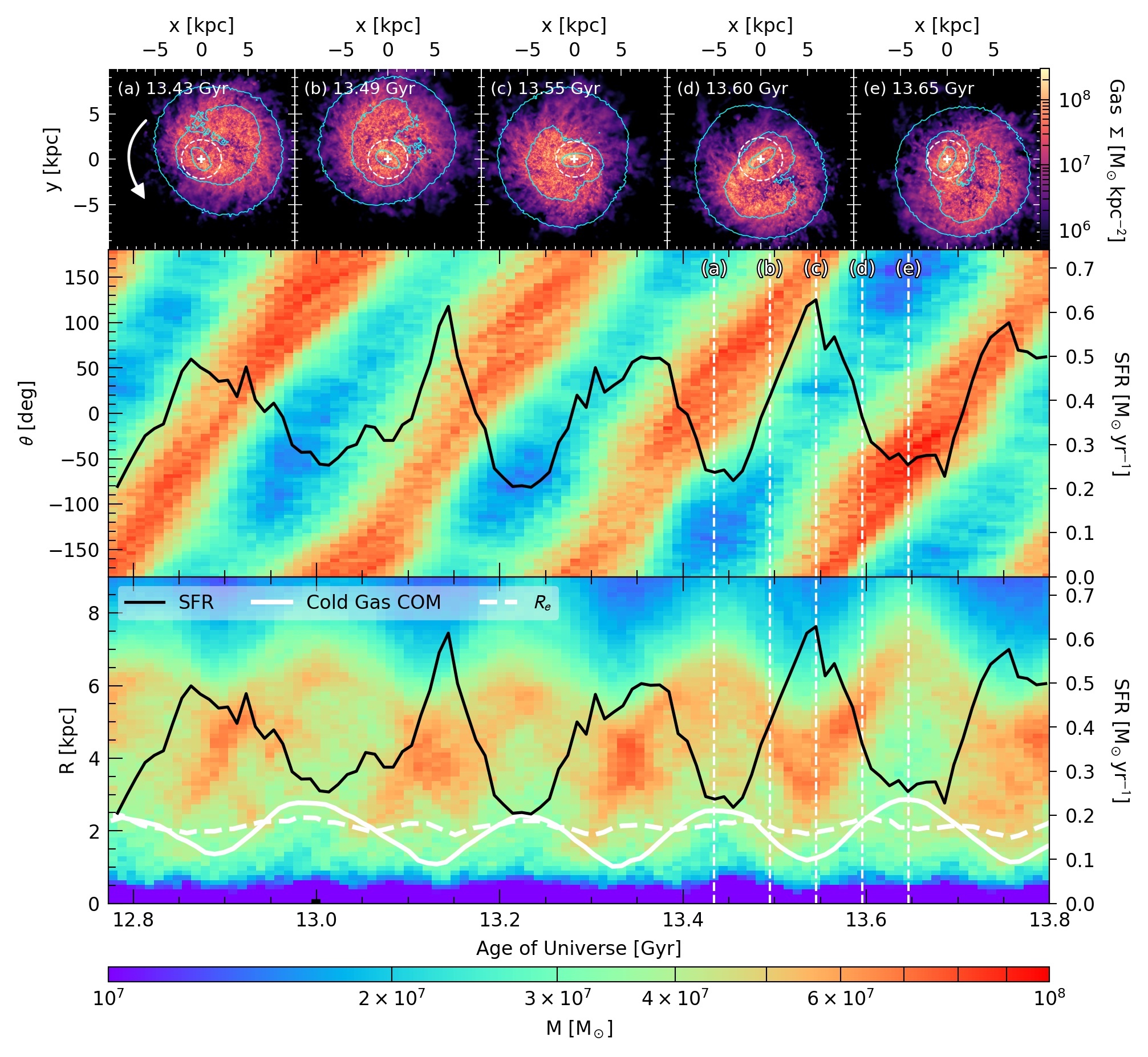}
\centering
\caption{\label{fig:orbit_theta_r_vs_time}
Gas dynamics of \galaxyname{r568} over the last $1$ Gyr of the high snapshot cadence simulation. The top row shows a projected density map of the cold gas spanning one cyclic SFH oscillation. The top row also shows a contour of the stellar projected density ($\log \Sigma_\star = \{6,7,8\}$ $M_\odot\,$kpc$^{-2}$) in cyan and $R_{\rm e}$ at that snapshot (white dashed line). The second and third row of panels show cold gas particle mass histograms, for angular ($\theta$) and galactocentric radial ($R$) positions binned at each snapshot ($10$ Myr temporal bins). The galaxies are centered using their usual SSC and oriented face-on using the AM vector of the cold gas at \zzero as a fixed reference frame at all times. The mass histogram panels include a normalized version of the ${\rm SFR}_{10}$ at each snapshot for reference, with the SFR values shown in the right axis. The radial histogram shows $R_{\rm e}$ (white dashed line) and the cold gas COM (solid white line) at each timestep. The snapshots that correspond to the projected density maps (first row of panels) are shown as vertical dashed lines in the histograms. These panels show that the galaxy's lopsidedness (which is apparent in $\theta$) carries the bulk of gas between apsis, creating density fluctuations relative to its mean, which results in cyclic fluctuations in SFR. Because the radial frequency is key for the star formation, the SFR shares the radial frequency of the gas, but with a $\sim \pi$ phase offset. Similarly, because of the lopsided mass distribution, the COM of the cold gas is offset from the galaxy center, showing the same frequency (phase shifted)  as the SFR. Lastly, $R_{\rm e}$ remains relatively stable throughout the window.
\vspace*{\fill}
}
\end{figure*}

The top row of Figure \ref{fig:orbit_theta_r_vs_time} shows a time series of projected gas density maps for \galaxyname{r568} (within $10$ kpc), which has one of the most periodic star formation histories in our sample (see top row of Figure \ref{fig:bursty_vs_cyclic} for the SFH and Figure \ref{fig:two_galaxies_maps} for morphology). Iso-contours of the stellar surface density are overlaid in cyan to show the stellar disk, which remains centered on the galaxy throughout. The second and third rows show mass histograms of the cold gas ($\log(T/{\rm K}) < 3.8$) in the same galaxy as a function of time, in position angle from the $+x$ axis ($\theta$) and in radius ($R$) from the total-mass shrinking-sphere center, respectively. As mentioned in Section \ref{sec:Simulation}, the shrinking-sphere center is an iterative estimate of the galactic center, obtained by repeatedly computing the center of mass of all particles within a sphere and shrinking its radius until the center of the galaxy's density peak is identified (i.e., the nucleus of the galaxy, central bulge or bar where one is present). This provides a center that is robust to substructure and to asymmetries in the outer disk. The histogram panels also show an overlay of the $10$ Myr SFR in black for reference, with SFR values on the right axis. 

The top row of Figure \ref{fig:orbit_theta_r_vs_time} starts when the bulk of the disk is maximally distant from the center ($13.43$ Gyr). At this stage, the SFR is near minimum in its cycle (SFR$_{10} \sim 0.30\ M_{\odot}\ {\rm yr}^{-1}$). Over the following $\sim 0.1$ Gyr the overdensity moves inward, compressing gas near the galactic center, and the SFR rises to a peak at $\sim 13.55$ Gyr (SFR$_{10} \sim 0.77\ M_{\odot}\ {\rm yr}^{-1}$). The overdensity then moves back outward, the central gas is rarefied, and the SFR falls to a minimum at $13.69$ Gyr. It is important to note that the gas disk is sloshing, not expanding and contracting. The radial change comes from the bulk of the cold gas moving between its furthest and closest approach in the disk. The top row of snapshots show the overall gas morphology does not change dramatically as it oscillates in radial and SFR phase.  

The $\theta$ mass distribution in the middle row of Figure \ref{fig:orbit_theta_r_vs_time} shows the lopsided disk's angular frequency. An important feature to note is that star formation responds to the radial frequency but not necessarily the angular frequency. This is apparent from the two moments of maximum displacement in the first and last panels in the top row of Figure \ref{fig:orbit_theta_r_vs_time} (at $\sim13.43$ and $\sim13.65$ Gyr). Both show the cold gas at maximum displacement, but at position angles or apsidal lines that differ. Furthermore, the mass distribution histograms indicate an offset in bulk $\theta$ at each SFR peak.

\begin{figure*}[p]
\includegraphics[width=\textwidth]{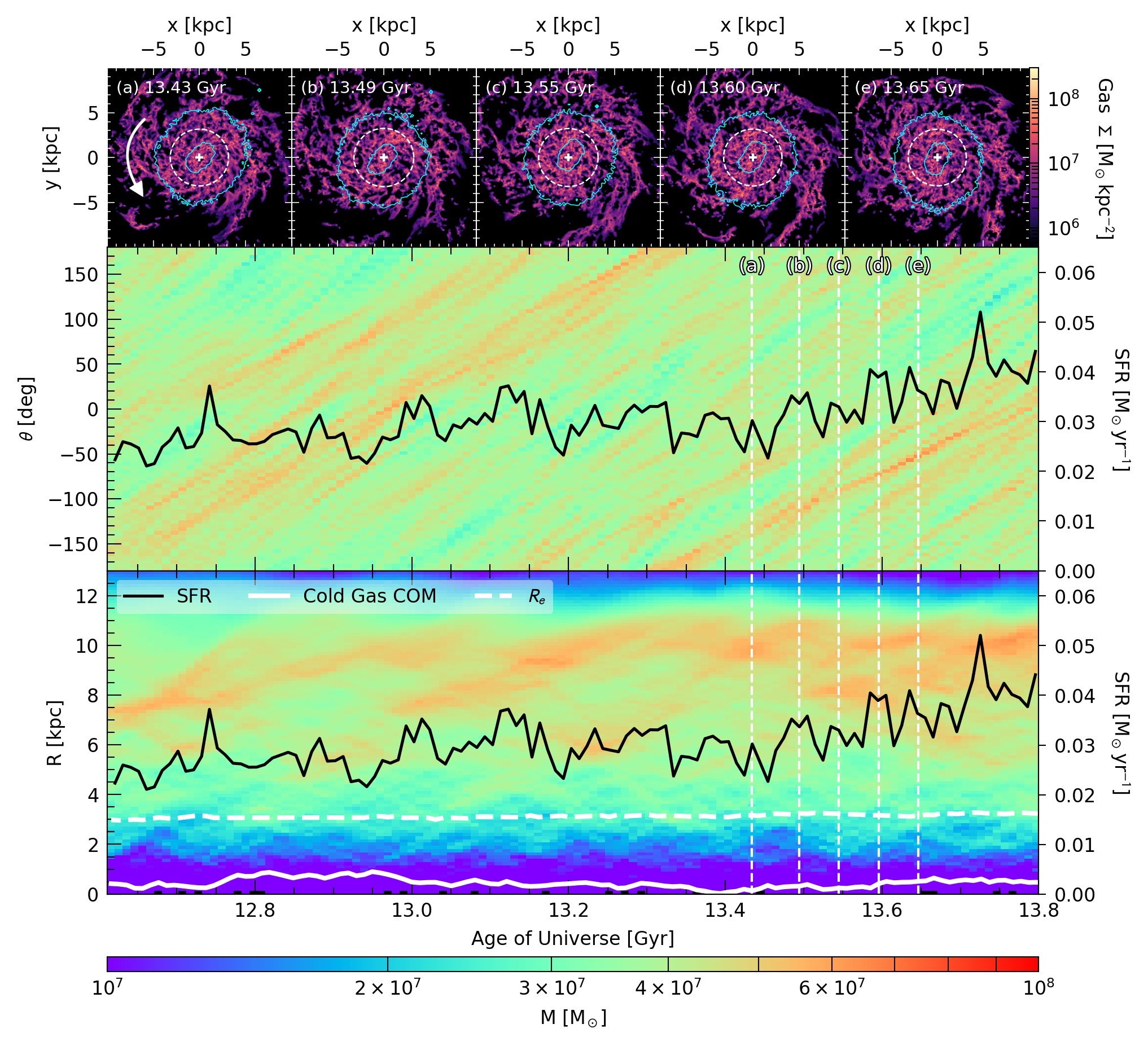}
\centering
\caption{\label{fig:r571_orbit_theta_r_vs_time}
Same panels and setup as Figure \ref{fig:orbit_theta_r_vs_time}, but showing the gas dynamics of \galaxyname{r571}, an axisymmetric disk, over the last $1$ Gyr of the high snapshot cadence simulation. The axisymmetric distribution of the cold gas in the disk results in a smooth distribution in $\theta$, with relatively dense regions (weak arms) appearing as strips as they traverse the disk. This mass distribution keeps the COM of the cold gas being near the SSC of the galaxy. The stellar contours in the top panels also show an axisymmetric distribution for this galaxy. Because the cold gas stays smoothly distributed, the minor fluctuation in the otherwise smooth SFH of \galaxyname{r571} are not coupled to the gas dynamics.
\vspace*{\fill}
}
\end{figure*}

\begin{figure*}[p]
\includegraphics[width=\textwidth]{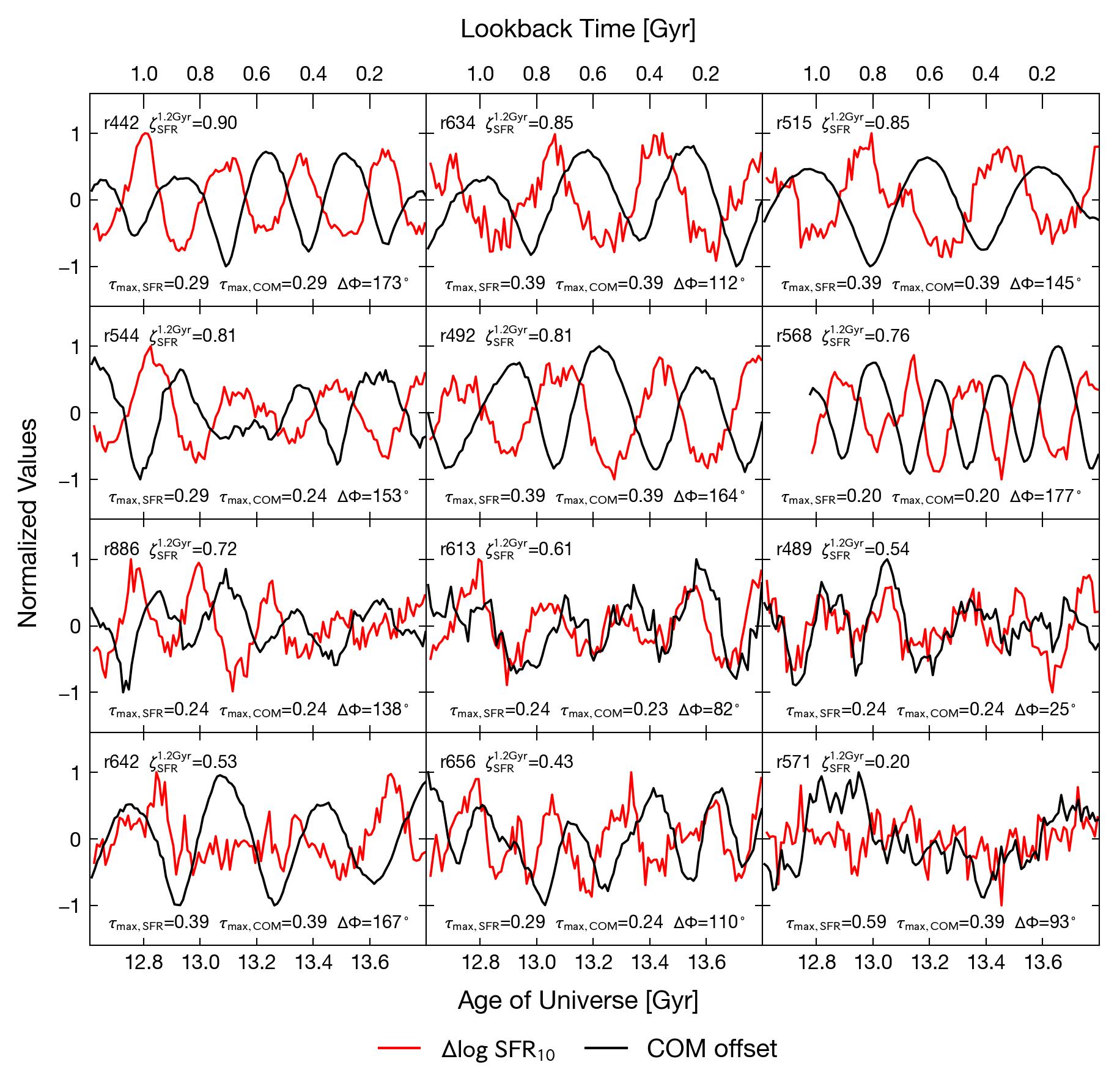}
\centering
\caption{\label{fig:grid_sfr_com}
Star formation and cold gas displacement for all galaxies in the high-cadence reruns over the final $1.2$ Gyr. Each panel shows one simulated galaxy. Red lines show the mean subtracted $\Delta\log{\rm SFR}_{10}$. Black lines show the distance between the cold gas COM (within $30$ kpc) and the galactic center (SSC). Each curve is normalized by its own maximum absolute value, so the panels compare the timing and phase of the two components. Panels are ordered by decreasing $\zeta_{\rm SFR}^{\rm 1.2\,Gyr}$ (labeled in each panel). In the most periodic galaxies, SFR peaks coincide with minima in the COM offset, when the bulk of the cold gas is closest to the galactic center. \galaxyname{r571}, which has a smooth SFH and an axisymmetric disk, is shown for reference.
}
\end{figure*}

In dwarf galaxies with lopsided gas disks, the asymmetric gas distribution displaces the gas center of mass (COM) from the shrinking-sphere center. We adopt the COM of the cold gas ($\log(T/{\rm K}) < 3.8$) as our tracer of this displacement because the cold gas is predominantly bound within the disk, responds to the disk potential and to feedback on short timescales, and contains the star-forming component. The stellar COM, by contrast, is contaminated by old stars that formed before the disk settled, and only becomes a reliable dynamical tracer once the stellar disk has built up sufficient mass \citep{Geda2026}. The bottom panel of Figure \ref{fig:orbit_theta_r_vs_time} shows the galactocentric radius of the cold gas COM as a function of time in white. It oscillates about a mean radius with period $T_{\rm r}$, tracks the bulk radial motion seen in the $R$ histogram, and is phase shifted by $\sim \pi$ relative to the SFR. The galaxy's stellar $R_{\rm e}$ (dashed white line) remains relatively stable over this interval ($\Delta R_{\rm e} \lesssim 0.5$ kpc).

For comparison, Figure \ref{fig:r571_orbit_theta_r_vs_time} shows the same angular and radial histograms for \galaxyname{r571}, a galaxy with a symmetric gas disk. The angular distribution of its cold gas is comparatively smooth, with only scattered clumps, and the radial distribution shows clumpy features but no bulk oscillation of the kind seen in Figure \ref{fig:orbit_theta_r_vs_time}. Owing to the symmetry of the gas disk, its cold gas COM stays close to the center (near $R=0$). This galaxy has a smooth SFH, with an SFR an order of magnitude below that of \galaxyname{r568}.

These results indicate that the radial oscillation of the star forming gas sets the dominant SFH timescale in the most cyclic systems, with $\tau_{\rm max} \approx T_{\rm r}$. In these galaxies the cold gas overdensity roughly completes one radial oscillation per azimuthal period ($T_{\rm r}/T_{\rm c} \sim 1$). This is why the more easily measured $T_{\rm c}$ can stand in for $T_{\rm r}$ when predicting $\tau_{\rm max}$ in Section \ref{sec:Angular Orbital Frequencies}. Residual differences between $T_{\rm r}$ and $T_{\rm c}$ (apsidal precession), together with assuming an axisymmetric mass distribution enclosed within $R_{\rm e}$ in approximating $T_{\rm c}(R_{\rm e})$, and the choice of $R = R_{\rm e}$, contribute scatter to the $\tau_{\rm max}$--$T_{\rm c}$ relation. We discuss these results further in Section \ref{sec:Discussion}.

To test whether the coupling between gas sloshing and star formation seen in \galaxyname{r568} holds across our sample, Figure \ref{fig:grid_sfr_com} compares the SFH with the cold gas COM offset for every galaxy in the high-cadence reruns (Section \ref{sec:high_cadence_reruns}) over the final $1.2$ Gyr. Since the window for the high-cadence reruns is $\sim 1$ Gyr, for each galaxy we subtract a linear fit in time from both $\log {\rm SFR}_{10}$ and the COM offset to remove secular trends over the window (i.e., subtract the mean). We normalize the mean subtracted COM and SFR series by its maximum absolute value so the two can be compared on a common scale. The panels are ordered by $\zeta_{\rm SFR}^{\rm 1.2\,Gyr}$, so the most periodic systems appear first. In the most cyclic galaxies, the COM offset oscillates at nearly the same period as the SFR and roughly in anti-phase with it, as found for \galaxyname{r568}. To quantify this, each panel lists the dominant Fourier period of the SFR ($\tau_{\rm max,SFR}$) and of the COM offset ($\tau_{\rm max,COM}$), together with the phase difference between the two ($\Delta \phi$). A phase difference near $\Delta \phi \sim 180^\circ$ indicates the anti-phase coupling seen in \galaxyname{r568}. We note that \galaxyname{r571} is our non-cyclic example and that for some systems, though they underwent cyclic episodes within the last $4$ Gyr, they may be dampening within the $1.2$ Gyr sampled by the high-cadence reruns (e.g., \galaxyname{r489} is phasing out of the episode near \zzero).

\begin{figure}[t]
\includegraphics[width=8.5cm]{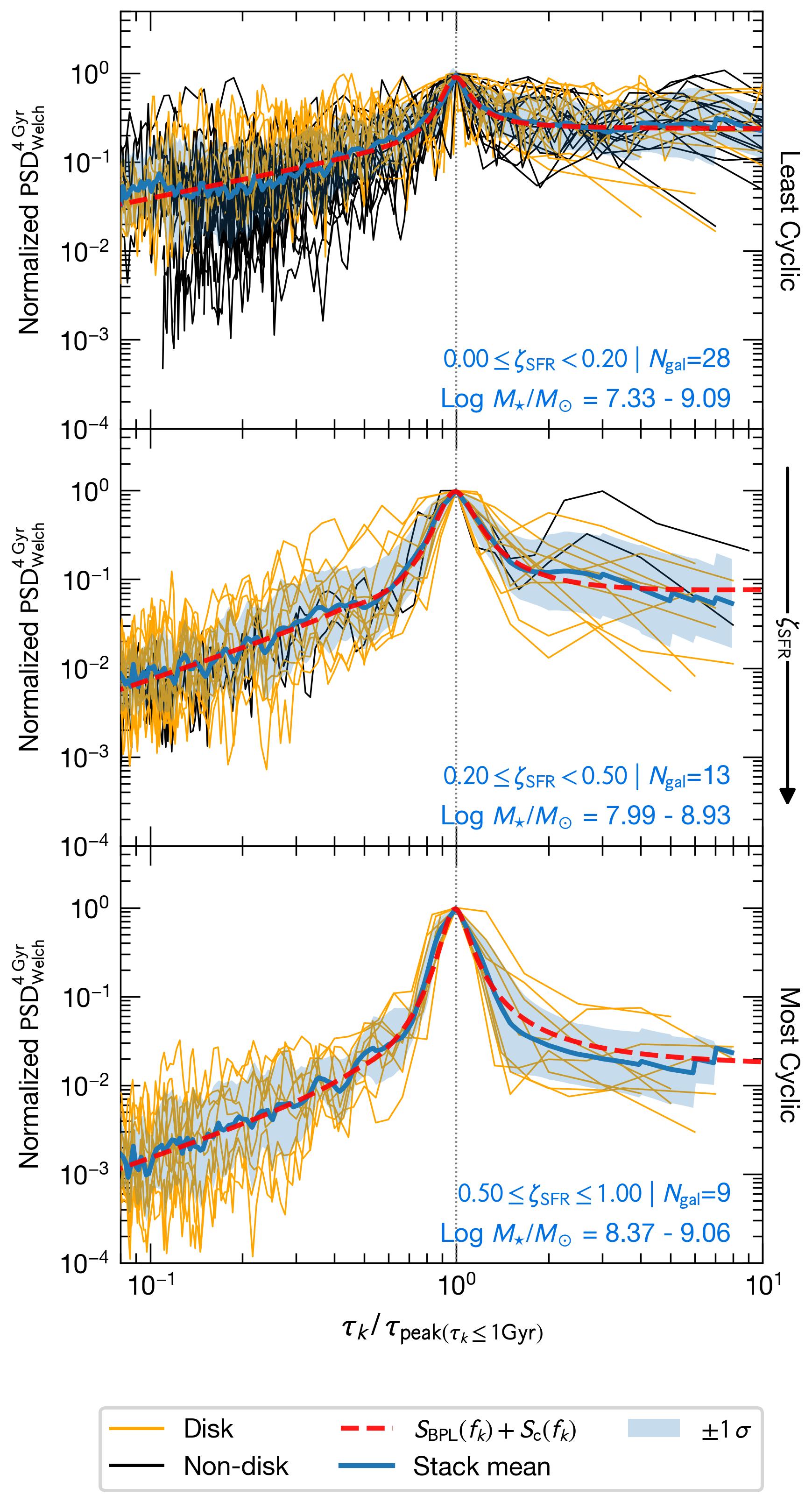}
\centering
\caption{\label{fig:norm_amp_spec_cyclic_lorentz}
Normalized \psdwelch of the SFHs of our final sample over the last 4 Gyr, binned by \cyc as in Figure~\ref{fig:norm_amp_spec_cyclic}. Each PSD is normalized by its peak power and peak timescale $\tau_{\rm peak}$ before stacking, so the cyclic bumps align at $\tau_k/\tau_{\rm peak} = 1$ ($\tau_{\rm peak} \sim \tau_{\rm max}$). The blue line and band show the mean and one standard deviation of the stacked PSDs in each bin. The red line is a fit to the mean with a two-component model, a broken power law for the continuum ($S_{\rm BPL}$, Equation \ref{eq:bpl}) plus a Lorentzian for the cyclic bump ($S_{\rm c}$, Equation \ref{eq:lorentzian}). The PSD of cyclic SFHs can be understood as an excitation of a Lorentzian component ($S_{\rm c}$) driven by the orbital dynamics of the galaxy's gas disk.}
\end{figure}

\section{Discussion}
\label{sec:Discussion}

In Section \ref{sec:Power Spectral Density Distributions} we showed that the stacked \psdwelch of the disk and non-disk sub-samples are similar, both following a broken power law on timescales $\tau_k \leq 1$ Gyr with flat slopes on long timescales ($\beta_{\rm low} < 0.01$), steep slopes on short timescales ($\beta_{\rm high} = 2.05$ for disks and $2.59$ for non-disks), and break timescales of $\sim 100$ Myr. This shape resembles a damped random walk, in agreement with analytic expectations and previous simulation studies \citep{Tacchella2020, Iyer2020}. The power spectra of individual galaxies, however, reveal that a subset of our sample has quasi-periodic (cyclic) SFHs dominated by a single timescale, $\tau_{\rm max}$. We quantified this behavior with the cyclicity parameter, \cyc, defined as the fraction of SFR variance carried by the dominant frequency. The most cyclic galaxies (\cyc $\gtrsim 0.5$; $18\%$ of our sample) are exclusively disks with \lmstar $> 8.3$ and are not necessarily the burstiest systems. For disks, $\tau_{\rm max}$ scales linearly with $R_{\rm e}$, pointing to a link between cyclic star formation and disk dynamics.

In Section \ref{sec:Cyclic SFH and Disk Dynamics} we examined the link between disk dynamics and star formation. The most cyclic galaxies host lopsided stellar and gas disks resembling Magellanic-type irregulars, with the cold gas showing systematically stronger $m{=}1$ asymmetry ($A_{m=1}/A_{m=0}$) than the stars. Across the disk sub-sample, $\tau_{\rm max}$ is comparable to the circular orbital period at the effective radius, $T_{\rm c}(R_{\rm e})$. Using high-cadence reruns of our most cyclic galaxies, we showed that the lopsided cold gas disk sloshes about the galactic center as the bulk of the gas moves between its maximum and minimum radius in its orbit. Star formation peaks when the gas is compressed near the galactic center and declines as it moves outward and is rarefied. This causes the offset of the cold gas center of mass from the galactic center to oscillate roughly in anti-phase with the SFR. The dominant SFH timescale in these systems is therefore set by the radial oscillation period of the gas ($\tau_{\rm max} \approx T_{\rm r}$), which is close to the azimuthal period ($T_{\rm r}/T_{\rm c} \sim 1$). In this section, we discuss the implications of these results and observational signatures of cyclic star formation.

\subsection{The Power Spectra of Cyclic Dwarfs}
\label{sec:The Power Spectra of Dwarf SFHs}

\subsubsection{Shape of Cyclic PSDs}
\label{sec:Shape of Cyclic PSDs}

The overall shape of the PSDs of cyclic galaxies can be modeled by adding a Lorentzian component centered at orbital timescale to the broken power-law continuum. In particular, the bump or quasi-periodic peak of cyclic galaxies, which spans a small range of frequencies, can be described by narrow Lorentzian component ($S_{\rm c}$):

\begin{equation}
\label{eq:lorentzian}
    S_{\rm c}(f_k) = 
    \frac{S_{0}\ (\gamma/2)^{2}}{(f_k - f_{c})^{2} + (\gamma/2)^{2}},
\end{equation}

\noindent where $S_{0}$ is the peak amplitude, $f_{c}$ the center frequency or offset, and $\gamma$ the full width at half maximum. Because the cyclic star formation in our most cyclic galaxies is a consequence of gas disk orbital dynamics, the $S_{\rm c}$ component is centered near the full orbital time of gas disk (see Section \ref{sec:Dynamics of Cyclic Galaxies}):

\begin{equation}
 f_{c} = \frac{1}{\tau_{\rm max}} \approx \frac{1}{T_{\rm r}} \sim \frac{1}{T_{c}(R_{e})}.
\end{equation}

To illustrate the two-component nature of cyclic PSDs, in Figure \ref{fig:norm_amp_spec_cyclic_lorentz} we stack the normalized \psdwelch of our simulated galaxies, binned by cyclicity, over the last $4$ Gyr of their respective SFHs and fit the resulting mean spectrum with $S_{\rm BPL}(f_k) + S_{\rm c}(f_k)$ using \texttt{astropy}. Each PSD is normalized by its peak power and peak timescale $\tau_{\rm peak} \sim \tau_{\rm max}$ prior to stacking, which aligns the cyclic bumps at $\tau_k/\tau_{\rm peak} = 1$. The two-component model reproduces both the continuum and the quasi-periodic peak. Therefore the cyclic SFH can be understood as a Lorentzian in the PSD ($S_{\rm c}(f_k)$) driven by the orbital dynamics of the galaxy's gas disk. In principle, the continuum break $\tau_{b}$ and the cycle timescale $\tau_{\rm max}$ are free to differ. A broken power-law only fit does not account for the quasi-periodic peak, so the break may be dragged toward $\tau_{\rm max}$ and $\tau_{\rm max}$ could be misread as the break timescale of the cyclic galaxy. This would conflate two physically different timescales, where $\tau_{b}$ reflects the transition to timescales dominated by stochastic formation and disruption of giant molecular clouds \citep[GMC,][]{Tacchella2020}, while $\tau_{\rm max}$ reflects the orbital dynamics of the star forming gas disk. Further detailed exploration of the functional form of cyclic PSDs is left for future work. 

\subsubsection{Cyclic PSDs in the Context of SFH Models}
\label{sec:Cyclic PSDs in the Context of SFH Models}

Cyclic SFHs can be placed in the context of two common statistical descriptions of SFH variability, damped random walks and stochastically driven damped harmonic oscillators. The extended regulator model  \citep{Caplar2019, Tacchella2020, Iyer2024} describes the SFH as a sum of damped random walks corresponding to gas inflow, gas cycling, and GMC formation and destruction ($\leq 1$ Gyr). These processes contribute to a Lorentzian, or a product of Lorentzians, centered at zero frequency ($f_c = 0$). The sum of these components produce a broken power-law like continuum at $\leq 1$ Gyr timescales. The cyclic bump in particular requires a component peaked at nonzero frequency, which lies outside the family of damped random walks considered by the extended regulator model \citep{Iyer2024}. Though the bump is well described by the stochastically driven, damped simple harmonic oscillator (SHO) kernel of \cite{ForemanMackey2017}, a single SHO term is not sufficient on its own.  Thus, the PSDs of our simulated galaxies are well described by the regulator model, with cyclic galaxies requiring an additional Lorentzian component (Equation \ref{eq:lorentzian}), driven by the orbital dynamics of the gas disk and centered at radial frequency of their star forming gas disks.

\subsection{What Causes Cyclic Behavior?}
\label{sec:What Causes Cyclic Behavior}

Our analysis in Section \ref{sec:Cyclic SFH and Disk Dynamics} shows that cyclic star formation in dwarf galaxies results from a coupling between the SFR and the dynamics of the star forming gas disk. Cyclic episodes occupy only part of the SFHs of the galaxies that express them, usually at late times for galaxies in our mass range. In the simulations, we often see external dynamical perturbations from merging, companion, or splashback satellites occur around the onset of cyclic behavior \citep{1989MNRAS.239..549W, 1994ApJ...421..481W, 2015A&A...575A..59S, 2016A&A...587A..24S, 2016A&A...595A..56S}. Close encounters with these satellites, typically with  $\log(M_{\rm halo}/{\rm M}_{\odot}) \gtrsim 8$, can perturb classical disks onto more eccentric orbits and trigger cyclic star formation, and in some cases a close fly-by or an interaction with a more massive satellite briefly suppresses an ongoing episode. The nature of these interactions varies enough from galaxy to galaxy, that more work is needed to isolate which aspects of an interaction matter most for triggering cyclic behavior. We defer a detailed study of these triggers to future work.

Beyond triggering cyclic SFH, why cyclic episodes can persist for multiple orbital periods is of particular interest as well. Stellar feedback is one possibility that could play a role in allowing the episodes to persist. Each cycle compresses the gas and drives a burst, and each burst contributes feedback. In a shallow potential, this feedback may suppress circularization of the gas, which may lead to feedback coupling that leads to prolonged sustained cyclic star forming episodes (i.e., prolonged sloshing modes). 

Another dynamical trigger, and a possible cause of persistence, is an instability in the distribution function, most likely of the dark matter. \cite{2023MNRAS.525.4962W} shows that features in the halo density profile can produce an inflection in the distribution function, so that it no longer satisfies Antonov's stability criterion \citep[$df/dE < 0$, ][]{1962spss.book.....A} and becomes unstable to an $l{=}1$ mode. \cite{2025ApJ...993..150D} show that such an inflection develops in ideal spherically symmetric systems whose density profile turns over quickly from a steep outer slope to a shallow inner one. The resulting $l{=}1$ mode in the dark matter halo acts on the disk as a precessing dipole \citep{2023MNRAS.525.4962W}, which could produce and sustain the lopsided gas disks of our most cyclic galaxies. Lastly, numerical effects may play a role in the lifetimes of cyclic SFR episodes, though the triggers are physically identifiable. In a follow up work, we will test these possible drivers by examining disk dynamics, star forming regions, and dark matter morphologies in detail. We will also explore the galaxy mass ranges that are most susceptible.  

\subsection{Observational Signatures}
\label{sec:Observational Signatures}

In this section we discuss possible observational signatures of galaxies that express cyclic star formation episodes. Since the most cyclic galaxies are disk galaxies, a morphological sample selection is needed to filter out non-disk galaxies. One possible approach is to take advantage of the relatively compact sizes of non-disk galaxies and apply a $R_e > 2$ kpc cut for $8.5 <$ \lmstar $< 9.5$ (or similar filter accounting for mass--size). This negates the need for morphological classification at the expense of filtering out compact gas disks that may be cyclic. 

\subsubsection{Offset Gas Disks}
\label{sec:Offset Gas Disks}

The most direct signature of a cyclic galaxy is a star forming gas disk whose center does not coincide with the photometric or kinematic center of the stellar body. In our mass range the stellar disk is still building its mass at late times \citep{Geda2026}, and a significant fraction of the stellar light may lie in the older stellar population that does not share the gas orbit, so the $m{=}1$ distortion is expressed most strongly in the gas or possibly H$\alpha$ regions. We expect the offset to be measurable in $21$ cm maps against the galaxy's optical isophotal center, and for the ${\rm H}_{\rm I}$ distribution to appear as an $m{=}1$ term in an azimuthal harmonic decomposition \citep[e.g., using Fourier-Laguerre basis expansions,][]{2025MNRAS.539..661G}. It is possible to detect the $m{=}1$ mode in stellar disks that have had sustained long cyclic episodes, which builds up a stellar population that inherits the cold gas disk's kinematics. Regardless of the component considered, measuring the offset requires a sizable sample of dwarf disk galaxies viewed close to face-on or with mild inclinations. It is also possible to detect lopsidedness in edge-on systems whenever the orbital semi-major axis of the gas is close to perpendicular to the line of sight (\galaxyname{NGC 55} for example). We expect the most common contaminants to be systems actively undergoing a major merger that have not yet coalesced, which can appear extended and lopsided, often alongside elevated \logssfr. Since our final sample consists of isolated galaxies, an isolation criterion is also needed to restrict environmental effects.   

\subsubsection{Star Forming Regions}
\label{sec:Star Forming Regions}

The last column of Figure \ref{fig:two_galaxies_maps} shows the young stellar population (age $<100$ Myr) for a stable (smooth SFR) and lopsided (cyclic) galaxy. Dwarf galaxies with stable disks in our simulated sample have their star forming regions distributed across the disk. In contrast, strongly lopsided disks undergoing large-amplitude cyclic SFR episodes have concentrated star-forming regions near the center. The compression that triggers a burst in lopsided disks occurs when the bulk of the star-forming gas makes its closest approach to the nucleus (see Figure \ref{fig:orbit_theta_r_vs_time}). At peak SFR the star-forming region therefore lies slightly offset from the isophotal center of the continuum (offset H$\alpha$ regions). At the SFR trough, the star-forming gas is at its maximum distance from the nucleus and is spread over the disk, so the H$\alpha$ is diffuse. These trends are observed in galaxies from cosmic noon to the present \citep[e.g.,][]{2013ApJ...779..135W, 2015ApJ...800...39G, 2015ApJ...810L..15S, 2023ApJ...951..147S, 2023ApJ...955..106M, Mintz2024, 2025ApJ...979...14S, 2026ApJ..1007..176S}. For example, \citet{Mintz2024} measure nonparametric morphological statistics (Gini, $M_{20}$, and asymmetry) of both the stellar continuum and the H$\alpha$ emission for dwarfs at $8 \lesssim \log(M_\star/M_\odot) \lesssim 9$ in the Merian Survey \citep{2025ApJ...993..110D}. They find that the lowest-mass galaxies with the highest sSFR have H$\alpha$ emission that is compact and slightly offset from the center of the continuum.

\subsubsection{The 0.2 Gyr Timescale}
\label{sec:The 0.2 Gyr Timescale}

Because observations capture real dwarf galaxies at a single moment in their star forming cycle, the cycle of any one galaxy cannot be followed. The timescales are instead inferred statistically from ensembles of galaxies observed at all phases, and the scatter in star formation indicators that probe different timescales, such as H$\alpha$ and FUV, constrains the burst amplitude and duration. This allows us to compare simulated star formation timescales to the corresponding timescales inferred from observations. Building on \citet{Weisz2012}, \citet{Emami2019} fit the H$\alpha$ and FUV luminosities of Local Volume galaxies with bursts that rise and decline exponentially and find e-folding times that increase with mass, from $\tau \approx 100$ Myr at $8 <$ \lmstar $< 8.5$ to $\tau \gtrsim 300$ Myr at \lmstar $> 8.5$. These timescales are of the same order as our simulated final sample (see Figure \ref{fig:peak_tau_vs_three_panels}). 

At cosmic noon, \cite{Mintz2026} use a continuum selected sample of $43$ JWST PRISM spectra of low-mass galaxies (\lmstar $<9.5$) and find numerous strong Balmer breaks that anticorrelate with H$\alpha$ equivalent width. Modeling the population with periodic bursts, they constrain burst timescales of $\sim 200$ Myr. The prevalence of long-timescales of burst may be related in some part to cyclic star formation. If the mechanism that stimulates cyclic star formation needs only a gas disk and a perturber, sloshing disks and cyclic episodes could be more common among dwarfs at higher redshifts. Nevertheless, establishing predictions for possible cyclic episodes and their physical mechanism at high redshifts requires further work focused on low-mass systems and their environments in the early universe. 

For disk dwarf galaxies, $\tau_{\rm max}$ scaling with $R_{\rm e}$ raises an interesting prediction (see Figure \ref{fig:tau_vs_re_tdyn}). For a disk at the compact/extended boundary of \cite{Geda2026}, $R_{\rm e} = 2$ kpc, and a typical $V_{c}(R_{\rm e}) \sim 50\ {\rm km\,s^{-1}}$ for our disk sample: 

\begin{equation} 
\label{eq:torbapprox} 
\tau_{\rm max} \sim T_{c} = \frac{2\pi\,(2\ {\rm kpc})}{50\,{\rm km\,s^{-1}}} = 0.25\,{\rm Gyr}, 
\end{equation} 

\noindent which is similar to the periods recovered by H$\alpha$/FUV analyses. However, a definitive observational test of $\tau_{\rm max}$ scaling with $R_{\rm e}$ requires a sample of dwarf disk galaxies at fixed mass. Dividing these disks into size bins of $R_{\rm e}$ and repeating the H$\alpha$/FUV fit in each bin would yield a burst timescale per size bin. If the timescale is set by the orbital period, it should grow with $R_{\rm e}$ across the bins, whereas if it is set by mass alone the bins should return the same value. The test does not require separating lopsided disks from stable ones. A lopsided disk raises the amplitude of the variation at $\tau_{\rm max}$ but does not change $\tau_{\rm max}$ itself, so every disk in the sample contributes to the measurement. We expect the clearest trend in the bins with $R_{\rm e} > 2$ kpc, since nearly every dwarf above this size in our simulated \zzero sample is an angular momentum supported disk \citep{Geda2026}.

\section{Summary and Conclusions}
\label{sec:Summary and Conclusions}

In this work we measure the timescales of SFR variability in $50$ isolated dwarf galaxies from the Massive Dwarfs simulations ($7.3 <$ \lmstar $< 9.1$) and link periodic star formation to the dynamics of their gas disks. The key findings of this work are summarized below.

\begin{enumerate}

\item At the population level, the stacked \psdwelch of both the disk and non-disk sub-samples follow a broken power law on timescales $\tau_k \leq 1$ Gyr, flat on long timescales and steep ($\beta_{\rm high} \approx 2$--$2.6$) on short timescales, with a break near $100$ Myr. This is the shape of a damped random walk and agrees with analytic expectations and previous simulations \citep{Tacchella2020, Iyer2020} (see Figure \ref{fig:sfr_power_spec} and Section \ref{sec:The PSD of Disks versus Non-Disks}).

\item Individually, galaxies show smooth, bursty, and cyclic SFHs (see Figure \ref{fig:bursty_vs_cyclic}). We quantify cyclic star formation with \cyc, the fraction of SFR variance from the dominant frequency (Equation \ref{eq:zeta}). The most cyclic galaxies (\cyc $\gtrsim 0.5$, $18\%$ of the sample) are exclusively disks with \lmstar $> 8.3$ and are not necessarily the burstiest galaxies (see Figures \ref{fig:norm_amp_spec_cyclic} and \ref{fig:peak_tau_vs_three_panels}, and Section \ref{sec:Bursty versus Cyclic Star Formation}).

\item For disk galaxies, the dominant fluctuation timescale $\tau_{\rm max}$ scales linearly with $R_{\rm e}$ and is approximated by the circular orbital period at that radius, $T_{\rm c}(R_{\rm e})$ (see Figure \ref{fig:tau_vs_re_tdyn} and Sections \ref{sec:Dominant Fluctuation Timescales} and \ref{sec:Angular Orbital Frequencies}).

\item The most cyclic galaxies host lopsided, Magellanic-type stellar and gas disks. The $m{=}1$ amplitude increases with \cyc and is systematically larger in the cold gas than in the stars (see Figures \ref{fig:two_galaxies_maps} and \ref{fig:bfe_disks}, and Section \ref{sec:Morphologies of Cyclic Galaxies}).

\item In lopsided galaxies, the lopsided cold gas disk sloshes about the galactic center. Star formation peaks when the bulk of the gas is compressed near the center and declines as it moves outward, so the cold gas COM offset oscillates radially in anti-phase with the SFR and $\tau_{\rm max} \approx T_{\rm r} \approx T_{\rm c}$. An axisymmetric disk shows no such coupling (see Figures \ref{fig:orbit_theta_r_vs_time}, \ref{fig:r571_orbit_theta_r_vs_time}, and \ref{fig:grid_sfr_com}, and Section \ref{sec:Radial Orbital Frequencies}).

\item The PSD of a cyclic galaxy is described by the broken power-law continuum plus a narrow Lorentzian centered at the orbital frequency of the gas disk (Equation \ref{eq:lorentzian}, see Figure \ref{fig:norm_amp_spec_cyclic_lorentz} and Sections \ref{sec:Shape of Cyclic PSDs} and \ref{sec:Cyclic PSDs in the Context of SFH Models}).

\end{enumerate}

Looking ahead, a follow-up study will use the basis function expansions introduced here to follow the orbits of the star-forming gas directly, identify the perturbations that set lopsided disks in motion, and determine why cyclic episodes persist for several orbital periods. The observational tests above can be carried out with existing H$\alpha$, FUV, and ${\rm H}_{\rm I}$ surveys of nearby dwarfs, and would establish whether the cycles we find in simulations are present in real galaxies.

\clearpage 

\begin{acknowledgments}
\label{sec:Acknowledgments}

The authors of this work thank Romain Teyssier for useful conversations on numerics and energy distribution functions. We thank Shaunak Modak for useful conversations about disk dynamics which lead to high-cadence output simulations, and Abby Mintz for helpful discussions on PSDs and H$\alpha$ morphologies. R.G. thanks Alan Kahn for donating compute and visualization equipment used in this work. A.M.B. acknowledges support from NASA Grant 80NSSC24K0894 and grant FI-CCA-Research-00011826 from the Simons Foundation. B.W.K. acknowledges support provided by NASA through a grant from the Space Telescope Science Institute, through grant HST AR-17547. This work used Stampede2 at the Texas Advanced Computing Center (TACC) through allocation MCA94P018 from the Advanced Cyberinfrastructure Coordination Ecosystem: Services $\&$ Support (ACCESS) program, which is supported by U.S. National Science Foundation grants 2138259, 2138286, 2138307, 2137603, and 2138296. Some of the simulations were performed using resources made available by the Flatiron Institute. The Flatiron Institute is a division of the Simons Foundation. This work was supported by NSERC (National Sciences and Engineering Research Council) of Canada. This research has made use of NASA's Astrophysics Data System. Grammarly was used to suggest grammar and language improvements.

\software{\texttt{pynbody} \citep{pynbody}, \texttt{astropy}  \citep{astropy:2013, astropy:2018, astropy:2022}, \expname \citep{Weinberg1999, Petersen2022, Petersen2024}, \texttt{matplotlib} \citep{Hunter:2007}, \texttt{numpy} \citep{numpy}, \texttt{python} \citep{python}, \texttt{PetroFit} \citep{2022AJ....163..202G}, \texttt{scipy} \citep{2020SciPy-NMeth}, \texttt{ChaNGa} \citep{Jetley2008, Jetley2010, Menon2015}, and \texttt{Cython} \citep{cython:2011}, \texttt{\href{https://www.tomwagg.com/software-citation-station/}{The Software Citation Station}} \citep{software-citation-station-paper, software-citation-station-zenodo}}
\end{acknowledgments}

\appendix
\restartappendixnumbering

\section{Table of Galaxies} 
\label{appendix:Table of Galaxies} 

Table \ref{table:overview} lists the \zzero properties of every galaxy in the Massive Dwarfs and Extended Marvelous Dwarfs suites, including the three galaxies excluded from our final sample (Section \ref{sec:Final Sample}). Galaxies are ordered by stellar mass. The disk classification follows Section \ref{sec:Disk Galaxy Classification} and \cyc is measured over the last $4$ Gyr of each SFH as described in Section \ref{sec:Bursty versus Cyclic Star Formation}.

\begin{longtable}[c]{l *{8}{c}}
\caption{\label{table:overview} Properties of the galaxy sample. Columns show: (1) galaxy name (simulation name); (2) total mass $M_{\rm total}$; (3) stellar mass $M_{\star}$; (4) stellar half-light radius $R_{e}$; (5) virial radius $R_{\rm vir}$; (6) \zzero star formation rate averaged over 10 Myr (\logsfr[10]); (7) \zzero specific star formation rate over 10 Myr \logssfr[10]; (8) \cyc ("cyclicness", see Equation \ref{eq:zeta}) for \logsfr[10] within the last $4$ Gyr of the simulation; and (9) flag for presence of a disk. Excluded galaxies are indicated with a footnote.}\\

\hline \hline
\multicolumn{1}{c}{Name} & \multicolumn{1}{c}{$M_{\rm total}$} & \multicolumn{1}{c}{$M_{\star}$} & \multicolumn{1}{c}{$R_{e}$} & \multicolumn{1}{c}{$R_{\rm vir}$} & \multicolumn{1}{c}{${\rm SFR}_{10}^{z_0}$} & \multicolumn{1}{c}{${\rm sSFR}_{10}^{z_0}$} & \multicolumn{1}{c}{$\zeta_{{\rm SFR}_{10}^{z_0}}^{4 {\rm Gyr}}$} & \multicolumn{1}{c}{Disk}\\
\multicolumn{1}{c}{(1)} & \multicolumn{1}{c}{(2)} & \multicolumn{1}{c}{(3)} & \multicolumn{1}{c}{(4)} & \multicolumn{1}{c}{(5)} & \multicolumn{1}{c}{(6)} & \multicolumn{1}{c}{(7)} & \multicolumn{1}{c}{(8)} & \multicolumn{1}{c}{(9)}\\
\hline
 & log $M_{\odot}$ & log $M_{\odot}$ & kpc & kpc & log $M_{\odot}$ yr$^{-1}$ & log yr$^{-1}$ &  & \\
\hline

\endfirsthead
\multicolumn{9}{c}{\tablename\ \thetable\ \textit{(continued)}}\\

\hline \hline
\multicolumn{1}{c}{Name} & \multicolumn{1}{c}{$M_{\rm total}$} & \multicolumn{1}{c}{$M_{\star}$} & \multicolumn{1}{c}{$R_{e}$} & \multicolumn{1}{c}{$R_{\rm vir}$} & \multicolumn{1}{c}{${\rm SFR}_{10}^{z_0}$} & \multicolumn{1}{c}{${\rm sSFR}_{10}^{z_0}$} & \multicolumn{1}{c}{$\zeta_{{\rm SFR}_{10}^{z_0}}^{4 {\rm Gyr}}$} & \multicolumn{1}{c}{Disk}\\ 
\multicolumn{1}{c}{(1)} & \multicolumn{1}{c}{(2)} & \multicolumn{1}{c}{(3)} & \multicolumn{1}{c}{(4)} & \multicolumn{1}{c}{(5)} & \multicolumn{1}{c}{(6)} & \multicolumn{1}{c}{(7)} & \multicolumn{1}{c}{(8)} & \multicolumn{1}{c}{(9)}\\
\hline
 & log $M_{\odot}$ & log $M_{\odot}$ & kpc & kpc & log $M_{\odot}$ yr$^{-1}$ & log yr$^{-1}$ &  & \\
\hline

\endhead
\hline
\multicolumn{9}{r}{\footnotesize \textit{Continued on next page}}\\
\endfoot
\hline
\multicolumn{9}{l}{\footnotesize $^a$ Actively merging at redshift \zzero\ and excluded from the final sample (\galaxyname{r615} and \galaxyname{r918}).}\\
\multicolumn{9}{l}{\footnotesize $^b$ \galaxyname{r1223}'s star formation history was not available.}\\
\endlastfoot
r615$^a$ & $11.22$ & $9.25$ & $1.84$ & $151.19$ & $0.48$ & $-0.48$ & - & \nodiskcolor{No}\\
r468 & $10.88$ & $9.09$ & $0.85$ & $113.23$ & $-0.47$ & $-1.43$ & $0.1331$ & \nodiskcolor{No}\\
r568 & $10.84$ & $9.06$ & $2.26$ & $109.81$ & $-0.35$ & $-1.31$ & $0.6796$ & \diskcolor{Yes}\\
r488 & $10.92$ & $9.02$ & $1.39$ & $119.55$ & $-0.25$ & $-1.21$ & $0.1962$ & \nodiskcolor{No}\\
r442 & $10.92$ & $8.95$ & $2.43$ & $116.53$ & $-0.65$ & $-1.60$ & $0.7288$ & \diskcolor{Yes}\\
r431 & $10.99$ & $8.95$ & $3.91$ & $123.13$ & $-0.56$ & $-1.52$ & $0.5045$ & \diskcolor{Yes}\\
r502 & $10.90$ & $8.93$ & $1.29$ & $116.57$ & $-0.39$ & $-1.34$ & $0.2252$ & \nodiskcolor{No}\\
r492 & $10.90$ & $8.85$ & $3.78$ & $115.50$ & $-0.43$ & $-1.37$ & $0.6139$ & \diskcolor{Yes}\\
r515 & $10.94$ & $8.82$ & $4.93$ & $118.51$ & $-0.45$ & $-1.39$ & $0.5987$ & \diskcolor{Yes}\\
r556 & $10.77$ & $8.81$ & $1.20$ & $104.36$ & $-0.77$ & $-1.71$ & $0.0929$ & \nodiskcolor{No}\\
r555 & $10.79$ & $8.72$ & $0.85$ & $105.76$ & $-0.76$ & $-1.70$ & $0.1857$ & \nodiskcolor{No}\\
r489 & $10.86$ & $8.68$ & $2.25$ & $113.57$ & $-0.37$ & $-1.31$ & $0.3955$ & \nodiskcolor{No}\\
r544 & $10.86$ & $8.68$ & $3.22$ & $111.01$ & $-0.54$ & $-1.48$ & $0.7522$ & \diskcolor{Yes}\\
r656 & $10.71$ & $8.54$ & $2.32$ & $98.67$ & $-0.97$ & $-1.91$ & $0.4085$ & \diskcolor{Yes}\\
r613 & $10.78$ & $8.52$ & $2.70$ & $105.68$ & $-1.00$ & $-1.93$ & $0.4318$ & \diskcolor{Yes}\\
r597 & $10.76$ & $8.52$ & $2.85$ & $103.13$ & $-1.06$ & $-1.99$ & $0.7034$ & \diskcolor{Yes}\\
r523 & $10.82$ & $8.51$ & $3.02$ & $107.60$ & $-1.12$ & $-2.05$ & $0.4085$ & \diskcolor{Yes}\\
r642 & $10.73$ & $8.47$ & $3.36$ & $100.44$ & $-1.13$ & $-2.06$ & $0.5226$ & \diskcolor{Yes}\\
r569 & $10.74$ & $8.46$ & $3.49$ & $102.25$ & $-1.15$ & $-2.07$ & $0.4051$ & \diskcolor{Yes}\\
r918$^a$ & $10.80$ & $8.45$ & $2.62$ & $106.36$ & $-1.24$ & $-2.16$ & - & \nodiskcolor{No}\\
r634 & $10.72$ & $8.43$ & $2.66$ & $101.06$ & $-1.22$ & $-2.14$ & $0.2826$ & \diskcolor{Yes}\\
r571 & $10.81$ & $8.42$ & $3.19$ & $106.63$ & $-1.32$ & $-2.25$ & $0.1101$ & \diskcolor{Yes}\\
r618 & $10.73$ & $8.42$ & $0.88$ & $101.06$ & $-1.15$ & $-2.07$ & $0.1278$ & \nodiskcolor{No}\\
r614 & $10.73$ & $8.41$ & $3.87$ & $100.72$ & $-1.35$ & $-2.28$ & $0.3027$ & \diskcolor{Yes}\\
r563 & $10.73$ & $8.39$ & $1.43$ & $101.76$ & $-1.06$ & $-1.98$ & $0.1949$ & \nodiskcolor{No}\\
r886 & $10.65$ & $8.37$ & $2.06$ & $96.24$ & $-1.19$ & $-2.12$ & $0.6342$ & \diskcolor{Yes}\\
r552 & $10.73$ & $8.36$ & $3.03$ & $100.24$ & $-0.96$ & $-1.89$ & $0.3302$ & \diskcolor{Yes}\\
r758 & $10.61$ & $8.24$ & $0.96$ & $91.23$ & $-1.40$ & $-2.32$ & $0.1226$ & \nodiskcolor{No}\\
r751 & $10.70$ & $8.23$ & $3.42$ & $100.48$ & $-1.20$ & $-2.12$ & $0.3947$ & \diskcolor{Yes}\\
r700 & $10.64$ & $8.17$ & $2.42$ & $94.96$ & $-1.21$ & $-2.12$ & $0.3483$ & \diskcolor{Yes}\\
r741 & $10.59$ & $8.03$ & $2.56$ & $90.57$ & $-1.71$ & $-2.61$ & $0.1740$ & \diskcolor{Yes}\\
r718 & $10.61$ & $8.02$ & $3.30$ & $94.48$ & $-1.94$ & $-2.84$ & $0.2590$ & \diskcolor{Yes}\\
r719 & $10.57$ & $7.99$ & $1.45$ & $89.41$ & $-1.72$ & $-2.62$ & $0.2154$ & \diskcolor{Yes}\\
r749 & $10.61$ & $7.92$ & $1.77$ & $92.93$ & $-1.92$ & $-2.82$ & $0.1396$ & \diskcolor{Yes}\\
r716 & $10.59$ & $7.92$ & $0.86$ & $92.49$ & $-1.64$ & $-2.54$ & $0.0841$ & \nodiskcolor{No}\\
r761 & $10.59$ & $7.91$ & $2.81$ & $91.29$ & $-1.64$ & $-2.54$ & $0.1831$ & \diskcolor{Yes}\\
r916 & $10.47$ & $7.85$ & $1.35$ & $82.41$ & $-1.86$ & $-2.75$ & $0.1374$ & \diskcolor{Yes}\\
r852 & $10.54$ & $7.83$ & $0.78$ & $86.73$ & $-1.72$ & $-2.61$ & $0.1133$ & \nodiskcolor{No}\\
r753 & $10.50$ & $7.81$ & $1.14$ & $83.77$ & $-2.41$ & $-3.30$ & $0.0684$ & \nodiskcolor{No}\\
r850 & $10.49$ & $7.78$ & $0.62$ & $84.50$ & $-2.02$ & $-2.92$ & $0.0890$ & \nodiskcolor{No}\\
r968 & $10.37$ & $7.75$ & $0.99$ & $76.45$ & $-2.59$ & $-3.48$ & $0.0451$ & \diskcolor{Yes}\\
r977 & $10.44$ & $7.73$ & $1.73$ & $81.42$ & $-2.34$ & $-3.23$ & $0.0585$ & \diskcolor{Yes}\\
r707 & $10.48$ & $7.64$ & $0.42$ & $83.09$ & $-2.46$ & $-3.34$ & $0.0611$ & \nodiskcolor{No}\\
r832 & $10.48$ & $7.59$ & $0.66$ & $84.94$ & $-1.95$ & $-2.83$ & $0.0958$ & \nodiskcolor{No}\\
r1412 & $10.27$ & $7.54$ & $1.16$ & $70.95$ & $-2.13$ & $-3.00$ & $0.0629$ & \diskcolor{Yes}\\
r1129 & $10.35$ & $7.54$ & $1.12$ & $76.18$ & $-2.41$ & $-3.29$ & $0.0296$ & \diskcolor{Yes}\\
r1000 & $10.42$ & $7.53$ & $1.36$ & $79.16$ & $-2.53$ & $-3.40$ & $0.0405$ & \diskcolor{Yes}\\
r1901 & $10.33$ & $7.52$ & $0.46$ & $74.00$ & $-2.56$ & $-3.43$ & $0.1009$ & \nodiskcolor{No}\\
r1023 & $10.19$ & $7.52$ & $0.75$ & $66.21$ & $-2.77$ & $-3.65$ & $0.1093$ & \nodiskcolor{No}\\
r1074 & $10.43$ & $7.51$ & $1.95$ & $81.37$ & $-2.27$ & $-3.15$ & $0.1045$ & \diskcolor{Yes}\\
r1199 & $10.30$ & $7.34$ & $0.68$ & $73.44$ & $-2.75$ & $-3.61$ & $0.1352$ & \nodiskcolor{No}\\
r2472 & $10.16$ & $7.33$ & $1.09$ & $64.85$ & $-2.72$ & $-3.59$ & $0.0481$ & \diskcolor{Yes}\\
r1223$^b$ & $10.24$ & $6.58$ & $0.48$ & $72.23$ & $-$ & $-$ & - & \nodiskcolor{No}\\
\end{longtable}

\clearpage

\section{Parameters Used in Analysis}
\label{appendix:Parameters}

\subsection{Welch PSD Parameters}
\label{appendix:Welch PSD Parameters}

Motivated by \cite{Iyer2020}, we also adopt Welch's method\footnote{Implemented in \texttt{scipy.signal.welch}} \citep{1967ITAuE..15...70W} which reduces spectral variance by averaging over overlapping windowed segments. Welch's method \citep{1967ITAuE..15...70W} divides a time series into $K$ overlapping segments, detrends and windows each one, and averages their periodograms. Averaging reduces the variance of the PSD at each frequency, at the cost of frequency resolution, since the frequency spacing is set by the segment length ($\Delta f = 1/(n_{\rm seg} \Delta t)$). We set the segment length equal to the full time series ($n_{\rm seg} = N$, $K = 1$), so our \psdwelch is a single windowed and detrended periodogram at the finest frequency resolution the data allow. We make this choice because the peaks of cyclic SFHs are narrow. With a single segment the overlap parameter has no effect, and we instead reduce the variance of the PSD by averaging over galaxies when stacking the spectra of a sub-sample. The Hann window tapers the series smoothly to zero at both ends, which suppresses leakage of power from the strongly populated low frequencies into high frequencies. For spectra like ours, this leakage would otherwise flatten the measured high-frequency slope. Table \ref{tab:welch_params} summarizes the adopted parameters.

\begin{table}[ht]
\centering
\caption{Parameters adopted for the Welch PSD estimation \citep{1967ITAuE..15...70W}, implemented in \texttt{scipy.signal.welch}.}
\label{tab:welch_params}
\begin{tabular}{lll}
\hline
Parameter & Value & Description \\
\hline
Window          & Hann              & Taper applied to each segment \\
Detrending      & Linear            & Removed within each segment \\
$n_{\rm seg}$   & $N$               & Segment length (full time series) \\
\hline
\end{tabular}
\end{table}

\subsection{BFE Parameters}
\label{appendix:BFE Parameters}

\begin{figure*}[tp]
\includegraphics[width=\textwidth]{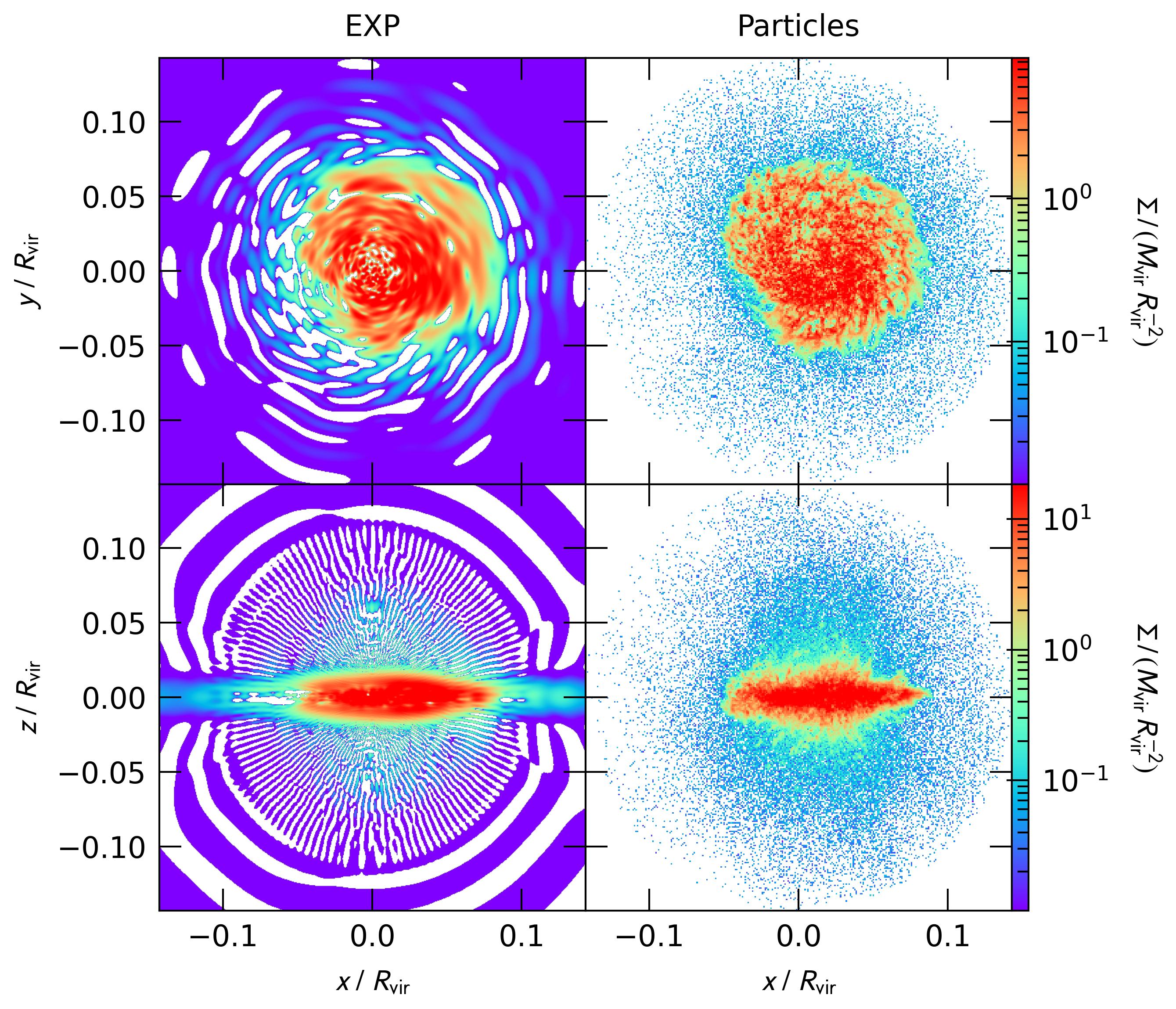}
\centering
\caption{\label{fig:exp_cold_gas_example}
Example of the cylindrical basis function expansion of the cold gas ($\log(T/{\rm K}) < 3.8$) in \galaxyname{r492} (a lopsided galaxy) at \zzero. The left column shows the surface density reconstructed from the expansion coefficients, using the basis parameters of Table \ref{tab:exp_params}, and the right column shows the surface density of the particles themselves. The top row is face-on and the bottom row is edge-on, with lengths in units of $R_{\rm vir}$ and surface densities in units of $M_{\rm vir}\,R_{\rm vir}^{-2}$. The expansion recovers the structure of the disk, including its lopsidedness and vertical extent, wherever the particle density is well sampled. The ringing visible outside the disk arises where the particle density is poorly sampled; these regions contain a negligible fraction of the cold gas mass and do not affect the coefficient power used in this work.
}
\end{figure*}

Table \ref{tab:exp_params} lists the parameters of the \texttt{cylinder} basis described in Section \ref{sec:Basis Function Expansions}. The scale length $a$ and scale height $h$ set the target density of Equation \ref{eq:target_density} and are measured separately for each galaxy, in units of the virial radius, which we adopt as the length unit of the expansion. The fiducial orders $l_{\rm max}^{\rm fid}$ and $n_{\rm max}^{\rm fid}$ set the size of the spherical basis from which the cylindrical functions are constructed. $m_{\rm max}$ and $n_{\rm max}$ set the number of azimuthal and radial functions retained in the expansion, of which $n_{\rm odd}$ per azimuthal order are antisymmetric about the midplane. The remaining entries control the grid on which the basis functions are tabulated and the numerical integration used to construct them. Figure \ref{fig:exp_cold_gas_example} shows an example expansion for one of the lopsided disk galaxies.

\begin{table}[ht]
\centering
\caption{Parameters of the \texttt{EXP} cylindrical basis. Lengths are in units of $R_{\rm vir}$.}
\label{tab:exp_params}
\begin{tabular}{lll}
\hline
Parameter & Value & Description \\
\hline
$a$ & $r_{1/2}^{xy}/R_{\rm vir}$ & Disk scale length \\
$h$ & $r_{1/2}^{z}/R_{\rm vir}$  & Disk scale height \\
$l_{\rm max}^{\rm fid}$ & $128$ & Fiducial spherical harmonic order \\
$n_{\rm max}^{\rm fid}$ & $72$  & Fiducial spherical radial order \\
$m_{\rm max}$ & $16$  & Max azimuthal order \\
$n_{\rm max}$ & $128$ & Radial functions per $m$ \\
$n_{\rm odd}$ & $12$  & anti-symmetric functions per $m$ \\
$n_R \times n_z$ & $512 \times 256$ & Meridional tabulation grid \\
$n_{\rm knot}^{R}$ & $200$ & Radial quadrature knots \\
$n_{\rm knot}^{\theta}$ & $80$ & Colatitude quadrature knots \\
$n_{\rm knot}^{\phi}$ & $1$ & Azimuthal quadrature knots \\
\hline
\end{tabular}
\end{table}

\clearpage
\bibliography{main}{}

@ARTICLE{Cruz2026,
       author = {{Cruz}, Akaxia and {Brooks}, Alyson M. and {Lisanti}, Mariangela and {Peter}, Annika H.~G. and {Geda}, Robel and {Quinn}, Thomas and {Tremmel}, Michael and {Munshi}, Ferah and {Keller}, Ben and {Wadsley}, James},
        title = "{Galaxy Size and Rotation Curve Diversity in {\ensuremath{\Lambda}}CDM with Baryons}",
      journal = {\apj},
         year = 2026,
        month = jul,
       volume = {1005},
       number = {2},
          eid = {157},
        pages = {157},
          doi = {10.3847/1538-4357/ae69ce},
archivePrefix = {arXiv},
       eprint = {2510.11800},
 primaryClass = {astro-ph.GA},
       adsurl = {https://ui.adsabs.harvard.edu/abs/2026ApJ..1005..157C}
}

@ARTICLE{Keith2025,
       author = {{Keith}, Blake and {Munshi}, Ferah and {Brooks}, Alyson M. and {Van Nest}, Jordan and {Engelhardt}, Anna and {Cruz}, Akaxia and {Keller}, Ben and {Quinn}, Thomas and {Wadsley}, James},
        title = "{A MARVEL-ous study of how well galaxy shapes reflect Dark Matter halo shapes in Cold Dark Matter Simulations}",
      journal = {arXiv e-prints},
         year = 2025,
        month = jan,
          eid = {arXiv:2501.16317},
        pages = {arXiv:2501.16317},
          doi = {10.48550/arXiv.2501.16317},
archivePrefix = {arXiv},
       eprint = {2501.16317},
 primaryClass = {astro-ph.GA},
       adsurl = {https://ui.adsabs.harvard.edu/abs/2025arXiv250116317K}
}

@ARTICLE{2025MNRAS.541.2180R,
       author = {{Ruan}, Dilys and {Brooks}, Alyson M. and {Cruz}, Akaxia and {Peter}, Annika H.~G. and {Keller}, Benjamin W. and {Quinn}, Thomas and {Wadsley}, James and {Adams}, Elizabeth A.~K.},
        title = "{Predictions for detecting a turndown in the baryonic Tully─Fisher relation}",
      journal = {\mnras},
         year = 2025,
        month = aug,
       volume = {541},
       number = {3},
        pages = {2180-2196},
          doi = {10.1093/mnras/staf1099},
archivePrefix = {arXiv},
       eprint = {2503.16607},
 primaryClass = {astro-ph.GA},
       adsurl = {https://ui.adsabs.harvard.edu/abs/2025MNRAS.541.2180R}
}

@ARTICLE{2025ApJ...993..230P,
       author = {{Piacitelli}, Daniel R. and {Brooks}, Alyson M. and {Christensen}, Charlotte and {Sanchez}, N. Nicole and {Faerman}, Yakov and {Shen}, Sijing and {Cruz}, Akaxia and {Keller}, Ben and {Quinn}, Thomas R. and {Wadsley}, James},
        title = "{Marvelous Metals: Surveying the Circumgalactic Medium of Simulated Dwarf Galaxies}",
      journal = {\apj},
         year = 2025,
        month = nov,
       volume = {993},
       number = {2},
          eid = {230},
        pages = {230},
          doi = {10.3847/1538-4357/ae06a0},
archivePrefix = {arXiv},
       eprint = {2505.08861},
 primaryClass = {astro-ph.GA},
       adsurl = {https://ui.adsabs.harvard.edu/abs/2025ApJ...993..230P}
}

@ARTICLE{Geda2026,
       author = {{Geda}, Robel and {Cruz}, Akaxia and {Wright}, Anna C. and {Greene}, Jenny E. and {Brooks}, Alyson and {Quinn}, Thomas and {Wadsley}, James and {Keller}, Ben},
        title = "{The Formation of Dwarf Galaxy Disks}",
      journal = {\apj},
         year = 2026,
        month = jun,
       volume = {1004},
       number = {1},
          eid = {110},
        pages = {110},
          doi = {10.3847/1538-4357/ae64fd},
archivePrefix = {arXiv},
       eprint = {2510.26875},
 primaryClass = {astro-ph.GA},
       adsurl = {https://ui.adsabs.harvard.edu/abs/2026ApJ..1004..110G}
}

@ARTICLE{Tremmel2015,
       author = {{Tremmel}, M. and {Governato}, F. and {Volonteri}, M. and {Quinn}, T.~R.},
        title = "{Off the beaten path: a new approach to realistically model the orbital decay of supermassive black holes in galaxy formation simulations}",
      journal = {\mnras},
         year = 2015,
        month = aug,
       volume = {451},
       number = {2},
        pages = {1868-1874},
          doi = {10.1093/mnras/stv1060},
archivePrefix = {arXiv},
       eprint = {1501.07609},
 primaryClass = {astro-ph.GA},
       adsurl = {https://ui.adsabs.harvard.edu/abs/2015MNRAS.451.1868T}
}

@ARTICLE{Tremmel2017,
       author = {{Tremmel}, M. and {Karcher}, M. and {Governato}, F. and {Volonteri}, M. and {Quinn}, T.~R. and {Pontzen}, A. and {Anderson}, L. and {Bellovary}, J.},
        title = "{The Romulus cosmological simulations: a physical approach to the formation, dynamics and accretion models of SMBHs}",
      journal = {\mnras},
         year = 2017,
        month = sep,
       volume = {470},
       number = {1},
        pages = {1121-1139},
          doi = {10.1093/mnras/stx1160},
archivePrefix = {arXiv},
       eprint = {1607.02151},
 primaryClass = {astro-ph.GA},
       adsurl = {https://ui.adsabs.harvard.edu/abs/2017MNRAS.470.1121T}
}

@ARTICLE{2019MNRAS.482.2913B,
       author = {{Bellovary}, Jillian M. and {Cleary}, Colleen E. and {Munshi}, Ferah and {Tremmel}, Michael and {Christensen}, Charlotte R. and {Brooks}, Alyson and {Quinn}, Thomas R.},
        title = "{Multimessenger signatures of massive black holes in dwarf galaxies}",
      journal = {\mnras},
         year = 2019,
        month = jan,
       volume = {482},
       number = {3},
        pages = {2913-2923},
          doi = {10.1093/mnras/sty2842},
archivePrefix = {arXiv},
       eprint = {1806.00471},
 primaryClass = {astro-ph.GA},
       adsurl = {https://ui.adsabs.harvard.edu/abs/2019MNRAS.482.2913B}
}

@ARTICLE{Booth2009,
       author = {{Booth}, C.~M. and {Schaye}, Joop},
        title = "{Cosmological simulations of the growth of supermassive black holes and feedback from active galactic nuclei: method and tests}",
      journal = {\mnras},
         year = 2009,
        month = sep,
       volume = {398},
       number = {1},
        pages = {53-74},
          doi = {10.1111/j.1365-2966.2009.15043.x},
archivePrefix = {arXiv},
       eprint = {0904.2572},
 primaryClass = {astro-ph.CO},
       adsurl = {https://ui.adsabs.harvard.edu/abs/2009MNRAS.398...53B}
}

@INPROCEEDINGS{Jetley2008,
  author={Jetley, Pritish and Gioachin, Filippo and Mendes, Celso and Kale, Laxmikant V. and Quinn, Thomas},
  booktitle={2008 IEEE International Symposium on Parallel and Distributed Processing}, 
  title={Massively parallel cosmological simulations with ChaNGa}, 
  year={2008},
  volume={},
  number={},
  pages={1-12},
  doi={10.1109/IPDPS.2008.4536319}
}

@INPROCEEDINGS{Jetley2010,
  author={Jetley, Pritish and Wesolowski, Lukasz and Gioachin, Filippo and Kalé, Laxmikant V. and Quinn, Thomas R.},
  booktitle={SC '10: Proceedings of the 2010 ACM/IEEE International Conference for High Performance Computing, Networking, Storage and Analysis}, 
  title={Scaling Hierarchical N-body Simulations on GPU Clusters}, 
  year={2010},
  volume={},
  number={},
  pages={1-11},
  doi={10.1109/SC.2010.49}
}

@ARTICLE{Menon2015,
       author = {{Menon}, Harshitha and {Wesolowski}, Lukasz and {Zheng}, Gengbin and {Jetley}, Pritish and {Kale}, Laxmikant and {Quinn}, Thomas and {Governato}, Fabio},
        title = "{Adaptive techniques for clustered N-body cosmological simulations}",
      journal = {Computational Astrophysics and Cosmology},
         year = 2015,
        month = mar,
       volume = {2},
          eid = {1},
        pages = {1},
          doi = {10.1186/s40668-015-0007-9},
archivePrefix = {arXiv},
       eprint = {1409.1929},
 primaryClass = {astro-ph.IM},
       adsurl = {https://ui.adsabs.harvard.edu/abs/2015ComAC...2....1M}
}

@ARTICLE{Regan2020,
       author = {{Regan}, John A. and {Wise}, John H. and {Woods}, Tyrone E. and {Downes}, Turlough P. and {O'Shea}, Brian W. and {Norman}, Michael L.},
        title = "{The Formation of Very Massive Stars in Early Galaxies and Implications for Intermediate Mass Black Holes}",
      journal = {The Open Journal of Astrophysics},
         year = 2020,
        month = dec,
       volume = {3},
       number = {1},
          eid = {15},
        pages = {15},
          doi = {10.21105/astro.2008.08090},
archivePrefix = {arXiv},
       eprint = {2008.08090},
 primaryClass = {astro-ph.GA},
       adsurl = {https://ui.adsabs.harvard.edu/abs/2020OJAp....3E..15R}
}

@ARTICLE{Regan2023,
       author = {{Regan}, John},
        title = "{Massive Star Formation in Overdense Regions of the Early Universe}",
      journal = {The Open Journal of Astrophysics},
         year = 2023,
        month = apr,
       volume = {6},
          eid = {12},
        pages = {12},
          doi = {10.21105/astro.2210.04899},
archivePrefix = {arXiv},
       eprint = {2210.04899},
 primaryClass = {astro-ph.GA},
       adsurl = {https://ui.adsabs.harvard.edu/abs/2023OJAp....6E..12R}
}

@ARTICLE{Wise2019,
       author = {{Wise}, John H. and {Regan}, John A. and {O'Shea}, Brian W. and {Norman}, Michael L. and {Downes}, Turlough P. and {Xu}, Hao},
        title = "{Formation of massive black holes in rapidly growing pre-galactic gas clouds}",
      journal = {\nat},
         year = 2019,
        month = jan,
       volume = {566},
       number = {7742},
        pages = {85-88},
          doi = {10.1038/s41586-019-0873-4},
archivePrefix = {arXiv},
       eprint = {1901.07563},
 primaryClass = {astro-ph.GA},
       adsurl = {https://ui.adsabs.harvard.edu/abs/2019Natur.566...85W}
}

@ARTICLE{2022MNRAS.510..959I,
       author = {{Ivkovic}, Mladen and {Teyssier}, Romain},
        title = "{ACACIA: a new method to produce on-the-fly merger trees in the RAMSES code}",
      journal = {\mnras},
         year = 2022,
        month = feb,
       volume = {510},
       number = {1},
        pages = {959-979},
          doi = {10.1093/mnras/stab3329},
archivePrefix = {arXiv},
       eprint = {1812.06708},
 primaryClass = {astro-ph.GA},
       adsurl = {https://ui.adsabs.harvard.edu/abs/2022MNRAS.510..959I}
}

@article{10.1093/mnras/staf016,
    author = {Geda, Robel and Teyssier, Romain},
    title = {Constructing merger trees of density peaks using phase-space watershed segmentation algorithm},
    journal = {Monthly Notices of the Royal Astronomical Society},
    pages = {staf016},
    year = {2025},
    month = {01},
    issn = {0035-8711},
    doi = {10.1093/mnras/staf016},
    url = {https://doi.org/10.1093/mnras/staf016},
    eprint = {https://academic.oup.com/mnras/advance-article-pdf/doi/10.1093/mnras/staf016/61430228/staf016.pdf},
}

@ARTICLE{1998ApJ...495...80B,
       author = {{Bryan}, Greg L. and {Norman}, Michael L.},
        title = "{Statistical Properties of X-Ray Clusters: Analytic and Numerical Comparisons}",
      journal = {\apj},
         year = 1998,
        month = mar,
       volume = {495},
       number = {1},
        pages = {80-99},
          doi = {10.1086/305262},
archivePrefix = {arXiv},
       eprint = {astro-ph/9710107},
 primaryClass = {astro-ph},
       adsurl = {https://ui.adsabs.harvard.edu/abs/1998ApJ...495...80B}
}

@ARTICLE{2012ApJ...757...85G,
       author = {{Geha}, M. and {Blanton}, M.~R. and {Yan}, R. and {Tinker}, J.~L.},
        title = "{A Stellar Mass Threshold for Quenching of Field Galaxies}",
      journal = {\apj},
         year = 2012,
        month = sep,
       volume = {757},
       number = {1},
          eid = {85},
        pages = {85},
          doi = {10.1088/0004-637X/757/1/85},
archivePrefix = {arXiv},
       eprint = {1206.3573},
 primaryClass = {astro-ph.CO},
       adsurl = {https://ui.adsabs.harvard.edu/abs/2012ApJ...757...85G}
}

@ARTICLE{2001MNRAS.325..845K,
       author = {{Knebe}, Alexander and {Green}, Andrew and {Binney}, James},
        title = "{Multi-level adaptive particle mesh (MLAPM): a c code for cosmological simulations}",
      journal = {\mnras},
         year = 2001,
        month = aug,
       volume = {325},
       number = {2},
        pages = {845-864},
          doi = {10.1046/j.1365-8711.2001.04532.x},
archivePrefix = {arXiv},
       eprint = {astro-ph/0103503},
 primaryClass = {astro-ph},
       adsurl = {https://ui.adsabs.harvard.edu/abs/2001MNRAS.325..845K}
}

@ARTICLE{Wadsley2004,
       author = {{Wadsley}, J.~W. and {Stadel}, J. and {Quinn}, T.},
        title = "{Gasoline: a flexible, parallel implementation of TreeSPH}",
      journal = {\na},
         year = 2004,
        month = feb,
       volume = {9},
       number = {2},
        pages = {137-158},
          doi = {10.1016/j.newast.2003.08.004},
archivePrefix = {arXiv},
       eprint = {astro-ph/0303521},
 primaryClass = {astro-ph},
       adsurl = {https://ui.adsabs.harvard.edu/abs/2004NewA....9..137W}
}

@ARTICLE{Wadsley2017,
       author = {{Wadsley}, James W. and {Keller}, Benjamin W. and {Quinn}, Thomas R.},
        title = "{Gasoline2: a modern smoothed particle hydrodynamics code}",
      journal = {\mnras},
         year = 2017,
        month = oct,
       volume = {471},
       number = {2},
        pages = {2357-2369},
          doi = {10.1093/mnras/stx1643},
archivePrefix = {arXiv},
       eprint = {1707.03824},
 primaryClass = {astro-ph.IM},
       adsurl = {https://ui.adsabs.harvard.edu/abs/2017MNRAS.471.2357W}
}

@ARTICLE{2010MNRAS.407.1581S,
       author = {{Shen}, S. and {Wadsley}, J. and {Stinson}, G.},
        title = "{The enrichment of the intergalactic medium with adiabatic feedback - I. Metal cooling and metal diffusion}",
      journal = {\mnras},
         year = 2010,
        month = sep,
       volume = {407},
       number = {3},
        pages = {1581-1596},
          doi = {10.1111/j.1365-2966.2010.17047.x},
archivePrefix = {arXiv},
       eprint = {0910.5956},
 primaryClass = {astro-ph.CO},
       adsurl = {https://ui.adsabs.harvard.edu/abs/2010MNRAS.407.1581S}
}

@ARTICLE{2012MNRAS.425.3058C,
       author = {{Christensen}, Charlotte and {Quinn}, Thomas and {Governato}, Fabio and {Stilp}, Adrienne and {Shen}, Sijing and {Wadsley}, James},
        title = "{Implementing molecular hydrogen in hydrodynamic simulations of galaxy formation}",
      journal = {\mnras},
         year = 2012,
        month = oct,
       volume = {425},
       number = {4},
        pages = {3058-3076},
          doi = {10.1111/j.1365-2966.2012.21628.x},
archivePrefix = {arXiv},
       eprint = {1205.5567},
 primaryClass = {astro-ph.CO},
       adsurl = {https://ui.adsabs.harvard.edu/abs/2012MNRAS.425.3058C}
}

@ARTICLE{2006MNRAS.373.1074S,
       author = {{Stinson}, Greg and {Seth}, Anil and {Katz}, Neal and {Wadsley}, James and {Governato}, Fabio and {Quinn}, Tom},
        title = "{Star formation and feedback in smoothed particle hydrodynamic simulations - I. Isolated galaxies}",
      journal = {\mnras},
         year = 2006,
        month = dec,
       volume = {373},
       number = {3},
        pages = {1074-1090},
          doi = {10.1111/j.1365-2966.2006.11097.x},
archivePrefix = {arXiv},
       eprint = {astro-ph/0602350},
 primaryClass = {astro-ph},
       adsurl = {https://ui.adsabs.harvard.edu/abs/2006MNRAS.373.1074S}
}

@ARTICLE{2014MNRAS.442.3013K,
       author = {{Keller}, B.~W. and {Wadsley}, J. and {Benincasa}, S.~M. and {Couchman}, H.~M.~P.},
        title = "{A superbubble feedback model for galaxy simulations}",
      journal = {\mnras},
         year = 2014,
        month = aug,
       volume = {442},
       number = {4},
        pages = {3013-3025},
          doi = {10.1093/mnras/stu1058},
archivePrefix = {arXiv},
       eprint = {1405.2625},
 primaryClass = {astro-ph.GA},
       adsurl = {https://ui.adsabs.harvard.edu/abs/2014MNRAS.442.3013K}
}

@ARTICLE{2014A&A...571A..16P,
       author = {{Planck Collaboration} and {Ade}, P.~A.~R. and {Aghanim}, N. and {Armitage-Caplan}, C. and {Arnaud}, M. and {Ashdown}, M. and {Atrio-Barandela}, F. and {Aumont}, J. and {Baccigalupi}, C. and {Banday}, A.~J. and {Barreiro}, R.~B. and {Bartlett}, J.~G. and {Battaner}, E. and {Benabed}, K. and {Beno{\^\i}t}, A. and {Benoit-L{\'e}vy}, A. and {Bernard}, J. -P. and {Bersanelli}, M. and {Bielewicz}, P. and {Bobin}, J. and {Bock}, J.~J. and {Bonaldi}, A. and {Bond}, J.~R. and {Borrill}, J. and {Bouchet}, F.~R. and {Bridges}, M. and {Bucher}, M. and {Burigana}, C. and {Butler}, R.~C. and {Calabrese}, E. and {Cappellini}, B. and {Cardoso}, J. -F. and {Catalano}, A. and {Challinor}, A. and {Chamballu}, A. and {Chary}, R. -R. and {Chen}, X. and {Chiang}, H.~C. and {Chiang}, L. -Y. and {Christensen}, P.~R. and {Church}, S. and {Clements}, D.~L. and {Colombi}, S. and {Colombo}, L.~P.~L. and {Couchot}, F. and {Coulais}, A. and {Crill}, B.~P. and {Curto}, A. and {Cuttaia}, F. and {Danese}, L. and {Davies}, R.~D. and {Davis}, R.~J. and {de Bernardis}, P. and {de Rosa}, A. and {de Zotti}, G. and {Delabrouille}, J. and {Delouis}, J. -M. and {D{\'e}sert}, F. -X. and {Dickinson}, C. and {Diego}, J.~M. and {Dolag}, K. and {Dole}, H. and {Donzelli}, S. and {Dor{\'e}}, O. and {Douspis}, M. and {Dunkley}, J. and {Dupac}, X. and {Efstathiou}, G. and {Elsner}, F. and {En{\ss}lin}, T.~A. and {Eriksen}, H.~K. and {Finelli}, F. and {Forni}, O. and {Frailis}, M. and {Fraisse}, A.~A. and {Franceschi}, E. and {Gaier}, T.~C. and {Galeotta}, S. and {Galli}, S. and {Ganga}, K. and {Giard}, M. and {Giardino}, G. and {Giraud-H{\'e}raud}, Y. and {Gjerl{\o}w}, E. and {Gonz{\'a}lez-Nuevo}, J. and {G{\'o}rski}, K.~M. and {Gratton}, S. and {Gregorio}, A. and {Gruppuso}, A. and {Gudmundsson}, J.~E. and {Haissinski}, J. and {Hamann}, J. and {Hansen}, F.~K. and {Hanson}, D. and {Harrison}, D. and {Henrot-Versill{\'e}}, S. and {Hern{\'a}ndez-Monteagudo}, C. and {Herranz}, D. and {Hildebrandt}, S.~R. and {Hivon}, E. and {Hobson}, M. and {Holmes}, W.~A. and {Hornstrup}, A. and {Hou}, Z. and {Hovest}, W. and {Huffenberger}, K.~M. and {Jaffe}, A.~H. and {Jaffe}, T.~R. and {Jewell}, J. and {Jones}, W.~C. and {Juvela}, M. and {Keih{\"a}nen}, E. and {Keskitalo}, R. and {Kisner}, T.~S. and {Kneissl}, R. and {Knoche}, J. and {Knox}, L. and {Kunz}, M. and {Kurki-Suonio}, H. and {Lagache}, G. and {L{\"a}hteenm{\"a}ki}, A. and {Lamarre}, J. -M. and {Lasenby}, A. and {Lattanzi}, M. and {Laureijs}, R.~J. and {Lawrence}, C.~R. and {Leach}, S. and {Leahy}, J.~P. and {Leonardi}, R. and {Le{\'o}n-Tavares}, J. and {Lesgourgues}, J. and {Lewis}, A. and {Liguori}, M. and {Lilje}, P.~B. and {Linden-V{\o}rnle}, M. and {L{\'o}pez-Caniego}, M. and {Lubin}, P.~M. and {Mac{\'\i}as-P{\'e}rez}, J.~F. and {Maffei}, B. and {Maino}, D. and {Mandolesi}, N. and {Maris}, M. and {Marshall}, D.~J. and {Martin}, P.~G. and {Mart{\'\i}nez-Gonz{\'a}lez}, E. and {Masi}, S. and {Massardi}, M. and {Matarrese}, S. and {Matthai}, F. and {Mazzotta}, P. and {Meinhold}, P.~R. and {Melchiorri}, A. and {Melin}, J. -B. and {Mendes}, L. and {Menegoni}, E. and {Mennella}, A. and {Migliaccio}, M. and {Millea}, M. and {Mitra}, S. and {Miville-Desch{\^e}nes}, M. -A. and {Moneti}, A. and {Montier}, L. and {Morgante}, G. and {Mortlock}, D. and {Moss}, A. and {Munshi}, D. and {Murphy}, J.~A. and {Naselsky}, P. and {Nati}, F. and {Natoli}, P. and {Netterfield}, C.~B. and {N{\o}rgaard-Nielsen}, H.~U. and {Noviello}, F. and {Novikov}, D. and {Novikov}, I. and {O'Dwyer}, I.~J. and {Osborne}, S. and {Oxborrow}, C.~A. and {Paci}, F. and {Pagano}, L. and {Pajot}, F. and {Paladini}, R. and {Paoletti}, D. and {Partridge}, B. and {Pasian}, F. and {Patanchon}, G. and {Pearson}, D. and {Pearson}, T.~J. and {Peiris}, H.~V. and {Perdereau}, O. and {Perotto}, L. and {Perrotta}, F. and {Pettorino}, V. and {Piacentini}, F. and {Piat}, M. and {Pierpaoli}, E. and {Pietrobon}, D. and {Plaszczynski}, S. and {Platania}, P. and {Pointecouteau}, E.},
        title = "{Planck 2013 results. XVI. Cosmological parameters}",
      journal = {\aap},
         year = 2014,
        month = nov,
       volume = {571},
          eid = {A16},
        pages = {A16},
          doi = {10.1051/0004-6361/201321591},
archivePrefix = {arXiv},
       eprint = {1303.5076},
 primaryClass = {astro-ph.CO},
       adsurl = {https://ui.adsabs.harvard.edu/abs/2014A&A...571A..16P}
}

@ARTICLE{2004MNRAS.351..399G,
       author = {{Gill}, Stuart P.~D. and {Knebe}, Alexander and {Gibson}, Brad K.},
        title = "{The evolution of substructure - I. A new identification method}",
      journal = {\mnras},
         year = 2004,
        month = jun,
       volume = {351},
       number = {2},
        pages = {399-409},
          doi = {10.1111/j.1365-2966.2004.07786.x},
archivePrefix = {arXiv},
       eprint = {astro-ph/0404258},
 primaryClass = {astro-ph},
       adsurl = {https://ui.adsabs.harvard.edu/abs/2004MNRAS.351..399G}
}

@ARTICLE{Iyer2020,
       author = {{Iyer}, Kartheik G. and {Tacchella}, Sandro and {Genel}, Shy and {Hayward}, Christopher C. and {Hernquist}, Lars and {Brooks}, Alyson M. and {Caplar}, Neven and {Dav{\'e}}, Romeel and {Diemer}, Benedikt and {Forbes}, John C. and {Gawiser}, Eric and {Somerville}, Rachel S. and {Starkenburg}, Tjitske K.},
        title = "{The diversity and variability of star formation histories in models of galaxy evolution}",
      journal = {\mnras},
         year = 2020,
        month = oct,
       volume = {498},
       number = {1},
        pages = {430-463},
          doi = {10.1093/mnras/staa2150},
archivePrefix = {arXiv},
       eprint = {2007.07916},
 primaryClass = {astro-ph.GA},
       adsurl = {https://ui.adsabs.harvard.edu/abs/2020MNRAS.498..430I}
}

@ARTICLE{1967ITAuE..15...70W,
       author = {{Welch}, P.},
        title = "{The use of fast Fourier transform for the estimation of power spectra: A method based on time averaging over short, modified periodograms}",
      journal = {IEEE Transactions on Audio and Electroacoustics},
         year = 1967,
        month = jun,
       volume = {15},
       number = {2},
        pages = {70-73},
          doi = {10.1109/TAU.1967.1161901},
       adsurl = {https://ui.adsabs.harvard.edu/abs/1967ITAuE..15...70W}
}

@ARTICLE{KadoFong2024,
       author = {{Kado-Fong}, Erin and {Geha}, Marla and {Mao}, Yao-Yuan and {de los Reyes}, Mithi A.~C. and {Wechsler}, Risa H. and {Weiner}, Benjamin and {Asali}, Yasmeen and {Kallivayalil}, Nitya and {Nadler}, Ethan O. and {Tollerud}, Erik J. and {Wang}, Yunchong},
        title = "{SAGAbg. II. The Low-mass Star-forming Sequence Evolves Significantly between 0.05 < z < 0.21}",
      journal = {\apj},
         year = 2024,
        month = nov,
       volume = {976},
       number = {1},
          eid = {83},
        pages = {83},
          doi = {10.3847/1538-4357/ad8137},
archivePrefix = {arXiv},
       eprint = {2409.12221},
 primaryClass = {astro-ph.GA},
       adsurl = {https://ui.adsabs.harvard.edu/abs/2024ApJ...976...83K}
}

@ARTICLE{Tacchella2020,
       author = {{Tacchella}, Sandro and {Forbes}, John C. and {Caplar}, Neven},
        title = "{Stochastic modelling of star-formation histories II: star-formation variability from molecular clouds and gas inflow}",
      journal = {\mnras},
         year = 2020,
        month = sep,
       volume = {497},
       number = {1},
        pages = {698-725},
          doi = {10.1093/mnras/staa1838},
archivePrefix = {arXiv},
       eprint = {2006.09382},
 primaryClass = {astro-ph.GA},
       adsurl = {https://ui.adsabs.harvard.edu/abs/2020MNRAS.497..698T}
}

@ARTICLE{Caplar2019,
       author = {{Caplar}, Neven and {Tacchella}, Sandro},
        title = "{Stochastic modelling of star-formation histories I: the scatter of the star-forming main sequence}",
      journal = {\mnras},
         year = 2019,
        month = aug,
       volume = {487},
       number = {3},
        pages = {3845-3869},
          doi = {10.1093/mnras/stz1449},
archivePrefix = {arXiv},
       eprint = {1901.07556},
 primaryClass = {astro-ph.GA},
       adsurl = {https://ui.adsabs.harvard.edu/abs/2019MNRAS.487.3845C}
}

@ARTICLE{ElBadry16,
       author = {{El-Badry}, Kareem and {Wetzel}, Andrew and {Geha}, Marla and {Hopkins}, Philip F. and {Kere{\v{s}}}, Dusan and {Chan}, T.~K. and {Faucher-Gigu{\`e}re}, Claude-Andr{\'e}},
        title = "{Breathing FIRE: How Stellar Feedback Drives Radial Migration, Rapid Size Fluctuations, and Population Gradients in Low-mass Galaxies}",
      journal = {\apj},
         year = 2016,
        month = apr,
       volume = {820},
       number = {2},
          eid = {131},
        pages = {131},
          doi = {10.3847/0004-637X/820/2/131},
archivePrefix = {arXiv},
       eprint = {1512.01235},
 primaryClass = {astro-ph.GA},
       adsurl = {https://ui.adsabs.harvard.edu/abs/2016ApJ...820..131E}
}

@ARTICLE{Weinberg1999,
       author = {{Weinberg}, Martin D.},
        title = "{An Adaptive Algorithm for N-Body Field Expansions}",
      journal = {\aj},
         year = 1999,
        month = jan,
       volume = {117},
       number = {1},
        pages = {629-637},
          doi = {10.1086/300669},
archivePrefix = {arXiv},
       eprint = {astro-ph/9805357},
 primaryClass = {astro-ph},
       adsurl = {https://ui.adsabs.harvard.edu/abs/1999AJ....117..629W}
}

@ARTICLE{Petersen2022,
       author = {{Petersen}, Michael S. and {Weinberg}, Martin D. and {Katz}, Neal},
        title = "{EXP: N-body integration using basis function expansions}",
      journal = {\mnras},
         year = 2022,
        month = mar,
       volume = {510},
       number = {4},
        pages = {6201-6217},
          doi = {10.1093/mnras/stab3639},
archivePrefix = {arXiv},
       eprint = {2104.14577},
 primaryClass = {astro-ph.GA},
       adsurl = {https://ui.adsabs.harvard.edu/abs/2022MNRAS.510.6201P}
}

@ARTICLE{Petersen2024,
       author = {{Petersen}, Michael and {Weinberg}, Martin},
        title = "{EXP: a Python/C++ package for basis function expansion methods in galactic dynamics}",
      journal = {The Journal of Open Source Software},
         year = 2025,
        month = may,
       volume = {10},
       number = {109},
          eid = {7302},
        pages = {7302},
          doi = {10.21105/joss.07302},
archivePrefix = {arXiv},
       eprint = {2505.06403},
 primaryClass = {astro-ph.GA},
       adsurl = {https://ui.adsabs.harvard.edu/abs/2025JOSS...10.7302P}
}

@ARTICLE{2013ApJ...766...56M,
       author = {{Munshi}, Ferah and {Governato}, F. and {Brooks}, A.~M. and {Christensen}, C. and {Shen}, S. and {Loebman}, S. and {Moster}, B. and {Quinn}, T. and {Wadsley}, J.},
        title = "{Reproducing the Stellar Mass/Halo Mass Relation in Simulated {\ensuremath{\Lambda}}CDM Galaxies: Theory versus Observational Estimates}",
      journal = {\apj},
         year = 2013,
        month = mar,
       volume = {766},
       number = {1},
          eid = {56},
        pages = {56},
          doi = {10.1088/0004-637X/766/1/56},
archivePrefix = {arXiv},
       eprint = {1209.1389},
 primaryClass = {astro-ph.CO},
       adsurl = {https://ui.adsabs.harvard.edu/abs/2013ApJ...766...56M}
}

@ARTICLE{1929RSPSA.125...54F,
       author = {{Fisher}, R.~A.},
        title = "{Tests of Significance in Harmonic Analysis}",
      journal = {Proceedings of the Royal Society of London Series A},
         year = 1929,
        month = aug,
       volume = {125},
       number = {796},
        pages = {54-59},
          doi = {10.1098/rspa.1929.0151},
       adsurl = {https://ui.adsabs.harvard.edu/abs/1929RSPSA.125...54F}
}

@article{astropy:2013,
Adsurl = {http://adsabs.harvard.edu/abs/2013A%26A...558A..33A},
Archiveprefix = {arXiv},
Author = {{Astropy Collaboration} and {Robitaille}, T.~P. and {Tollerud}, E.~J. and {Greenfield}, P. and {Droettboom}, M. and {Bray}, E. and {Aldcroft}, T. and {Davis}, M. and {Ginsburg}, A. and {Price-Whelan}, A.~M. and {Kerzendorf}, W.~E. and {Conley}, A. and {Crighton}, N. and {Barbary}, K. and {Muna}, D. and {Ferguson}, H. and {Grollier}, F. and {Parikh}, M.~M. and {Nair}, P.~H. and {Unther}, H.~M. and {Deil}, C. and {Woillez}, J. and {Conseil}, S. and {Kramer}, R. and {Turner}, J.~E.~H. and {Singer}, L. and {Fox}, R. and {Weaver}, B.~A. and {Zabalza}, V. and {Edwards}, Z.~I. and {Azalee Bostroem}, K. and {Burke}, D.~J. and {Casey}, A.~R. and {Crawford}, S.~M. and {Dencheva}, N. and {Ely}, J. and {Jenness}, T. and {Labrie}, K. and {Lim}, P.~L. and {Pierfederici}, F. and {Pontzen}, A. and {Ptak}, A. and {Refsdal}, B. and {Servillat}, M. and {Streicher}, O.},
Doi = {10.1051/0004-6361/201322068},
Eid = {A33},
Eprint = {1307.6212},
Journal = {\aap},
Month = oct,
Pages = {A33},
Primaryclass = {astro-ph.IM},
Title = {{Astropy: A community Python package for astronomy}},
Volume = 558,
Year = 2013}

@ARTICLE{astropy:2018,
       author = {{Astropy Collaboration} and {Price-Whelan}, A.~M. and
         {Sip{\H{o}}cz}, B.~M. and {G{\"u}nther}, H.~M. and {Lim}, P.~L. and
         {Crawford}, S.~M. and {Conseil}, S. and {Shupe}, D.~L. and
         {Craig}, M.~W. and {Dencheva}, N. and {Ginsburg}, A. and {Vand
        erPlas}, J.~T. and {Bradley}, L.~D. and {P{\'e}rez-Su{\'a}rez}, D. and
         {de Val-Borro}, M. and {Aldcroft}, T.~L. and {Cruz}, K.~L. and
         {Robitaille}, T.~P. and {Tollerud}, E.~J. and {Ardelean}, C. and
         {Babej}, T. and {Bach}, Y.~P. and {Bachetti}, M. and {Bakanov}, A.~V. and
         {Bamford}, S.~P. and {Barentsen}, G. and {Barmby}, P. and
         {Baumbach}, A. and {Berry}, K.~L. and {Biscani}, F. and {Boquien}, M. and
         {Bostroem}, K.~A. and {Bouma}, L.~G. and {Brammer}, G.~B. and
         {Bray}, E.~M. and {Breytenbach}, H. and {Buddelmeijer}, H. and
         {Burke}, D.~J. and {Calderone}, G. and {Cano Rodr{\'\i}guez}, J.~L. and
         {Cara}, M. and {Cardoso}, J.~V.~M. and {Cheedella}, S. and {Copin}, Y. and
         {Corrales}, L. and {Crichton}, D. and {D'Avella}, D. and {Deil}, C. and
         {Depagne}, {\'E}. and {Dietrich}, J.~P. and {Donath}, A. and
         {Droettboom}, M. and {Earl}, N. and {Erben}, T. and {Fabbro}, S. and
         {Ferreira}, L.~A. and {Finethy}, T. and {Fox}, R.~T. and
         {Garrison}, L.~H. and {Gibbons}, S.~L.~J. and {Goldstein}, D.~A. and
         {Gommers}, R. and {Greco}, J.~P. and {Greenfield}, P. and
         {Groener}, A.~M. and {Grollier}, F. and {Hagen}, A. and {Hirst}, P. and
         {Homeier}, D. and {Horton}, A.~J. and {Hosseinzadeh}, G. and {Hu}, L. and
         {Hunkeler}, J.~S. and {Ivezi{\'c}}, {\v{Z}}. and {Jain}, A. and
         {Jenness}, T. and {Kanarek}, G. and {Kendrew}, S. and {Kern}, N.~S. and
         {Kerzendorf}, W.~E. and {Khvalko}, A. and {King}, J. and {Kirkby}, D. and
         {Kulkarni}, A.~M. and {Kumar}, A. and {Lee}, A. and {Lenz}, D. and
         {Littlefair}, S.~P. and {Ma}, Z. and {Macleod}, D.~M. and
         {Mastropietro}, M. and {McCully}, C. and {Montagnac}, S. and
         {Morris}, B.~M. and {Mueller}, M. and {Mumford}, S.~J. and {Muna}, D. and
         {Murphy}, N.~A. and {Nelson}, S. and {Nguyen}, G.~H. and
         {Ninan}, J.~P. and {N{\"o}the}, M. and {Ogaz}, S. and {Oh}, S. and
         {Parejko}, J.~K. and {Parley}, N. and {Pascual}, S. and {Patil}, R. and
         {Patil}, A.~A. and {Plunkett}, A.~L. and {Prochaska}, J.~X. and
         {Rastogi}, T. and {Reddy Janga}, V. and {Sabater}, J. and
         {Sakurikar}, P. and {Seifert}, M. and {Sherbert}, L.~E. and
         {Sherwood-Taylor}, H. and {Shih}, A.~Y. and {Sick}, J. and
         {Silbiger}, M.~T. and {Singanamalla}, S. and {Singer}, L.~P. and
         {Sladen}, P.~H. and {Sooley}, K.~A. and {Sornarajah}, S. and
         {Streicher}, O. and {Teuben}, P. and {Thomas}, S.~W. and
         {Tremblay}, G.~R. and {Turner}, J.~E.~H. and {Terr{\'o}n}, V. and
         {van Kerkwijk}, M.~H. and {de la Vega}, A. and {Watkins}, L.~L. and
         {Weaver}, B.~A. and {Whitmore}, J.~B. and {Woillez}, J. and
         {Zabalza}, V. and {Astropy Contributors}},
        title = "{The Astropy Project: Building an Open-science Project and Status of the v2.0 Core Package}",
      journal = {\aj},
         year = 2018,
        month = sep,
       volume = {156},
       number = {3},
          eid = {123},
        pages = {123},
          doi = {10.3847/1538-3881/aabc4f},
archivePrefix = {arXiv},
       eprint = {1801.02634},
 primaryClass = {astro-ph.IM},
       adsurl = {https://ui.adsabs.harvard.edu/abs/2018AJ....156..123A}
}

@ARTICLE{astropy:2022,
       author = {{Astropy Collaboration} and {Price-Whelan}, Adrian M. and {Lim}, Pey Lian and {Earl}, Nicholas and {Starkman}, Nathaniel and {Bradley}, Larry and {Shupe}, David L. and {Patil}, Aarya A. and {Corrales}, Lia and {Brasseur}, C.~E. and {N{"o}the}, Maximilian and {Donath}, Axel and {Tollerud}, Erik and {Morris}, Brett M. and {Ginsburg}, Adam and {Vaher}, Eero and {Weaver}, Benjamin A. and {Tocknell}, James and {Jamieson}, William and {van Kerkwijk}, Marten H. and {Robitaille}, Thomas P. and {Merry}, Bruce and {Bachetti}, Matteo and {G{"u}nther}, H. Moritz and {Aldcroft}, Thomas L. and {Alvarado-Montes}, Jaime A. and {Archibald}, Anne M. and {B{'o}di}, Attila and {Bapat}, Shreyas and {Barentsen}, Geert and {Baz{'a}n}, Juanjo and {Biswas}, Manish and {Boquien}, M{'e}d{'e}ric and {Burke}, D.~J. and {Cara}, Daria and {Cara}, Mihai and {Conroy}, Kyle E. and {Conseil}, Simon and {Craig}, Matthew W. and {Cross}, Robert M. and {Cruz}, Kelle L. and {D'Eugenio}, Francesco and {Dencheva}, Nadia and {Devillepoix}, Hadrien A.~R. and {Dietrich}, J{"o}rg P. and {Eigenbrot}, Arthur Davis and {Erben}, Thomas and {Ferreira}, Leonardo and {Foreman-Mackey}, Daniel and {Fox}, Ryan and {Freij}, Nabil and {Garg}, Suyog and {Geda}, Robel and {Glattly}, Lauren and {Gondhalekar}, Yash and {Gordon}, Karl D. and {Grant}, David and {Greenfield}, Perry and {Groener}, Austen M. and {Guest}, Steve and {Gurovich}, Sebastian and {Handberg}, Rasmus and {Hart}, Akeem and {Hatfield-Dodds}, Zac and {Homeier}, Derek and {Hosseinzadeh}, Griffin and {Jenness}, Tim and {Jones}, Craig K. and {Joseph}, Prajwel and {Kalmbach}, J. Bryce and {Karamehmetoglu}, Emir and {Ka{l}uszy{'n}ski}, Miko{l}aj and {Kelley}, Michael S.~P. and {Kern}, Nicholas and {Kerzendorf}, Wolfgang E. and {Koch}, Eric W. and {Kulumani}, Shankar and {Lee}, Antony and {Ly}, Chun and {Ma}, Zhiyuan and {MacBride}, Conor and {Maljaars}, Jakob M. and {Muna}, Demitri and {Murphy}, N.~A. and {Norman}, Henrik and {O'Steen}, Richard and {Oman}, Kyle A. and {Pacifici}, Camilla and {Pascual}, Sergio and {Pascual-Granado}, J. and {Patil}, Rohit R. and {Perren}, Gabriel I. and {Pickering}, Timothy E. and {Rastogi}, Tanuj and {Roulston}, Benjamin R. and {Ryan}, Daniel F. and {Rykoff}, Eli S. and {Sabater}, Jose and {Sakurikar}, Parikshit and {Salgado}, Jes{'u}s and {Sanghi}, Aniket and {Saunders}, Nicholas and {Savchenko}, Volodymyr and {Schwardt}, Ludwig and {Seifert-Eckert}, Michael and {Shih}, Albert Y. and {Jain}, Anany Shrey and {Shukla}, Gyanendra and {Sick}, Jonathan and {Simpson}, Chris and {Singanamalla}, Sudheesh and {Singer}, Leo P. and {Singhal}, Jaladh and {Sinha}, Manodeep and {Sip{H{o}}cz}, Brigitta M. and {Spitler}, Lee R. and {Stansby}, David and {Streicher}, Ole and {{{S}}umak}, Jani and {Swinbank}, John D. and {Taranu}, Dan S. and {Tewary}, Nikita and {Tremblay}, Grant R. and {Val-Borro}, Miguel de and {Van Kooten}, Samuel J. and {Vasovi{'c}}, Zlatan and {Verma}, Shresth and {de Miranda Cardoso}, Jos{'e} Vin{'i}cius and {Williams}, Peter K.~G. and {Wilson}, Tom J. and {Winkel}, Benjamin and {Wood-Vasey}, W.~M. and {Xue}, Rui and {Yoachim}, Peter and {Zhang}, Chen and {Zonca}, Andrea and {Astropy Project Contributors}},
        title = "{The Astropy Project: Sustaining and Growing a Community-oriented Open-source Project and the Latest Major Release (v5.0) of the Core Package}",
      journal = {\apj},
         year = 2022,
        month = aug,
       volume = {935},
       number = {2},
          eid = {167},
        pages = {167},
          doi = {10.3847/1538-4357/ac7c74},
archivePrefix = {arXiv},
       eprint = {2206.14220},
 primaryClass = {astro-ph.IM},
       adsurl = {https://ui.adsabs.harvard.edu/abs/2022ApJ...935..167A}
}

@Article{Hunter:2007,
  Author    = {Hunter, J. D.},
  Title     = {Matplotlib: A 2D graphics environment},
  Journal   = {Computing in Science \& Engineering},
  Volume    = {9},
  Number    = {3},
  Pages     = {90--95},
  publisher = {IEEE COMPUTER SOC},
  doi       = {10.1109/MCSE.2007.55},
  year      = 2007
}

@article{numpy,
 title         = {Array programming with {NumPy}},
 author        = {Charles R. Harris and K. Jarrod Millman and St{\'{e}}fan J.
                 van der Walt and Ralf Gommers and Pauli Virtanen and David
                 Cournapeau and Eric Wieser and Julian Taylor and Sebastian
                 Berg and Nathaniel J. Smith and Robert Kern and Matti Picus
                 and Stephan Hoyer and Marten H. van Kerkwijk and Matthew
                 Brett and Allan Haldane and Jaime Fern{\'{a}}ndez del
                 R{\'{i}}o and Mark Wiebe and Pearu Peterson and Pierre
                 G{\'{e}}rard-Marchant and Kevin Sheppard and Tyler Reddy and
                 Warren Weckesser and Hameer Abbasi and Christoph Gohlke and
                 Travis E. Oliphant},
 year          = {2020},
 month         = sep,
 journal       = {Nature},
 volume        = {585},
 number        = {7825},
 pages         = {357--362},
 doi           = {10.1038/s41586-020-2649-2},
 publisher     = {Springer Science and Business Media {LLC}},
 url           = {https://doi.org/10.1038/s41586-020-2649-2}
}

@book{python,
  author    = {Van Rossum, Guido and Drake, Fred L.},
  title     = {Python 3 Reference Manual},
  year      = {2009},
  isbn      = {1441412697},
  publisher = {CreateSpace},
  address   = {Scotts Valley, CA}
}

@ARTICLE{2020SciPy-NMeth,
  author  = {Virtanen, Pauli and Gommers, Ralf and Oliphant, Travis E. and
            Haberland, Matt and Reddy, Tyler and Cournapeau, David and
            Burovski, Evgeni and Peterson, Pearu and Weckesser, Warren and
            Bright, Jonathan and {van der Walt}, St{\'e}fan J. and
            Brett, Matthew and Wilson, Joshua and Millman, K. Jarrod and
            Mayorov, Nikolay and Nelson, Andrew R. J. and Jones, Eric and
            Kern, Robert and Larson, Eric and Carey, C J and
            Polat, {\.I}lhan and Feng, Yu and Moore, Eric W. and
            {VanderPlas}, Jake and Laxalde, Denis and Perktold, Josef and
            Cimrman, Robert and Henriksen, Ian and Quintero, E. A. and
            Harris, Charles R. and Archibald, Anne M. and
            Ribeiro, Ant{\^o}nio H. and Pedregosa, Fabian and
            {van Mulbregt}, Paul and {SciPy 1.0 Contributors}},
  title   = {{{SciPy} 1.0: Fundamental Algorithms for Scientific
            Computing in Python}},
  journal = {Nature Methods},
  year    = {2020},
  volume  = {17},
  pages   = {261--272},
  adsurl  = {https://rdcu.be/b08Wh},
  doi     = {10.1038/s41592-019-0686-2},
}

@ARTICLE{cython:2011, 
author={S. Behnel and R. Bradshaw and C. Citro and L. Dalcin and D. S. Seljebotn and K. Smith},
journal={Computing in Science Engineering},
title={Cython: The Best of Both Worlds},
year={2011},
volume={13},
number={2},
pages={31-39},
doi={10.1109/MCSE.2010.118},
ISSN={1521-9615},
month={March}}

@misc{pynbody,
        author = {{Pontzen}, Andrew and {Ro{\v{s}}kar}, Rok and {Stinson}, Greg and {Woods}, Rory},
        title = "{pynbody: N-Body/SPH analysis for python}",
        howpublished = {Astrophysics Source Code Library, record ascl:1305.002},
        year = 2013,
        month = may,
        eid = {ascl:1305.002},
        archivePrefix = {ascl},
        eprint = {1305.002},
        adsurl = {https://ui.adsabs.harvard.edu/abs/2013ascl.soft05002P}
}

@ARTICLE{software-citation-station-paper,
       author = {{Wagg}, Tom and {Broekgaarden}, Floor S.},
        title = "{Streamlining and standardizing software citations with The Software Citation Station}",
      journal = {arXiv e-prints},
         year = 2024,
        month = jun,
          eid = {arXiv:2406.04405},
        pages = {arXiv:2406.04405},
archivePrefix = {arXiv},
       eprint = {2406.04405},
 primaryClass = {astro-ph.IM},
       adsurl = {https://ui.adsabs.harvard.edu/abs/2024arXiv240604405W}
}

@misc{software-citation-station-zenodo,
  author       = {Tom Wagg and
                  Floor Broekgaarden and
                  Kayhan Gültekin},
  title        = {TomWagg/software-citation-station: v1.2},
  month        = aug,
  year         = 2024,
  publisher    = {Zenodo},
  version      = {v1.2},
  doi          = {10.5281/zenodo.13225824},
  url          = {https://doi.org/10.5281/zenodo.13225824},
}

@ARTICLE{Iyer2024,
       author = {{Iyer}, Kartheik G. and {Speagle}, Joshua S. and {Caplar}, Neven and {Forbes}, John C. and {Gawiser}, Eric and {Leja}, Joel and {Tacchella}, Sandro},
        title = "{Stochastic Modeling of Star Formation Histories. III. Constraints from Physically Motivated Gaussian Processes}",
      journal = {\apj},
         year = 2024,
        month = jan,
       volume = {961},
       number = {1},
          eid = {53},
        pages = {53},
          doi = {10.3847/1538-4357/acff64},
archivePrefix = {arXiv},
       eprint = {2208.05938},
 primaryClass = {astro-ph.GA},
       adsurl = {https://ui.adsabs.harvard.edu/abs/2024ApJ...961...53I}
}

@ARTICLE{ForemanMackey2017,
       author = {{Foreman-Mackey}, Daniel and {Agol}, Eric and {Ambikasaran}, Sivaram and {Angus}, Ruth},
        title = "{Fast and Scalable Gaussian Process Modeling with Applications to Astronomical Time Series}",
      journal = {\aj},
         year = 2017,
        month = dec,
       volume = {154},
       number = {6},
          eid = {220},
        pages = {220},
          doi = {10.3847/1538-3881/aa9332},
archivePrefix = {arXiv},
       eprint = {1703.09710},
 primaryClass = {astro-ph.IM},
       adsurl = {https://ui.adsabs.harvard.edu/abs/2017AJ....154..220F}
}

@ARTICLE{2025ApJ...993..150D,
       author = {{Dattathri}, Shashank and {van den Bosch}, Frank C. and {Weinberg}, Martin D. and {Banik}, Uddipan},
        title = "{The Dipole Instability in Gravitational N-body Systems: A Natural Explanation for Lopsidedness and Off-centered Nuclei in Galaxies}",
      journal = {\apj},
         year = 2025,
        month = nov,
       volume = {993},
       number = {1},
          eid = {150},
        pages = {150},
          doi = {10.3847/1538-4357/ae0623},
archivePrefix = {arXiv},
       eprint = {2505.23905},
 primaryClass = {astro-ph.GA},
       adsurl = {https://ui.adsabs.harvard.edu/abs/2025ApJ...993..150D}
}

@ARTICLE{Mintz2024,
       author = {{Mintz}, Abby and {Greene}, Jenny E. and {Kado-Fong}, Erin and {Danieli}, Shany and {Li}, Jiaxuan and {Luo}, Yifei and {Leauthaud}, Alexie and {Baldassare}, Vivienne and {Huang}, Song and {Peter}, Annika H.~G. and {Bhattacharyya}, Joy and {Li}, Mingyu and {Pan}, Yue},
        title = "{A Nonparametric Morphological Analysis of H{\ensuremath{\alpha}} Emission in Bright Dwarfs Using the Merian Survey}",
      journal = {\apj},
         year = 2024,
        month = oct,
       volume = {974},
       number = {2},
          eid = {273},
        pages = {273},
          doi = {10.3847/1538-4357/ad6861},
archivePrefix = {arXiv},
       eprint = {2410.01886},
 primaryClass = {astro-ph.GA},
       adsurl = {https://ui.adsabs.harvard.edu/abs/2024ApJ...974..273M}
}

@ARTICLE{Weisz2012,
       author = {{Weisz}, Daniel R. and {Johnson}, Benjamin D. and {Johnson}, L. Clifton and {Skillman}, Evan D. and {Lee}, Janice C. and {Kennicutt}, Robert C. and {Calzetti}, Daniela and {van Zee}, Liese and {Bothwell}, Matthew S. and {Dalcanton}, Julianne J. and {Dale}, Daniel A. and {Williams}, Benjamin F.},
        title = "{Modeling the Effects of Star Formation Histories on H{\ensuremath{\alpha}} and Ultraviolet Fluxes in nearby Dwarf Galaxies}",
      journal = {\apj},
         year = 2012,
        month = jan,
       volume = {744},
       number = {1},
          eid = {44},
        pages = {44},
          doi = {10.1088/0004-637X/744/1/44},
archivePrefix = {arXiv},
       eprint = {1109.2905},
 primaryClass = {astro-ph.CO},
       adsurl = {https://ui.adsabs.harvard.edu/abs/2012ApJ...744...44W}
}

@ARTICLE{Emami2019,
       author = {{Emami}, Najmeh and {Siana}, Brian and {Weisz}, Daniel R. and {Johnson}, Benjamin D. and {Ma}, Xiangcheng and {El-Badry}, Kareem},
        title = "{A Closer Look at Bursty Star Formation with L $_{H{\ensuremath{\alpha}} }$ and L $_{UV}$ Distributions}",
      journal = {\apj},
         year = 2019,
        month = aug,
       volume = {881},
       number = {1},
          eid = {71},
        pages = {71},
          doi = {10.3847/1538-4357/ab211a},
archivePrefix = {arXiv},
       eprint = {1809.06380},
 primaryClass = {astro-ph.GA},
       adsurl = {https://ui.adsabs.harvard.edu/abs/2019ApJ...881...71E}
}

@ARTICLE{Mintz2026,
       author = {{Mintz}, Abby and {Setton}, David J. and {Greene}, Jenny E. and {Leja}, Joel and {Wang}, Bingjie and {Burnham}, Emilie and {Suess}, Katherine A. and {Atek}, Hakim and {Bezanson}, Rachel and {Brammer}, Gabriel and {Cutler}, Sam E. and {Dayal}, Pratika and {Feldmann}, Robert and {Furtak}, Lukas J. and {Glazebrook}, Karl and {Khullar}, Gourav and {Kokorev}, Vasily and {Labb{\'e}}, Ivo and {Matthee}, Jorryt and {Maseda}, Michael V. and {Miller}, Tim B. and {Mitsuhashi}, Ikki and {Nanayakkara}, Themiya and {Pan}, Richard and {Price}, Sedona H. and {Weaver}, John R. and {Whitaker}, Katherine E. and {Wu}, Belinda},
        title = "{Taking a Break at Cosmic Noon: Continuum-selected Low-mass Galaxies Require Long Burst Cycles}",
      journal = {\apj},
         year = 2026,
        month = feb,
       volume = {998},
       number = {2},
          eid = {257},
        pages = {257},
          doi = {10.3847/1538-4357/ae328c},
archivePrefix = {arXiv},
       eprint = {2506.16510},
 primaryClass = {astro-ph.GA},
       adsurl = {https://ui.adsabs.harvard.edu/abs/2026ApJ...998..257M}
}

@ARTICLE{patel18,
       author = {{Patel}, Shannon G. and {Kelson}, Daniel D. and {Diao}, Nicholas and {Tonnesen}, Stephanie and {Abramson}, Louis E.},
        title = "{Testing the Breathing Mode in Intermediate-mass Galaxies and Its Predicted Star Formation Rate-size Anti-correlation}",
      journal = {\apjl},
         year = 2018,
        month = oct,
       volume = {866},
       number = {2},
          eid = {L21},
        pages = {L21},
          doi = {10.3847/2041-8213/aae524},
archivePrefix = {arXiv},
       eprint = {1807.02118},
 primaryClass = {astro-ph.GA},
       adsurl = {https://ui.adsabs.harvard.edu/abs/2018ApJ...866L..21P}
}

@ARTICLE{Hopkins2023,
       author = {{Hopkins}, Philip F. and {Gurvich}, Alexander B. and {Shen}, Xuejian and {Hafen}, Zachary and {Grudi{\'c}}, Michael Y. and {Kurinchi-Vendhan}, Shalini and {Hayward}, Christopher C. and {Jiang}, Fangzhou and {Orr}, Matthew E. and {Wetzel}, Andrew and {Kere{\v{s}}}, Du{\v{s}}an and {Stern}, Jonathan and {Faucher-Gigu{\`e}re}, Claude-Andr{\'e} and {Bullock}, James and {Wheeler}, Coral and {El-Badry}, Kareem and {Loebman}, Sarah R. and {Moreno}, Jorge and {Boylan-Kolchin}, Michael and {Quataert}, Eliot},
        title = "{What causes the formation of discs and end of bursty star formation?}",
      journal = {\mnras},
         year = 2023,
        month = oct,
       volume = {525},
       number = {2},
        pages = {2241-2286},
          doi = {10.1093/mnras/stad1902},
archivePrefix = {arXiv},
       eprint = {2301.08263},
 primaryClass = {astro-ph.GA},
       adsurl = {https://ui.adsabs.harvard.edu/abs/2023MNRAS.525.2241H}
}

@ARTICLE{Benavides2025,
       author = {{Benavides}, Jos{\'e} A. and {Sales}, Laura V. and {Wetzel}, Andrew and {Moreno}, Jorge and {Feldmann}, Robert and {Mercado}, Francisco J. and {Bullock}, James S. and {Hopkins}, Philip F. and {Faucher-Guig{\`e}re}, Claude-Andr{\'e} and {Stern}, Jonathan and {Wheeler}, Coral and {Kere{\v{s}}}, Du{\v{s}}an},
        title = "{Disks no more: the morphology of low-mass simulated galaxies in FIREbox}",
      journal = {arXiv e-prints},
         year = 2025,
        month = aug,
          eid = {arXiv:2508.00991},
        pages = {arXiv:2508.00991},
          doi = {10.48550/arXiv.2508.00991},
archivePrefix = {arXiv},
       eprint = {2508.00991},
 primaryClass = {astro-ph.GA},
       adsurl = {https://ui.adsabs.harvard.edu/abs/2025arXiv250800991B}
}

@ARTICLE{1997PASP..109..745B,
       author = {{Bothun}, G. and {Impey}, C. and {McGaugh}, S.},
        title = "{Low-Surface-Brightness Galaxies: Hidden Galaxies Revealed}",
      journal = {\pasp},
         year = 1997,
        month = jul,
       volume = {109},
        pages = {745-758},
          doi = {10.1086/133941},
       adsurl = {https://ui.adsabs.harvard.edu/abs/1997PASP..109..745B}
}

@ARTICLE{2019ApJ...886...74M,
       author = {{McQuinn}, Kristen. B.~W. and {van Zee}, Liese and {Skillman}, Evan D.},
        title = "{Galactic Winds in Low-mass Galaxies}",
      journal = {\apj},
         year = 2019,
        month = nov,
       volume = {886},
       number = {1},
          eid = {74},
        pages = {74},
          doi = {10.3847/1538-4357/ab4c37},
archivePrefix = {arXiv},
       eprint = {1910.04167},
 primaryClass = {astro-ph.GA},
       adsurl = {https://ui.adsabs.harvard.edu/abs/2019ApJ...886...74M}
}

@ARTICLE{2022NatAs...6..659B,
       author = {{Battaglia}, Giuseppina and {Nipoti}, Carlo},
        title = "{Stellar dynamics and dark matter in Local Group dwarf galaxies}",
      journal = {Nature Astronomy},
         year = 2022,
        month = may,
       volume = {6},
        pages = {659-672},
          doi = {10.1038/s41550-022-01638-7},
archivePrefix = {arXiv},
       eprint = {2205.07821},
 primaryClass = {astro-ph.GA},
       adsurl = {https://ui.adsabs.harvard.edu/abs/2022NatAs...6..659B}
}

@ARTICLE{2022ApJ...940....8M,
       author = {{McQuinn}, Kristen. B.~W. and {Adams}, Elizabeth A.~K. and {Cannon}, John M. and {Fuson}, Jackson and {Skillman}, Evan D. and {Brooks}, Alyson and {Rhode}, Katherine L. and {Haynes}, Martha P. and {Inoue}, John L. and {Marine}, Joshua and {Salzer}, John. J. and {Talluri}, Anjana K.},
        title = "{The Turndown of the Baryonic Tully-Fisher Relation and Changing Baryon Fraction at Low Galaxy Masses}",
      journal = {\apj},
         year = 2022,
        month = nov,
       volume = {940},
       number = {1},
          eid = {8},
        pages = {8},
          doi = {10.3847/1538-4357/ac9285},
archivePrefix = {arXiv},
       eprint = {2203.10105},
 primaryClass = {astro-ph.GA},
       adsurl = {https://ui.adsabs.harvard.edu/abs/2022ApJ...940....8M}
}

@ARTICLE{2009ARA&A..47..371T,
       author = {{Tolstoy}, Eline and {Hill}, Vanessa and {Tosi}, Monica},
        title = "{Star-Formation Histories, Abundances, and Kinematics of Dwarf Galaxies in the Local Group}",
      journal = {\araa},
         year = 2009,
        month = sep,
       volume = {47},
       number = {1},
        pages = {371-425},
          doi = {10.1146/annurev-astro-082708-101650},
archivePrefix = {arXiv},
       eprint = {0904.4505},
 primaryClass = {astro-ph.CO},
       adsurl = {https://ui.adsabs.harvard.edu/abs/2009ARA&A..47..371T}
}

@ARTICLE{2010ApJ...724...49M,
       author = {{McQuinn}, Kristen B.~W. and {Skillman}, Evan D. and {Cannon}, John M. and {Dalcanton}, Julianne and {Dolphin}, Andrew and {Hidalgo-Rodr{\'\i}guez}, Sebastian and {Holtzman}, Jon and {Stark}, David and {Weisz}, Daniel and {Williams}, Benjamin},
        title = "{The Nature of Starbursts. II. The Duration of Starbursts in Dwarf Galaxies}",
      journal = {\apj},
         year = 2010,
        month = nov,
       volume = {724},
       number = {1},
        pages = {49-58},
          doi = {10.1088/0004-637X/724/1/49},
archivePrefix = {arXiv},
       eprint = {1009.2940},
 primaryClass = {astro-ph.CO},
       adsurl = {https://ui.adsabs.harvard.edu/abs/2010ApJ...724...49M}
}

@ARTICLE{2010ApJ...721..297M,
       author = {{McQuinn}, Kristen B.~W. and {Skillman}, Evan D. and {Cannon}, John M. and {Dalcanton}, Julianne and {Dolphin}, Andrew and {Hidalgo-Rodr{\'\i}guez}, Sebastian and {Holtzman}, Jon and {Stark}, David and {Weisz}, Daniel and {Williams}, Benjamin},
        title = "{The Nature of Starbursts. I. The Star Formation Histories of Eighteen Nearby Starburst Dwarf Galaxies}",
      journal = {\apj},
         year = 2010,
        month = sep,
       volume = {721},
       number = {1},
        pages = {297-317},
          doi = {10.1088/0004-637X/721/1/297},
archivePrefix = {arXiv},
       eprint = {1008.1589},
 primaryClass = {astro-ph.CO},
       adsurl = {https://ui.adsabs.harvard.edu/abs/2010ApJ...721..297M}
}

@ARTICLE{2011ApJ...739....5W,
       author = {{Weisz}, Daniel R. and {Dalcanton}, Julianne J. and {Williams}, Benjamin F. and {Gilbert}, Karoline M. and {Skillman}, Evan D. and {Seth}, Anil C. and {Dolphin}, Andrew E. and {McQuinn}, Kristen B.~W. and {Gogarten}, Stephanie M. and {Holtzman}, Jon and {Rosema}, Keith and {Cole}, Andrew and {Karachentsev}, Igor D. and {Zaritsky}, Dennis},
        title = "{The ACS Nearby Galaxy Survey Treasury. VIII. The Global Star Formation Histories of 60 Dwarf Galaxies in the Local Volume}",
      journal = {\apj},
         year = 2011,
        month = sep,
       volume = {739},
       number = {1},
          eid = {5},
        pages = {5},
          doi = {10.1088/0004-637X/739/1/5},
archivePrefix = {arXiv},
       eprint = {1101.1093},
 primaryClass = {astro-ph.CO},
       adsurl = {https://ui.adsabs.harvard.edu/abs/2011ApJ...739....5W}
}

@ARTICLE{2014ApJ...789...96A,
       author = {{Atek}, Hakim and {Kneib}, Jean-Paul and {Pacifici}, Camilla and {Malkan}, Matthew and {Charlot}, Stephane and {Lee}, Janice and {Bedregal}, Alejandro and {Bunker}, Andrew J. and {Colbert}, James W. and {Dressler}, Alan and {Hathi}, Nimish and {Lehnert}, Matthew and {Martin}, Crystal L. and {McCarthy}, Patrick and {Rafelski}, Marc and {Ross}, Nathaniel and {Siana}, Brian and {Teplitz}, Harry I.},
        title = "{Hubble Space Telescope Grism Spectroscopy of Extreme Starbursts across Cosmic Time: The Role of Dwarf Galaxies in the Star Formation History of the Universe}",
      journal = {\apj},
         year = 2014,
        month = jul,
       volume = {789},
       number = {2},
          eid = {96},
        pages = {96},
          doi = {10.1088/0004-637X/789/2/96},
archivePrefix = {arXiv},
       eprint = {1406.4132},
 primaryClass = {astro-ph.GA},
       adsurl = {https://ui.adsabs.harvard.edu/abs/2014ApJ...789...96A}
}

@ARTICLE{2022MNRAS.511.4464A,
       author = {{Atek}, Hakim and {Furtak}, Lukas J. and {Oesch}, Pascal and {van Dokkum}, Pieter and {Reddy}, Naveen and {Contini}, Thierry and {Illingworth}, Garth and {Wilkins}, Stephen},
        title = "{The star formation burstiness and ionizing efficiency of low-mass galaxies}",
      journal = {\mnras},
         year = 2022,
        month = apr,
       volume = {511},
       number = {3},
        pages = {4464-4479},
          doi = {10.1093/mnras/stac360},
archivePrefix = {arXiv},
       eprint = {2202.04081},
 primaryClass = {astro-ph.GA},
       adsurl = {https://ui.adsabs.harvard.edu/abs/2022MNRAS.511.4464A}
}

@ARTICLE{2023ApJ...950..125M,
       author = {{M{\'e}rida}, Rosa M. and {P{\'e}rez-Gonz{\'a}lez}, Pablo G. and {S{\'a}nchez-Bl{\'a}zquez}, Patricia and {Garc{\'\i}a-Argum{\'a}nez}, {\'A}ngela and {Annunziatella}, Marianna and {Costantin}, Luca and {Lumbreras-Calle}, Alejandro and {Alcalde-Pampliega}, Bel{\'e}n and {Barro}, Guillermo and {Espino-Briones}, N{\'e}stor and {Koekemoer}, Anton M.},
        title = "{Probing the Star Formation Main Sequence Down to {}10$^{8}$ M $_{{\ensuremath{\odot}}}$ at 1.0 < z < 3.0}",
      journal = {\apj},
         year = 2023,
        month = jun,
       volume = {950},
       number = {2},
          eid = {125},
        pages = {125},
          doi = {10.3847/1538-4357/acc7a3},
archivePrefix = {arXiv},
       eprint = {2303.16234},
 primaryClass = {astro-ph.GA},
       adsurl = {https://ui.adsabs.harvard.edu/abs/2023ApJ...950..125M}
}

@ARTICLE{2024Natur.629...53L,
       author = {{Looser}, Tobias J. and {D'Eugenio}, Francesco and {Maiolino}, Roberto and {Witstok}, Joris and {Sandles}, Lester and {Curtis-Lake}, Emma and {Chevallard}, Jacopo and {Tacchella}, Sandro and {Johnson}, Benjamin D. and {Baker}, William M. and {Suess}, Katherine A. and {Carniani}, Stefano and {Ferruit}, Pierre and {Arribas}, Santiago and {Bonaventura}, Nina and {Bunker}, Andrew J. and {Cameron}, Alex J. and {Charlot}, Stephane and {Curti}, Mirko and {de Graaff}, Anna and {Maseda}, Michael V. and {Rawle}, Tim and {Rix}, Hans-Walter and {Del Pino}, Bruno Rodr{\'\i}guez and {Smit}, Renske and {{\"U}bler}, Hannah and {Willott}, Chris and {Alberts}, Stacey and {Egami}, Eiichi and {Eisenstein}, Daniel J. and {Endsley}, Ryan and {Hausen}, Ryan and {Rieke}, Marcia and {Robertson}, Brant and {Shivaei}, Irene and {Williams}, Christina C. and {Boyett}, Kristan and {Chen}, Zuyi and {Ji}, Zhiyuan and {Jones}, Gareth C. and {Kumari}, Nimisha and {Nelson}, Erica and {Perna}, Michele and {Saxena}, Aayush and {Scholtz}, Jan},
        title = "{A recently quenched galaxy 700 million years after the Big Bang}",
      journal = {\nat},
         year = 2024,
        month = may,
       volume = {629},
       number = {8010},
        pages = {53-57},
          doi = {10.1038/s41586-024-07227-0},
archivePrefix = {arXiv},
       eprint = {2302.14155},
 primaryClass = {astro-ph.GA},
       adsurl = {https://ui.adsabs.harvard.edu/abs/2024Natur.629...53L}
}

@ARTICLE{2024MNRAS.533.1111E,
       author = {{Endsley}, Ryan and {Stark}, Daniel P. and {Whitler}, Lily and {Topping}, Michael W. and {Johnson}, Benjamin D. and {Robertson}, Brant and {Tacchella}, Sandro and {Alberts}, Stacey and {Baker}, William M. and {Bhatawdekar}, Rachana and {Boyett}, Kristan and {Bunker}, Andrew J. and {Cameron}, Alex J. and {Carniani}, Stefano and {Charlot}, Stephane and {Chen}, Zuyi and {Chevallard}, Jacopo and {Curtis-Lake}, Emma and {Danhaive}, A. Lola and {Egami}, Eiichi and {Eisenstein}, Daniel J. and {Hainline}, Kevin and {Helton}, Jakob M. and {Ji}, Zhiyuan and {Looser}, Tobias J. and {Maiolino}, Roberto and {Nelson}, Erica and {Pusk{\'a}s}, D{\'a}vid and {Rieke}, George and {Rieke}, Marcia and {Rix}, Hans-Walter and {Sandles}, Lester and {Saxena}, Aayush and {Simmonds}, Charlotte and {Smit}, Renske and {Sun}, Fengwu and {Williams}, Christina C. and {Willmer}, Christopher N.~A. and {Willott}, Chris and {Witstok}, Joris},
        title = "{The star-forming and ionizing properties of dwarf z 6-9 galaxies in JADES: insights on bursty star formation and ionized bubble growth}",
      journal = {\mnras},
         year = 2024,
        month = sep,
       volume = {533},
       number = {1},
        pages = {1111-1142},
          doi = {10.1093/mnras/stae1857},
archivePrefix = {arXiv},
       eprint = {2306.05295},
 primaryClass = {astro-ph.GA},
       adsurl = {https://ui.adsabs.harvard.edu/abs/2024MNRAS.533.1111E}
}

@ARTICLE{2025ApJ...979..193C,
       author = {{Cole}, Justin W. and {Papovich}, Casey and {Finkelstein}, Steven L. and {Bagley}, Micaela B. and {Dickinson}, Mark and {Iyer}, Kartheik G. and {Yung}, L.~Y. Aaron and {Ciesla}, Laure and {Amor{\'\i}n}, Ricardo O. and {Arrabal Haro}, Pablo and {Bhatawdekar}, Rachana and {Calabr{\`o}}, Antonello and {Cleri}, Nikko J. and {de la Vega}, Alexander and {Dekel}, Avishai and {Endsley}, Ryan and {Gawiser}, Eric and {Giavalisco}, Mauro and {Hathi}, Nimish P. and {Hirschmann}, Michaela and {Holwerda}, Benne W. and {Kartaltepe}, Jeyhan S. and {Koekemoer}, Anton M. and {Lucas}, Ray A. and {Mascia}, Sara and {Mobasher}, Bahram and {P{\'e}rez-Gonz{\'a}lez}, Pablo G. and {Rodighiero}, Giulia and {Ronayne}, Kaila and {Tacchella}, Sandro and {Weiner}, Benjamin J. and {Wilkins}, Stephen M.},
        title = "{CEERS: Increasing Scatter along the Star-forming Main Sequence Indicates Early Galaxies Form in Bursts}",
      journal = {\apj},
         year = 2025,
        month = feb,
       volume = {979},
       number = {2},
          eid = {193},
        pages = {193},
          doi = {10.3847/1538-4357/ad9a6a},
archivePrefix = {arXiv},
       eprint = {2312.10152},
 primaryClass = {astro-ph.GA},
       adsurl = {https://ui.adsabs.harvard.edu/abs/2025ApJ...979..193C}
}

@ARTICLE{2025MNRAS.544.4551S,
       author = {{Simmonds}, C. and {Tacchella}, S. and {McClymont}, W. and {Curtis-Lake}, E. and {D'Eugenio}, F. and {Hainline}, K. and {Johnson}, B.~D. and {Kravtsov}, A. and {Pusk{\'a}s}, D. and {Robertson}, B. and {Stoffers}, A. and {Willott}, C. and {Baker}, W.~M. and {Belokurov}, V.~A. and {Bhatawdekar}, R. and {Bunker}, A.~J. and {Carniani}, S. and {Chevallard}, J. and {Curti}, M. and {Duan}, Q. and {Helton}, J.~M. and {Ji}, Z. and {Looser}, T.~J. and {Maiolino}, R. and {Maseda}, M.~V. and {Shivaei}, I. and {Williams}, C.~C.},
        title = "{Bursting at the seams: the star-forming main sequence and its scatter at z = 3─9 using NIRCam photometry from JADES}",
      journal = {\mnras},
         year = 2025,
        month = dec,
       volume = {544},
       number = {4},
        pages = {4551-4575},
          doi = {10.1093/mnras/staf1950},
archivePrefix = {arXiv},
       eprint = {2508.04410},
 primaryClass = {astro-ph.GA},
       adsurl = {https://ui.adsabs.harvard.edu/abs/2025MNRAS.544.4551S}
}

@ARTICLE{2014MNRAS.441.2717K,
       author = {{Kauffmann}, Guinevere},
        title = "{Quantitative constraints on starburst cycles in galaxies with stellar masses in the range {}10$^{8}$-{}10$^{10}$ M$_{☉}$}",
      journal = {\mnras},
         year = 2014,
        month = jul,
       volume = {441},
       number = {3},
        pages = {2717-2724},
          doi = {10.1093/mnras/stu752},
       adsurl = {https://ui.adsabs.harvard.edu/abs/2014MNRAS.441.2717K}
}

@ARTICLE{2001MNRAS.322..231K,
       author = {{Kroupa}, Pavel},
        title = "{On the variation of the initial mass function}",
      journal = {\mnras},
         year = 2001,
        month = apr,
       volume = {322},
       number = {2},
        pages = {231-246},
          doi = {10.1046/j.1365-8711.2001.04022.x},
archivePrefix = {arXiv},
       eprint = {astro-ph/0009005},
 primaryClass = {astro-ph},
       adsurl = {https://ui.adsabs.harvard.edu/abs/2001MNRAS.322..231K}
}

@ARTICLE{1978IEEEP..66...51H,
       author = {{Harris}, Fredric J.},
        title = "{On the Use of Windows for Harmonic Analysis with the Discrete Fourier Transform}",
      journal = {IEEE Proceedings},
         year = 1978,
        month = jan,
       volume = {66},
        pages = {51-83},
          doi = {10.1109/PROC.1978.10837},
       adsurl = {https://ui.adsabs.harvard.edu/abs/1978IEEEP..66...51H}
}

@ARTICLE{2019MNRAS.484..915M,
       author = {{Matthee}, Jorryt and {Schaye}, Joop},
        title = "{The origin of scatter in the star formation rate-stellar mass relation}",
      journal = {\mnras},
         year = 2019,
        month = mar,
       volume = {484},
       number = {1},
        pages = {915-932},
          doi = {10.1093/mnras/stz030},
archivePrefix = {arXiv},
       eprint = {1805.05956},
 primaryClass = {astro-ph.GA},
       adsurl = {https://ui.adsabs.harvard.edu/abs/2019MNRAS.484..915M}
}

@ARTICLE{2019ApJ...879...11K,
       author = {{Katsianis}, Antonios and {Zheng}, Xianzhong and {Gonzalez}, Valentino and {Blanc}, Guillermo and {Lagos}, Claudia del P. and {Davies}, Luke J.~M. and {Camps}, Peter and {Tr{\v{c}}ka}, Ana and {Baes}, Maarten and {Schaye}, Joop and {Trayford}, James W. and {Theuns}, Tom and {Stalevski}, Marko},
        title = "{An Evolving and Mass-dependent {\ensuremath{\sigma}}sSFR-M $_{{\ensuremath{\star}}}$ Relation for Galaxies}",
      journal = {\apj},
         year = 2019,
        month = jul,
       volume = {879},
       number = {1},
          eid = {11},
        pages = {11},
          doi = {10.3847/1538-4357/ab1f8d},
archivePrefix = {arXiv},
       eprint = {1905.02023},
 primaryClass = {astro-ph.GA},
       adsurl = {https://ui.adsabs.harvard.edu/abs/2019ApJ...879...11K}
}

@ARTICLE{2021ApJ...906...96A,
       author = {{Applebaum}, Elaad and {Brooks}, Alyson M. and {Christensen}, Charlotte R. and {Munshi}, Ferah and {Quinn}, Thomas R. and {Shen}, Sijing and {Tremmel}, Michael},
        title = "{Ultrafaint Dwarfs in a Milky Way Context: Introducing the Mint Condition DC Justice League Simulations}",
      journal = {\apj},
         year = 2021,
        month = jan,
       volume = {906},
       number = {2},
          eid = {96},
        pages = {96},
          doi = {10.3847/1538-4357/abcafa},
archivePrefix = {arXiv},
       eprint = {2008.11207},
 primaryClass = {astro-ph.GA},
       adsurl = {https://ui.adsabs.harvard.edu/abs/2021ApJ...906...96A}
}

@ARTICLE{Stinson2007,
       author = {{Stinson}, G.~S. and {Dalcanton}, J.~J. and {Quinn}, T. and {Kaufmann}, T. and {Wadsley}, J.},
        title = "{Breathing in Low-Mass Galaxies: A Study of Episodic Star Formation}",
      journal = {\apj},
         year = 2007,
        month = sep,
       volume = {667},
       number = {1},
        pages = {170-175},
          doi = {10.1086/520504},
archivePrefix = {arXiv},
       eprint = {0705.4494},
 primaryClass = {astro-ph},
       adsurl = {https://ui.adsabs.harvard.edu/abs/2007ApJ...667..170S}
}

@ARTICLE{2003MNRAS.338...14P,
       author = {{Power}, C. and {Navarro}, J.~F. and {Jenkins}, A. and {Frenk}, C.~S. and {White}, S.~D.~M. and {Springel}, V. and {Stadel}, J. and {Quinn}, T.},
        title = "{The inner structure of {\ensuremath{\Lambda}}CDM haloes - I. A numerical convergence study}",
      journal = {\mnras},
         year = 2003,
        month = jan,
       volume = {338},
       number = {1},
        pages = {14-34},
          doi = {10.1046/j.1365-8711.2003.05925.x},
archivePrefix = {arXiv},
       eprint = {astro-ph/0201544},
 primaryClass = {astro-ph},
       adsurl = {https://ui.adsabs.harvard.edu/abs/2003MNRAS.338...14P}
}

@ARTICLE{2019ApJ...871...21G,
       author = {{Genel}, Shy and {Bryan}, Greg L. and {Springel}, Volker and {Hernquist}, Lars and {Nelson}, Dylan and {Pillepich}, Annalisa and {Weinberger}, Rainer and {Pakmor}, R{\"u}diger and {Marinacci}, Federico and {Vogelsberger}, Mark},
        title = "{A Quantification of the Butterfly Effect in Cosmological Simulations and Implications for Galaxy Scaling Relations}",
      journal = {\apj},
         year = 2019,
        month = jan,
       volume = {871},
       number = {1},
          eid = {21},
        pages = {21},
          doi = {10.3847/1538-4357/aaf4bb},
archivePrefix = {arXiv},
       eprint = {1807.07084},
 primaryClass = {astro-ph.GA},
       adsurl = {https://ui.adsabs.harvard.edu/abs/2019ApJ...871...21G}
}

@ARTICLE{1992ApJ...391..502K,
       author = {{Katz}, Neal},
        title = "{Dissipational Galaxy Formation. II. Effects of Star Formation}",
      journal = {\apj},
         year = 1992,
        month = jun,
       volume = {391},
        pages = {502},
          doi = {10.1086/171366},
       adsurl = {https://ui.adsabs.harvard.edu/abs/1992ApJ...391..502K}
}

@ARTICLE{1896RSPTA.187..253P,
       author = {{Pearson}, Karl},
        title = "{VII. Mathematical contributions to the theory of evolution.{\textemdash}III. Regression, heredity, and panmixia}",
      journal = {Philosophical Transactions of the Royal Society of London Series A},
         year = 1896,
        month = dec,
       volume = {187},
        pages = {253-318},
          doi = {10.1098/rsta.1896.0007},
       adsurl = {https://ui.adsabs.harvard.edu/abs/1896RSPTA.187..253P}
}

@ARTICLE{1989MNRAS.239..549W,
       author = {{Weinberg}, Martin D.},
        title = "{Self-gravitating response of a spherical galaxy to sinking satellites}",
      journal = {\mnras},
         year = 1989,
        month = aug,
       volume = {239},
        pages = {549-569},
          doi = {10.1093/mnras/239.2.549},
       adsurl = {https://ui.adsabs.harvard.edu/abs/1989MNRAS.239..549W}
}

@ARTICLE{1994ApJ...421..481W,
       author = {{Weinberg}, Martin D.},
        title = "{Weakly Damped Modes in Star Clusters and Galaxies}",
      journal = {\apj},
         year = 1994,
        month = feb,
       volume = {421},
        pages = {481},
          doi = {10.1086/173665},
archivePrefix = {arXiv},
       eprint = {astro-ph/9306020},
 primaryClass = {astro-ph},
       adsurl = {https://ui.adsabs.harvard.edu/abs/1994ApJ...421..481W}
}

@ARTICLE{2023MNRAS.525.4962W,
       author = {{Weinberg}, Martin D.},
        title = "{New dipole instabilities in spherical stellar systems}",
      journal = {\mnras},
         year = 2023,
        month = nov,
       volume = {525},
       number = {4},
        pages = {4962-4975},
          doi = {10.1093/mnras/stad2591},
archivePrefix = {arXiv},
       eprint = {2209.06846},
 primaryClass = {astro-ph.GA},
       adsurl = {https://ui.adsabs.harvard.edu/abs/2023MNRAS.525.4962W}
}

@ARTICLE{1962spss.book.....A,
       author = {{Antonov}, V.~A.},
        title = "{Solution of the problem of stability of stellar system Emden's density law and the spherical distribution of velocities}",
      journal = {Vestnik Leningradskogo Universiteta},
         year = 1962,
       volume = 7,
        pages = {135},
         note = {English translation in IAU Symposium 113, Dynamics of Star Clusters, ed. J. Goodman \& P. Hut (Dordrecht: Reidel, 1985), 525},
       adsurl = {https://ui.adsabs.harvard.edu/abs/1962spss.book.....A}
}

@ARTICLE{2015A&A...575A..59S,
       author = {{Starkenburg}, T.~K. and {Helmi}, A.},
        title = "{Dark influences: imprints of dark satellites on dwarf galaxies}",
      journal = {\aap},
         year = 2015,
        month = mar,
       volume = {575},
          eid = {A59},
        pages = {A59},
          doi = {10.1051/0004-6361/201425082},
archivePrefix = {arXiv},
       eprint = {1411.1421},
 primaryClass = {astro-ph.GA},
       adsurl = {https://ui.adsabs.harvard.edu/abs/2015A&A...575A..59S}
}

@ARTICLE{2016A&A...587A..24S,
       author = {{Starkenburg}, T.~K. and {Helmi}, A. and {Sales}, L.~V.},
        title = "{Dark influences II. Gas and star formation in minor mergers of dwarf galaxies with dark satellites}",
      journal = {\aap},
         year = 2016,
        month = mar,
       volume = {587},
          eid = {A24},
        pages = {A24},
          doi = {10.1051/0004-6361/201527247},
archivePrefix = {arXiv},
       eprint = {1508.06281},
 primaryClass = {astro-ph.GA},
       adsurl = {https://ui.adsabs.harvard.edu/abs/2016A&A...587A..24S}
}

@ARTICLE{2016A&A...595A..56S,
       author = {{Starkenburg}, T.~K. and {Helmi}, A. and {Sales}, L.~V.},
        title = "{Dark influences. III. Structural characterization of minor mergers of dwarf galaxies with dark satellites}",
      journal = {\aap},
         year = 2016,
        month = oct,
       volume = {595},
          eid = {A56},
        pages = {A56},
          doi = {10.1051/0004-6361/201528066},
archivePrefix = {arXiv},
       eprint = {1601.00020},
 primaryClass = {astro-ph.GA},
       adsurl = {https://ui.adsabs.harvard.edu/abs/2016A&A...595A..56S}
}

@ARTICLE{2025MNRAS.539..661G,
       author = {{Ganapathy}, Ananya and {Petersen}, Michael S. and {Yaaqib}, Rashid and {Filion}, Carrie},
        title = "{Disc asymmetry characterization in JWST-observed galaxies at 1 < z < 4}",
      journal = {\mnras},
         year = 2025,
        month = may,
       volume = {539},
       number = {2},
        pages = {661-673},
          doi = {10.1093/mnras/staf549},
archivePrefix = {arXiv},
       eprint = {2411.11972},
 primaryClass = {astro-ph.GA},
       adsurl = {https://ui.adsabs.harvard.edu/abs/2025MNRAS.539..661G}
}

@ARTICLE{2025ApJ...993..110D,
       author = {{Danieli}, Shany and {Kado-Fong}, Erin and {Huang}, Song and {Luo}, Yifei and {Li}, Ting S. and {Kelvin}, Lee S. and {Leauthaud}, Alexie and {Greene}, Jenny E. and {Mintz}, Abby and {Lin}, Xiaojing and et al.},
        title = "{First Data Release of the Merian Survey: A Wide-field Imaging Survey of Dwarf Galaxies at z {\ensuremath{\sim}} 0.06─0.10}",
      journal = {\apj},
         year = 2025,
        month = nov,
       volume = {993},
       number = {1},
          eid = {110},
        pages = {110},
          doi = {10.3847/1538-4357/ae003e},
archivePrefix = {arXiv},
       eprint = {2410.01884},
 primaryClass = {astro-ph.GA},
       adsurl = {https://ui.adsabs.harvard.edu/abs/2025ApJ...993..110D}
}

@article{2022AJ....163..202G,
  author        = {{Geda}, Robel and {Crawford}, Steven M. and {Hunt}, Lucas and {Bershady}, Matthew and {Tollerud}, Erik and {Randriamampandry}, Solohery},
  title         = "{PetroFit: A Python Package for Computing Petrosian Radii and Fitting Galaxy Light Profiles}",
  journal       = {\aj},
  year          = 2022,
  month         = may,
  volume        = {163},
  number        = {5},
  eid           = {202},
  pages         = {202},
  doi           = {10.3847/1538-3881/ac5908},
  archiveprefix = {arXiv},
  eprint        = {2202.13493},
  primaryclass  = {astro-ph.IM},
  adsurl        = {https://ui.adsabs.harvard.edu/abs/2022AJ....163..202G}
}

@ARTICLE{1942ApJ....95..329S,
       author = {{Spitzer}, Jr., Lyman},
        title = "{The Dynamics of the Interstellar Medium. III. Galactic Distribution.}",
      journal = {\apj},
         year = 1942,
        month = may,
       volume = {95},
        pages = {329},
          doi = {10.1086/144407},
       adsurl = {https://ui.adsabs.harvard.edu/abs/1942ApJ....95..329S}
}

@ARTICLE{2019MNRAS.482.2244K,
       author = {{Keller}, B.~W. and {Wadsley}, J.~W. and {Wang}, L. and {Kruijssen}, J.~M.~D.},
        title = "{Chaos and variance in galaxy formation}",
      journal = {\mnras},
         year = 2019,
        month = jan,
       volume = {482},
       number = {2},
        pages = {2244-2261},
          doi = {10.1093/mnras/sty2859},
archivePrefix = {arXiv},
       eprint = {1803.05445},
 primaryClass = {astro-ph.GA},
       adsurl = {https://ui.adsabs.harvard.edu/abs/2019MNRAS.482.2244K}
}

@ARTICLE{2026ApJ..1007..176S,
       author = {{Sok}, Visal and {Muzzin}, Adam and {Forrest}, Ben and {Wilson}, Gillian and {Chen}, Jialu and {Tan}, Vivian Yun Yan and {Withers}, Sunna and {Abraham}, Roberto and {Brada{\v{c}}}, Maru{\v{s}}a and {Estrada-Carpenter}, Vicente and {Iyer}, Kartheik G. and {Martis}, Nicholas S. and {Noirot}, Ga{\"e}l and {Sarrouh}, Ghassan T.~E. and {Sawicki}, Marcin and {Willott}, Chris J. and {Henry}, Aur{\'e}lien and {Jagga}, Naadiyah and {Marchesini}, Danilo and {McConachie}, Ian and {Myers}, Katherine and {Nunes}, Nelson and {Robbins}, Luke},
        title = "{Metal-poor Gas Accretion Drives Giant Clump Formation at 0.6 < z < 2.6}",
      journal = {\apj},
         year = 2026,
        month = aug,
       volume = {1007},
       number = {2},
          eid = {176},
        pages = {176},
          doi = {10.3847/1538-4357/ae8f42},
archivePrefix = {arXiv},
       eprint = {2606.05284},
 primaryClass = {astro-ph.GA},
       adsurl = {https://ui.adsabs.harvard.edu/abs/2026ApJ..1007..176S}
}

@ARTICLE{2025ApJ...979...14S,
       author = {{Sok}, Visal and {Muzzin}, Adam and {Jablonka}, Pascale and {Tan}, Vivian Yun Yan and {Marsan}, Z. Cemile and {Marchesini}, Danilo and {Wilson}, Gillian and {Alcorn}, Leo Y.},
        title = "{An Indication of Gas Inflow in Clumpy Star-forming Galaxies near z {\ensuremath{\sim}} 1: Lower Gas-phase Metallicities in Clumpy Galaxies Compared to Nonclumpy Galaxies}",
      journal = {\apj},
         year = 2025,
        month = jan,
       volume = {979},
       number = {1},
          eid = {14},
        pages = {14},
          doi = {10.3847/1538-4357/ad96bb},
archivePrefix = {arXiv},
       eprint = {2410.21370},
 primaryClass = {astro-ph.GA},
       adsurl = {https://ui.adsabs.harvard.edu/abs/2025ApJ...979...14S}
}

@ARTICLE{2015ApJ...800...39G,
       author = {{Guo}, Yicheng and {Ferguson}, Henry C. and {Bell}, Eric F. and {Koo}, David C. and {Conselice}, Christopher J. and {Giavalisco}, Mauro and {Kassin}, Susan and {Lu}, Yu and {Lucas}, Ray and {Mandelker}, Nir and {McIntosh}, Daniel H. and {Primack}, Joel R. and {Ravindranath}, Swara and {Barro}, Guillermo and {Ceverino}, Daniel and {Dekel}, Avishai and {Faber}, Sandra M. and {Fang}, Jerome J. and {Koekemoer}, Anton M. and {Noeske}, Kai and {Rafelski}, Marc and {Straughn}, Amber},
        title = "{Clumpy Galaxies in CANDELS. I. The Definition of UV Clumps and the Fraction of Clumpy Galaxies at 0.5 < z < 3}",
      journal = {\apj},
         year = 2015,
        month = feb,
       volume = {800},
       number = {1},
          eid = {39},
        pages = {39},
          doi = {10.1088/0004-637X/800/1/39},
archivePrefix = {arXiv},
       eprint = {1410.7398},
 primaryClass = {astro-ph.GA},
       adsurl = {https://ui.adsabs.harvard.edu/abs/2015ApJ...800...39G}
}

@ARTICLE{2023ApJ...955..106M,
       author = {{Martin}, Alec and {Guo}, Yicheng and {Wang}, Xin and {Koekemoer}, Anton M. and {Rafelski}, Marc and {Teplitz}, Harry I. and {Windhorst}, Rogier A. and {Alavi}, Anahita and {Grogin}, Norman A. and {Prichard}, Laura and {Sunnquist}, Ben and {Ceverino}, Daniel and {Chartab}, Nima and {Conselice}, Christopher J. and {Dai}, Y. Sophia and {Dekel}, Avishai and {Gardner}, Jonathan P. and {Gawiser}, Eric and {Hathi}, Nimish P. and {Hayes}, Matthew J. and {Jansen}, Rolf A. and {Ji}, Zhiyuan and {Koo}, David C. and {Lucas}, Ray A. and {Mandelker}, Nir and {Mehta}, Vihang and {Mobasher}, Bahram and {Nedkova}, Kalina V. and {Primack}, Joel and {Ravindranath}, Swara and {Robertson}, Brant E. and {Rutkowski}, Michael J. and {Sattari}, Zahra and {Soto}, Emmaris and {Yung}, L.~Y. Aaron},
        title = "{UV-bright Star-forming Clumps and Their Host Galaxies in UVCANDELS at 0.5 {\ensuremath{\leq}} z {\ensuremath{\leq}} 1}",
      journal = {\apj},
         year = 2023,
        month = oct,
       volume = {955},
       number = {2},
          eid = {106},
        pages = {106},
          doi = {10.3847/1538-4357/aced3e},
archivePrefix = {arXiv},
       eprint = {2308.00041},
 primaryClass = {astro-ph.GA},
       adsurl = {https://ui.adsabs.harvard.edu/abs/2023ApJ...955..106M}
}

@ARTICLE{2015ApJ...810L..15S,
       author = {{S{\'a}nchez Almeida}, J. and {Elmegreen}, B.~G. and {Mu{\~n}oz-Tu{\~n}{\'o}n}, C. and {Elmegreen}, D.~M. and {P{\'e}rez-Montero}, E. and {Amor{\'\i}n}, R. and {Filho}, M.~E. and {Ascasibar}, Y. and {Papaderos}, P. and {V{\'\i}lchez}, J.~M.},
        title = "{Localized Starbursts in Dwarf Galaxies Produced by the Impact of Low-metallicity Cosmic Gas Clouds}",
      journal = {\apjl},
         year = 2015,
        month = sep,
       volume = {810},
       number = {2},
          eid = {L15},
        pages = {L15},
          doi = {10.1088/2041-8205/810/2/L15},
archivePrefix = {arXiv},
       eprint = {1509.00180},
 primaryClass = {astro-ph.GA},
       adsurl = {https://ui.adsabs.harvard.edu/abs/2015ApJ...810L..15S}
}

@ARTICLE{2013ApJ...779..135W,
       author = {{Wuyts}, Stijn and {F{\"o}rster Schreiber}, Natascha M. and {Nelson}, Erica J. and {van Dokkum}, Pieter G. and {Brammer}, Gabe and {Chang}, Yu-Yen and {Faber}, Sandra M. and {Ferguson}, Henry C. and {Franx}, Marijn and {Fumagalli}, Mattia and {Genzel}, Reinhard and {Grogin}, Norman A. and {Kocevski}, Dale D. and {Koekemoer}, Anton M. and {Lundgren}, Britt and {Lutz}, Dieter and {McGrath}, Elizabeth J. and {Momcheva}, Ivelina and {Rosario}, David and {Skelton}, Rosalind E. and {Tacconi}, Linda J. and {van der Wel}, Arjen and {Whitaker}, Katherine E.},
        title = "{A CANDELS-3D-HST synergy: Resolved Star Formation Patterns at 0.7 < z < 1.5}",
      journal = {\apj},
         year = 2013,
        month = dec,
       volume = {779},
       number = {2},
          eid = {135},
        pages = {135},
          doi = {10.1088/0004-637X/779/2/135},
archivePrefix = {arXiv},
       eprint = {1310.5702},
 primaryClass = {astro-ph.CO},
       adsurl = {https://ui.adsabs.harvard.edu/abs/2013ApJ...779..135W}
}

@ARTICLE{2023ApJ...951..147S,
       author = {{Sattari}, Zahra and {Mobasher}, Bahram and {Chartab}, Nima and {Kelson}, Daniel D. and {Teplitz}, Harry I. and {Rafelski}, Marc and {Grogin}, Norman A. and {Koekemoer}, Anton M. and {Wang}, Xin and {Windhorst}, Rogier A. and {Alavi}, Anahita and {Prichard}, Laura and {Sunnquist}, Ben and {Gardner}, Jonathan P. and {Gawiser}, Eric and {Hathi}, Nimish P. and {Hayes}, Matthew J. and {Ji}, Zhiyuan and {Mehta}, Vihang and {Robertson}, Brant E. and {Scarlata}, Claudia and {Yung}, L.~Y. Aaron and {Conselice}, Christopher J. and {Dai}, Y. Sophia and {Guo}, Yicheng and {Lucas}, Ray A. and {Martin}, Alec and {Ravindranath}, Swara},
        title = "{Fraction of Clumpy Star-forming Galaxies at 0.5 {\ensuremath{\leq}} z {\ensuremath{\leq}} 3 in UVCANDELS: Dependence on Stellar Mass and Environment}",
      journal = {\apj},
         year = 2023,
        month = jul,
       volume = {951},
       number = {2},
          eid = {147},
        pages = {147},
          doi = {10.3847/1538-4357/acd5d6},
archivePrefix = {arXiv},
       eprint = {2305.09021},
 primaryClass = {astro-ph.GA},
       adsurl = {https://ui.adsabs.harvard.edu/abs/2023ApJ...951..147S}
}
\bibliographystyle{aasjournalv7}

\end{document}